\documentclass{elsart}

\usepackage{graphicx}

\usepackage{amssymb}
\usepackage{amsmath}

\begin{document}

\begin{frontmatter}

\title{Superconductivity close to the Mott state: From condensed-matter systems to
superfluidity in optical lattices}

\author[Yale]{Karyn Le Hur} 
\ead{karyn.lehur@yale.edu}
and
\author[ETH]{T. Maurice Rice}

\address[Yale]{Department of Physics, Yale University, New Haven, CT 06520, USA}
\address[ETH]{Theoretical Physics, ETH Z\" urich, CH-8093 Z\" urich, Switzerland}

\begin{abstract}
Since the discovery of high-temperature superconductivity in 1986 by Bednorz and M\" uller, great efforts
have been devoted to finding out how and why it works. From the d-wave symmetry of the order parameter, the importance of antiferromagnetic fluctuations, and the presence of a 
mysterious pseudogap phase close to the Mott state, one can conclude that high-$T_c$ superconductors are clearly distinguishable from the well-understood BCS superconductors. 
The d-wave superconducting state can be understood through a Gutzwiller-type projected BCS wave-function. In this review article, we revisit the Hubbard model at half-filling and focus on the emergence of exotic superconductivity with d-wave symmetry in the vicinity of the Mott  state, starting from ladder systems and then studying the dimensional crossovers to higher dimensions. This allows to confirm that short-range antiferromagnetic fluctuations can mediate superconductivity with d-wave symmetry. Ladders are also nice prototype systems allowing to demonstrate the truncation of the Fermi surface and the emergence of a Resonating Valence Bond (RVB) state with preformed pairs in the vicinity of the Mott state. In two dimensions, a similar scenario emerges from renormalization group arguments. We also discuss theoretical predictions for the d-wave superconducting phase as well as the pseudogap phase, and address the crossover to the overdoped regime. Finally, cold atomic systems with tunable parameters also provide a complementary insight into this outstanding problem.
\end{abstract}

\end{frontmatter}

\pagebreak

{\bf \Large Content}

{\bf 1 Introduction} 

{\bf 2 The Ladder prototype}

2.1 From RVB physics to exotic superconductivity

2.2 Weak-coupling regime

2.3 D-Mott state from weak coupling

2.4 Emergent symmetries

{\bf 3 Quasi-1D solution and Truncation of the Fermi surface}

{\bf 4 Dimensional crossovers and Theories in two dimensions}

4.1 Antiferromagnetic long-range correlations

4.2 Truncated Fermi surface and d-wave superconductivity

4.3 Very large doping: Reminiscence of the Landau-Fermi liquid

4.4 RG in two dimensions, phenomenology of the pseudogap phase, two gaps

4.5 Gutzwiller projected d-wave superconducting state and mean-field theory

4.6 Pseudogap phase, generalized Luttinger sum rule and electron propagator

4.7 Overdoped cuprates and Breakdown of Landau-Fermi liquid

{\bf 5 Cold Atomic Fermions and Hubbard model in optical lattices} 

5.1 Light and Atomic Parameters

5.2 Plaquette models with ultracold fermions and d-wave superfluidity

5.3 Noise Correlations and Entanglement within a BCS pair

5.4 BCS-BEC crossover and pseudogap

{\bf Conclusion and Discussion}

{\bf Appendices with technical details}

\pagebreak

\section{Introduction}

Shortly after the discovery of high temperature superconductivity in cuprates \cite{BM}, Anderson \cite{Anderson} suggested that the key to this perplexing phenomenon resides in the large repulsive (positive) interactions in the copper oxide planes. As a matter of fact, at zero hole doping, the 
single-band Hubbard model captures the insulating behavior of the parent compounds. The origin
of this insulating behavior was described many years ago by Nevill Mott \cite{Mott} as a correlation effect. In the ground state, each Cu site would be in a $d^9$ configuration opening a charge gap
of the order of the Hubbard interaction $U$. Essentially, the electrons become localized as a result
of the Hubbard interaction, and their spins form an antiferromagnetic arrangement
(the N\' eel state) as a result of the virtual hopping of the antiparallel spins from one Cu ion to the next
(the parallel configuration being disallowed by the Pauli exclusion principle) \cite{Affleck}.  It is relevant to observe that the layer symmetry breaks the degeneracy of the $d$ orbitals down to a single $d_{x^2-y^2}$ orbital, so that orbital issues which cause complications in most other transition metal oxides do not play a role here. The energy gain due to this ordering is known as the superexchange energy $J\sim 4t^2/U$, where $t$ is the hopping energy between Cu ions.

The Hamiltonian of the Hubbard model takes the explicit form
\begin{equation}
H=-t\sum_{\langle i;j\rangle,s} \left(d^{\dagger}_{is} d_{js} +h.c.\right) +U \sum_i d^{\dagger}_{i\uparrow}
d_{i\uparrow} d^{\dagger}_{i\downarrow} d_{i\downarrow}.
\end{equation}
Here, $i$ and $j$ run over lattice sites, $\langle ... \rangle$ denotes nearest neighbors, and $s$ the spin 
of an electron. The first part describes the hopping of electrons between adjacent sites on the lattice (formed by the Cu ions) and the second part the on-site repulsion $U>0$ of electrons at the same site. One electron per lattice site corresponds to half-filling (undoped case).  The strong-coupling limit of the Hubbard model results in the $t-J$ model. The two-dimensional (2D) doped Hubbard model has so far resisted a definitive solution, primarily because its spins and holes are highly entangled with no obvious small parameter to separate them. Whether the 2D Hubbard model describes the physics of high-$T_c$ cuprates remains a great subject of interest.  In the cuprates, upon doping with holes the antiferromagnetism is rapidly destroyed
and above a certain level superconductivity irrefutably occurs with $d_{x^2-y^2}$ pairing symmetry, as conclusively demonstrated by phase sensitive experiments \cite{phase1,phase2}. In fact, the low-energy density of states can be directly accessed via  scanning tunnel spectroscopy experiments \cite{Davis}. 
Such higher angular momentum pairs have led to the speculation that the pairing in the cuprates have a different origin from that of the well-understood Bardeen-Cooper-Schrieffer (BCS) superconductors \cite{BCS}. To answer this question, one certainly needs to understand the normal phase at intermediate doping levels between the magnetic and d-wave superconducting phases (consult Fig. 1). 

\begin{figure}
\begin{center}
\includegraphics[width=9.3cm,height=8.1cm]{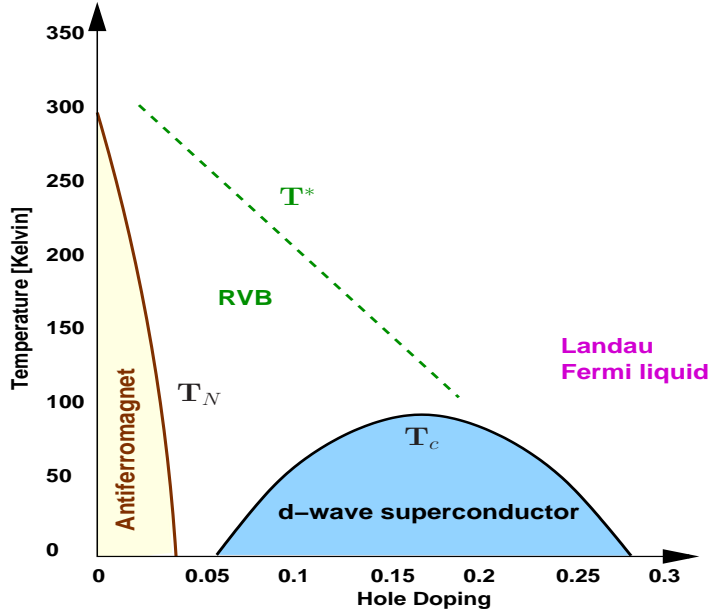}
\end{center}
\caption{Schematic phase diagram of high-$T_c$ superconductors.}
\end{figure}

There are experimental signs of a spin gap opening below a crossover temperature $T^*$ \cite{Alloul,Johnston}. The ultimate nature of the underlying quantum ground state in this portion of the phase diagram, commonly called the `pseudo-gap' regime, is an intriguing puzzle. In 1987, Anderson \cite{Anderson} postulated a description based on a spin liquid of spin singlets which is lightly hole doped. Rather than forming a fixed array of dimers, strong quantum fluctuations lead to a superposition of singlet configurations, {\it i.e.}, the valence bond singlets resonate between different configurations. This is the Resonating Valence Bond (RVB) theory that predicts the occurrence of
preformed pairs above the superconducting transition. This RVB concept explains the decrease of the
uniform spin susceptibility and the reduction of the specific heat below $T^*$. The holes (vacancies) are responsible for transport in the $ab$ plane; the conductivity spectral weight in the $ab$ plane is given by the hole concentration and is therefore unaffected by singlet formation. For $c$-axis
conductivity an electron is transported between planes; since an electron carries a spin-1/2, it is necessary to break a singlet. This qualitatively explains the strong reduction in c-axis transport at $T^*$.

This RVB concept provides a qualitative understanding of the pseudogap region. At a more quantitative level, it has been shown that this RVB theory gives an excellent description of the key features of the superconducting state \cite{Andersonetal,Zhang,Ogata}. On the other hand, the situation for the normal phase (above $T_c$) is less satisfactory; the RVB theory can be formulated in a gauge theory but then approximations are required to evaluate it \cite{Lee}. In fact, the instability of the Landau-Fermi liquid appears to be more than just an instability to superconductivity. For example, unlike the superconducting energy gap, which vanishes at isolated points on the Fermi surface (the d-wave nodes), the pseudogap vanishes along segments of the Fermi surface centered at these nodes, known as Fermi arcs \cite{Norman,Kapitulnik,Kallin,Kanigel}. This strange phase separation in momentum space is a key feature of the pseudogap phase. Nernst effects become appreciable above $T_c$ (but below a temperature $T_{onset}<T^*$) indicating the existence of
preformed pairs above $T_c$ \cite{Ong}. Additionally, as a result of the prominent Hubbard interaction $U$, Mott physics still plays a crucial role as observed in optical transport \cite{Timusk}. Therefore, the instability of the Landau-Fermi liquid in the pseudogap phase could be a precursor of the Mott insulating phase.

This has been a source of motivation to search for (doped) Mott insulators of the spin-liquid sort. 
Generally, spin liquids are more common in low dimensions where quantum fluctuations can suppress magnetism. Quasi-one dimensional (1D) ladder materials \cite{ladder1,ladder2} are promising in this regard and have received extensive attention, particularly the two-leg ladder \cite{Dagotto}. Interacting
fermions moving on a single chain form the celebrated Luttinger liquid. But when chains are assembled
to form ladders this gives rise to remarkable changes in the physics. The two-leg (2-leg) ladder at half-filling is an example of insulating spin liquid that can be regarded as a form of short-range RVB state; there is a charge gap and a spin gap in the strong-coupling \cite{largeU} as well as in the weak-coupling regime \cite{smallU}. This insulating spin liquid phase can be viewed as a quantum disordered
superconductor; it exhibits pairing, with an approximate d-wave symmetry. This phase is then referred to 
as the D-Mott phase \cite{Lin}, which exhibits an enlarged SO(8) symmetry in the weak-coupling regime. Moreover, upon doping, the holes bind in relative d-wave pairs and the two-leg ladder exhibits quasi-long range superconducting (d-wave) pairing correlations \cite{smallU,Leon}. This behavior is reminiscent of that seen in the underdoped cuprate superconductors. The glue for the pairing of holes in the two-leg ladder system may be viewed as the spin gap formation; the holes pair to avoid the breaking of the resonating singlets. On the other hand, it is instructive to notice that spinless fermions can also support unconventional (p-wave) superconductivity in the weak-interaction regime \cite{UrsKaryn}. This result shows that the interchain hopping not only favors single-particle hopping between chains, but it also strongly reinforces the Cooper channel in the ladder problem. 

The three-leg (3-leg) ladder is also very interesting. At half-filling, the strong coupling limit is equivalent
to the 3-leg Heisenberg spin-1/2 antiferromagnetic ladder. This has been well studied and reduces in the low-energy sector to an effective single-chain Heisenberg model with longer range (but unfrustrated
coupling) \cite{Frischmuth}. In fact, it is well established that even-leg ladders have a spin gap and odd-leg ladders have one gapless spinon mode \cite{Dagotto}: this is the well-known odd-even effect for
spin ladders. Interestingly, we have shown that the weak-coupling limit also shows a similar behavior
\cite{UKM}. Here, it is  certainly worth mentioning that while for (undoped) spin ladders, theory and experiment are in agreement \cite{Azuma}, superconductivity has up to now only been observed in a two-leg ladder material under high pressure \cite{ladder2}.
Upon doping the 3-leg ladder, both in the strong \cite{Maurice,Scalapino} and weak-coupling \cite{UKM} limit, one finds that the holes enter the channel with odd parity under reflection about the central leg and form a Luttinger liquid, whereas the even parity channels remain at the stoichiometric filling and continue to form an insulating spin liquid (RVB state). The result is a finite
region of hole doping where the original Fermi surface with 3 bands (or 6 Fermi points) is partially truncated to 2 Fermi points, but without a broken translational symmetry. The key is Umklapp scattering
which introduces a charge gap in even parity channels even away from half-filling. We have also shown that this truncation of the Fermi surface is robust to the inclusion of a small repulsive next-nearest 
neighbor hopping $t'$ \cite{JohnKaryn} which should be added to any realistic (1-band) description of the cuprate systems. The 3-leg ladder is a clear example of a partially truncated Fermi surface through the formation of an RVB state over part of the Fermi surface. However, in the slightly doped 3-leg ladder, the Fermi surface reduces to nodal points and not to Fermi arcs as in high-$T_c$ cuprates.

The ladders exhibit an odd-even effect whereas the 2D system has two gapless magnon modes. It is then interesting to know how the quasi-1D system evolves towards the 2D counterpart. In antiferromagnetic spin ladders, using the non-linear sigma model \cite{Sudip}, one can show that
the spin gap for even-leg ladders disappear as $\exp-N$, whereas at weak $U$, the spin gap disappears double-exponentially as a function of the number of legs $N$ \cite{Urs}. The next question is: ``what is the relation between antiferromagnetic fluctuations and superconductivity?''. The quasi-1D approach, in the limit of large number of legs $N\rightarrow +\infty$, also allows to rigorously show that short-range antiferromagnetic fluctuations stabilize d-wave superconductivity for not too large doping levels \cite{Urs}. A Kohn-Luttinger type attraction \cite{KohnLuttinger} is generated by antiferromagnetic processes like in two 2D Hubbard models \cite{Varma,Scalapino1,Scalapino2}. In two dimensions, renormalization group (RG) methods have been developed to treat weak to strong interactions. The antiferromagnetic and d-wave superconducting phase can be predicted from the examination of the RG flow, {\it e.g.}, with a Fermi surface containing the saddle points $(\pi,0)$ and $(0,\pi)$ (and Fermi surface nesting as well as van Hove singularities) \cite{Schulz,Dzya,Lederer}.  The theory has been pushed further through the functional renormalization group by Honerkamp {\it et al.} \cite{Honerkampetal}. 
A spin-density wave analysis in the ordered state and a discussion on the evolution of the Fermi surface in underdoped cuprates can be found in Ref. \cite{Chubukov}.

Even though a proper (rigorous) theory of the pseudogap phase is still missing, some progress has also been accomplished recently. First, the quasi-1D approach shows that one can still have a truncation of the Fermi surface when $N\rightarrow +\infty$; the antinodal points form an insulating spin liquid similar to the 2-leg ladder whereas the nodal regions are gapless. Moreover, this allows to prove that the antiferromagnetic (RVB) fluctuations (in the antinodal directions) will reinforce d-wave superconductivity at low energy in the vicinity of the nodal directions, resulting in d-wave superconductivity in the nodal regions. Second, by including a small repulsive next-nearest neighbor hopping $t'$, RG arguments show that a similar scenario can emerge in two dimensions close to half-filling due to the effect of umklapp scattering \cite{Furukawa}; this requires an interaction strength larger than $U_c$ where $U_c/t\propto \ln^{-2}(t/t')$. Those approaches suggest the appearance of two energy scales close to the Mott state: $T^*$ embodies the energy scale at which umklapp scatterings
still play a crucial role in the antinodal regions due to the proximity of the Mott state, producing an (insulating) RVB liquid, and $T_c$ represents the superconducting energy scale at which the Fermi arcs become superconducting as a result of the prominent antiferromagnetic fluctuations in the antinodal regions. However, a proper low-energy theory is lacking and has only been explored through numerical techniques \cite{Lauchli}. The main difficulty is to build an unbiased 2D low-energy theory that include all the (divergent) particle-particle and particle-hole channels,
namely, forward, backward (Cooper), umklapp, and antiferromagnetic processes. On the other hand,  a phenomenological theory of the pseudogap phase \cite{zhangnew,Yangnew} can be built by analogy with the ladder systems and an array of coupled ladders \cite{KTM}. The proposal of a Fermi surface consisting of 4 disconnected arcs has parallels to gauge theory calculations for the doped $t-J$ model \cite{Lee,WL}.  

Finally, it is also worth mentioning the recent development of Dynamical Mean Field Theory (DMFT) \cite{Ferrero,Georgesnew} and cluster type approaches \cite{Andre}.

Ultracold atoms in optical lattices are promising simulators of complex many-body Hamiltonians that arise in condensed-matter systems. They embody very clean systems which can be tuned in a very controlled manner from weak to strong coupling \cite{Bloch}. The Hubbard model has been recently
realized with both repulsive and attractive interactions \cite{fermion,Esslinger,Blochnew}. On the other hand, it has been shown for some time that the minimal 2D unit (the 4-site plaquette on the square lattice) can sustain d-wave pairing physics \cite{ST}. The single plaquette has been the starting point of various approaches to the Hubbard model in two dimensions \cite{Altman,Kivelson}. The hole pair that is created has a d-wave character and leads to d-wave superfluidity once the plaquettes are coupled. Cold atomic gases in optical lattices may allow to realize those ``plaquette'' systems \cite{Rey} and then to prove that d-wave superfluidity can resist to inhomogeneous hopping amplitudes (even though one needs to improve the cooling procedure of fermions to access the N\' eel as well as the d-wave superconducting state).

The outline of this review article is as follows. 

First, we focus on ladder systems that have a continuous crossover between weak and strong interactions and show many salient features such as RVB physics and truncated Fermi surface as well as d-wave superconductivity close to the Mott state for spinful electrons. For weak interactions, we underline the emergence of an enlarged symmetry in the low energy sector. Then, we discuss the dimensional crossovers to two dimensions and discuss analytical expressions for the antiferromagnetic Hamiltonian as well as the d-wave superconductor appearing upon doping. For large enough doping, one recovers a Fermi liquid when antiferromagnetic correlations disappear completely. We also describe the pseudogap phase emerging from the quasi-1D approach and show that at low energy the nodal regions are unstable towards d-wave superconductivity as a result of antiferromagnetic fluctuations (or the emergence of RVB physics in the antinodal directions). We continue by reviewing the RG results in two dimensions and mean-field approaches for the strong-coupling doped $t-J$ model. In particular, introducing a Gutzwiller projected BCS wavefunction and resorting to mean-field theory to solve the $t-J$ model, gives a relatively good description of the d-wave superconducting phase. We also suggest that the pseudogap phase might be understood by drawing a parallel with the $N$-leg Hubbard ladder in the weak-coupling regime. We summarize the theoretical results and draw a comparison with experiment, both in the superconducting phase and in the pseudogap phase. We also address the crossover to the overdoped limit and the origin of the breakdown of the Landau-Fermi liquid near the onset of superconductivity. On the other hand, we show that cold atomic systems in optical lattices allow to probe several features of Hubbard models. In particular, we investigate the d-wave pairing in 2D plaquette Hamiltonians which can be implemented in optical lattices.

 In all these Hubbard models --- from homogeneous (Hubbard model in two dimensions) to inhomogeneous models (plaquette and ladder models) ---  antiferromagnetic fluctuations favor a d-wave superconducting state.

In fact, the occurrence of d-wave superconductivity in electron systems developing prominent  spin fluctuations at the wavevector $\vec{Q}=(\pi,\pi)$ (in two dimensions) can be intuited as follows. First consider the BCS gap equation,
\begin{equation}
\Delta_{SC}(\vec{p}) = -\sum_{\vec{p'}} V_S(\vec{p}-\vec{p'})\frac{\Delta_{SC}(\vec{p'})}{2E_{\vec{p'}}},
\end{equation}
where $V_S$ embodies the singlet channel interaction\footnote{$\Delta_{SC}$ represents the superconducting gap and $E_{\vec{p}}=(\Delta_{SC}^2(\vec{p}) +\xi_{\vec{p}}^2)^{1/2}$ ($\xi_{\vec{p}}$ embodies the kinetic term where we have subtracted the chemical potential).} (which is modified by antiferromagnetic spin fluctuations) \cite{reviewS}. When the interaction is attractive $(V_S<0)$ and slowly varying over the Fermi surface, one can see that this inevitably results in s-wave superconductivity, {\it i.e.}, the gap has the same sign over all the Fermi surface. Now, let us consider repulsive interactions. Spin fluctuations will make the singlet channel interaction more positive at large momentum transfer $\vec{p}-\vec{p'} = \vec{Q} = (\pi,\pi)$. Now, suppose $\Delta_{SC}(\vec{p'})$ is positive near $(0,\pi)$ then this produces a negative gap at $(\pi,0)$, so that a $d_{x^2-y^2}$ gap develops. As shown below, this comes about from band-structure nesting effects in two dimensions close to half-filling or because of strong-coupling,  short-range valence bond correlations such as in the (2-leg) Hubbard or $t-J$ ladders. 
It is also interesting to note that in 1986 (approximately at the same time than the discovery of high-$T_c$ cuprates), Emery suggested that back-scattering from spin fluctuations could lead to the pairing of holes in the Bechgaard salts \cite{Emery}. In
addition, three papers in 1986 argued that antiferromagnetic
spin fluctuations might mediate d-wave pairing in heavy fermion
materials \cite{Varma,Scalapino1,Scal}. 

\section{The Ladder prototype}

Even though the Hubbard chain  does not support unconventional superconductivity for
repulsive interactions, the two-leg ladder supports unconventional superconductivity (p-wave
superconductivity for spinless electrons \cite{UrsKaryn} and d-wave superconductivity for spinful electrons). First, we give a justification to the occurrence of RVB physics and d-wave superconductivity
in the two-leg spin (or $t-J$) ladder. Then, we discuss the weak-coupling limit and introduce the D-Mott phase, that is very similar to the insulating RVB state emerging for large interactions, but it exhibits a large SO(8) symmetry \cite{Lin}. 

\subsection{From RVB physics to exotic superconductivity}

A pedagogical way to understand the emergence of RVB physics and d-wave superconductivity in ladders is through the large onsite interaction limit (or in the weak-interaction limit when the Mott gap in the chains exceeds the interchain tunneling amplitude \cite{KarynMott}). At half-filling, the system then is equivalent to a two-leg spin ladder described by a superexchange $J=4t^2/U>0$ along the chains and an intrachain superexchange $J_{\perp}=4t_{\perp}^2/U>0$ \cite{Dagotto}. Let us start with the limit $J_{\perp}\gg J$. To minimize energy, the spins of the two chains will form singlet bonds, resulting in an RVB-type wavefunction:
\begin{equation}
|RVB>\ = {P}_{D=0} \sum_{ i;j} F(i-j) (d^{\dagger}_{1i\uparrow} d^{\dagger}_{2j\downarrow} -d^{\dagger}_{1i\downarrow}d^{\dagger}_{2j\uparrow})|Vac>,
\end{equation}
where $d_{1(2)is}$ annihilates an electron in chain $i$ of the ladder, $|Vac\rangle$ refers to the vacuum,  and ${P}_{D=0}$ is the Gutzwiller's projector \cite{Gutzwiller} leaving only singly-occupied sites with spins,
\begin{equation}
{P}_{D=0} = \prod_{i}\prod_{\alpha=1,2} (1-n_{i\uparrow}^{\alpha} n_{i\downarrow}^{\alpha}).
\end{equation}
By construction, the function $F(i-j)$ decays exponentially for distances larger than $\xi \propto 1/J_{\perp}$.
From this simple analysis, one can conclude that in the limit $J_{\perp}>J$, the two-leg ladder has a Mott
gap of the order of $U$ and a spin gap $\sim J_{\perp}$. Additionally, the (gapped) $S=1$ magnon excitation, corresponding to break a singlet, has a dispersion of the form \cite{triplet} (of course, the magnon is massive due to the prominent spin gap $\sim J_{\perp}$):
\begin{equation}
\omega(k) = J_{\perp} + J\cos(k) + \frac{J^2}{4 J_{\perp}}(3-\cos(2k))+...\  .
\end{equation}
\begin{figure}
\begin{center}
\includegraphics[width=8cm,height=6.5cm]{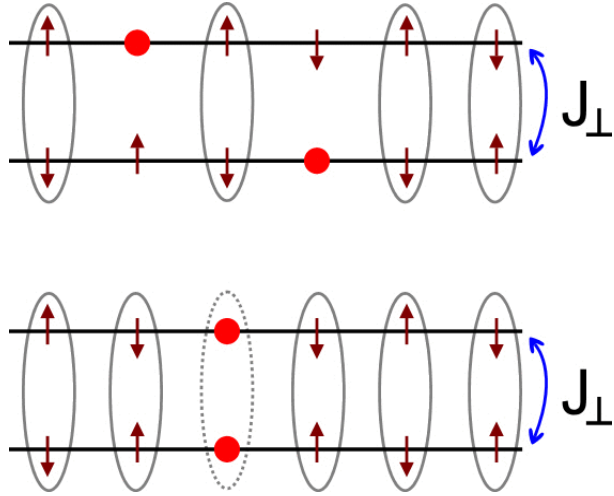}
\end{center}
\caption{Two-leg ladder: the holes pair to minimize the cost of magnetic energy.}
\end{figure}
Here, $k$ refers to the wavevector associated with the direction along the chains, and
for simplicity, {\it the lattice spacing will be fixed to $1$}. The minimum of energy is at $k=\pi$ emphasizing the antiferromagnetic correlations along the chains. To see
the connection with d-wave superconductivity, it is useful to rewrite this rung-singlet state in terms
of the {\it bonding} (1) and {\it antibonding} (2) operators:
\begin{equation}
d_{1/2 i s} = \frac{1}{\sqrt{2}}\left(\Psi_{1i s} \pm \Psi_{2i s}\right).
\end{equation}
Since the function $F$ decays exponentially with space, one may approximate $|RVB\rangle$ by the
state (the electron spins across the rungs of the ladder are locked into singlets):
\begin{equation}
|RVB>\ \sim \sum_{i} (\Psi^{\dagger}_{1i\uparrow} \Psi^{\dagger}_{1i\downarrow} -\Psi^{\dagger}_{2i\uparrow}\Psi^{\dagger}_{2i\downarrow})|Vac>.
\end{equation}
This is equivalent to add a singlet (Cooper) pair into the bonding and antibonding orbitals. This paired form is suggestive of d-wave superconductivity. When viewed in momentum space, the ground state of
a superconductor is a product of singlet pairs with zero center of mass momentum at different points
around the Fermi surface. In an s-wave superconductor the pairs are all added with the same sign, but
if the pairs are formed with a relative angular momentum ({\it e.g.}, d-wave) sign changes are expected.
The connection with d-wave superconductivity will be made more concrete in the next subsection when
representing the bonding and antibonding bands in the two-dimensional Brillouin zone. To conclude
this analysis at large $J_{\perp}/J\gg 1$, when equally doping the two chains of the ladder\footnote{Here, we consider homogeneous doping in the two chains (below, $\delta$ represents the hole density per site). On the other hand, a substantial asymmetry in the chemical potentials of the two chains is well-known to be pair-breaking \cite{Sigrist}.}, one expects
the emergence of d-wave superconductivity in the system to minimize the magnetic energy (see Fig. 2). 

It is also instructive to observe that close to half-filling superconducting correlations will exhibit a universal form. In the limit of large $J_{\perp}$ one may define the superconducting order parameter
as:
\begin{equation}
\Delta_i = (d^{\dagger}_{1i\uparrow} d^{\dagger}_{2i\downarrow} -d^{\dagger}_{1i\downarrow}d^{\dagger}_{2i\uparrow}).
\end{equation}
Owing to the fact that in one dimension, one cannot have a true symmetry breaking, in the d-wave superconducting phase one expects:
\begin{equation}
\langle \Delta^{\dagger}_i \Delta_j \rangle \propto |i-j|^{-\eta_{scd}},
\end{equation}
where $\eta_{scd}\ll 2$ (when superconducting fluctuations are prominent). Close to half-filling, the exponent $\eta_{scd}$ can be intuited as follows \cite{Schulzpair}. The low-energy Hilbert space consists of rungs of the ladder where either both sites are occupied and form a singlet or both sites are empty. Singly occupied sites or rungs with a triplet lie higher by an energy of order $J_{\perp}$. It is convenient to define the state where all rungs are occupied as the vaccum such that an empty rung can be visualized as the creation of a (hard-core) boson $b^{\dagger}_i=\Delta_i$. In second order perturbation theory in $t$ and $J$ then one obtains the low-energy Hamiltonian:
\begin{equation}
H_{eff} = -t_{eff} \sum_i (b^{\dagger}_i b_{i+1} +h.c.) + V_{eff} \sum_i n_i n_{i+1},
\end{equation}
where $n_i = b^{\dagger}_ i b_i\leq 1$, $t_{eff} = 8 t^2/(3 J_{\perp})$, and $V_{eff} = (16 t^2/3 - 3 J^2/8)/J_{\perp}$. In the dilute limit for the hole pairs (close to half-filling) one can neglect the effect of the intersite interaction 
$V_{eff}$ and the Hamiltonian turns into a model of free fermions through the Jordan-Wigner transformation. By analogy to the XY chain, then one predicts the universal correlations \cite{Schulzpair}:
\begin{equation}
\langle  \Delta^{\dagger}_i \Delta_j \rangle \propto |i-j|^{-1/2}.
\end{equation}
This power law close to half-filling (and the RVB picture at half-filling) appears for all values of 
$J_{\perp}/J \neq 0$ \cite{Tsvelik}. (Remember that superconductivity has up to now only been observed in a two-leg ladder material under high pressure \cite{ladder2}.)

\subsection{Weak-coupling regime}

Below, we demonstrate thoroughly that the physics still remains the same in the weak-coupling regime $U\ll (t_{\perp},t)$. We apply RG techniques in the interaction $U$ along the lines of our Ref. \cite{UrsKaryn}. In this limit, it is a good approach to diagonalize the kinetic term first; going to the momentum space, one gets two bands 
(the bonding and antibonding bands; see Eq. (6)). The
kinetic term takes the precise form:
\begin{equation}
H_0 = \sum_{i=1,2} \int dk E_i(k) \Psi^{\dagger}_i(k) \Psi_i(k),
\end{equation}
where the dispersion relations obey:
\begin{equation}
E_i(k) = \mp t_{\perp} -2t\cos(k).
\end{equation}
Band 1 is the bonding band and band 2 the antibonding band. See fig. 3.

By analogy with the two-dimensional case, the associated transverse momenta are denoted 
$k_{\perp}=(0,\pi)$\footnote{For the 2-leg ladder, we choose the convention $E_i(k)=-t_{\perp}\cos k_{\perp}-2t\cos k$ where $k_{\perp}=0$ for the bonding band and $k_{\perp}=\pi$ for the antibonding band. This allows us to make an analogy with the antinodal regions of the 2D half-filled Fermi surface.} (consult Fig. 4). Since we focus on the low-energy physics, we linearize the dispersion
relation $E_i$ around the Fermi momenta $\pm k_{Fi}$, which are fixed by the chemical potential
$\mu=E_i(k_{Fi})$ and the filling $n$ corresponding to the electron density per site. At half-filling $E_i(k_{Fi})=0$.

\begin{figure}
\begin{center}
\includegraphics[width=8cm,height=5.5cm]{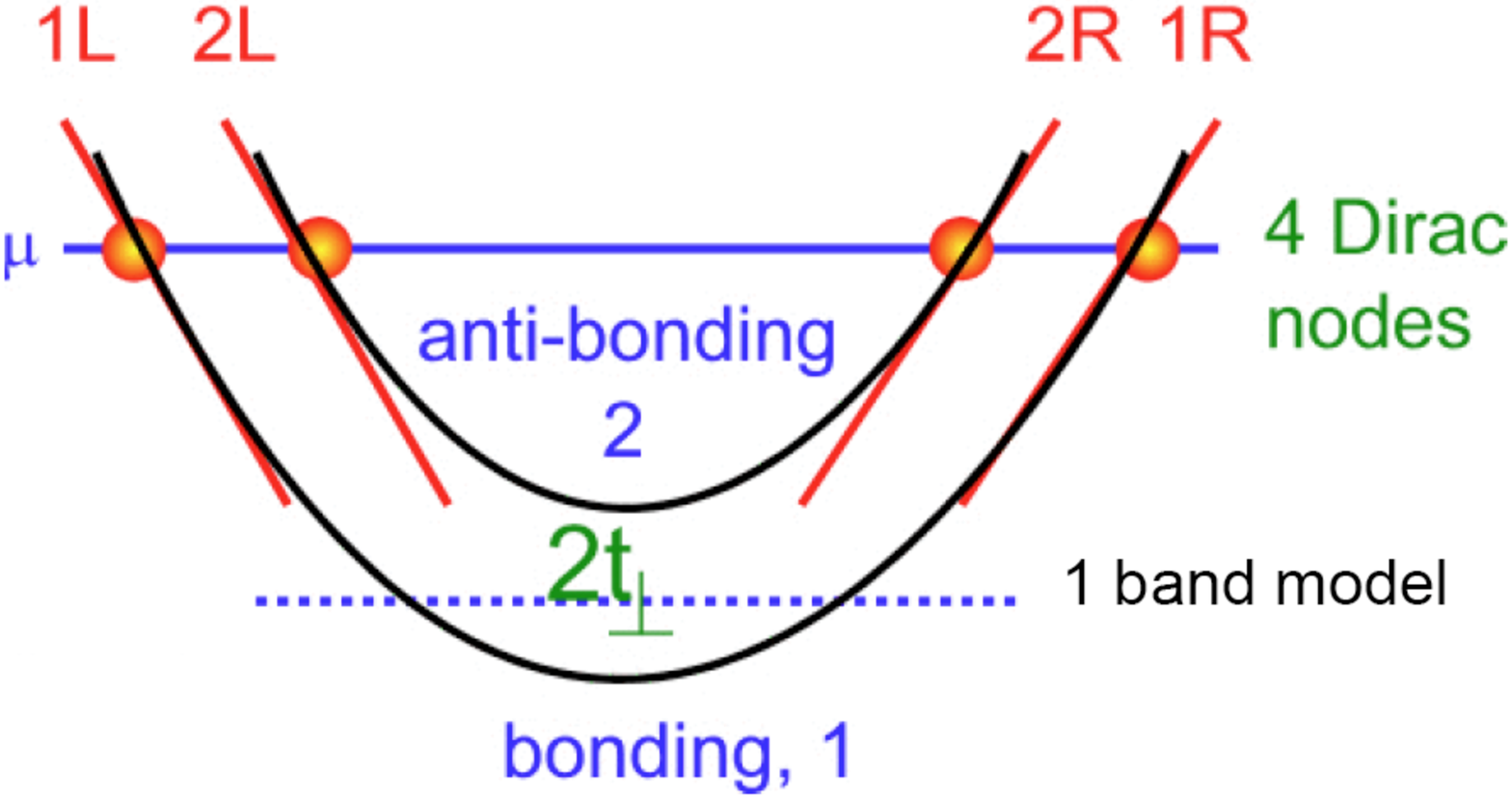}
\end{center}
\caption{Band structure of the two-leg ladder.}
\end{figure}

We first study the spinless fermion case where $(k_{F1}+k_{F2})=2n\pi$ and the RG equations can be solved analytically \cite{UrsKaryn}. The Fermi wavevectors are given by:
\begin{equation}
k_{Fi} = \pi n \pm \arcsin\left(\frac{t_{\perp}}{2t\sin(\pi n)}\right).
\end{equation}
The Fermi velocities thus obey:
\begin{equation}
v_1-v_2 = \frac{2t_{\perp}}{\tan(\pi n)}.
\end{equation}
It should be noted that for spinless ladders, the filling is $0\leq n \leq 1$ and the hole doping (hole density per site) away
from half-filling corresponds to $\delta=0.5-n$. Now, the effect of the interchain hopping is totally included in the two velocities $v_1$ and $v_2$ which are only equivalent at half-filling. In this subsection, we do not consider the half-filled case, $(k_{F1}+k_{F2})=\pi$ and focus on the system away from the insulating regime where the difference in the velocities (or a finite $t_{\perp}$ value) has the remarkable effect of driving the system to a superconducting state for repulsive interactions $U>0$.

Including all interactions allowed by symmetry (leaving away completely chiral interactions that do not affect the fixed point properties; see Appendix A), in momentum space, the Hamiltonian takes the form $H=H_0+H_{Int}$ where
\begin{equation}
H_0 = \sum_{i=1,2} v_i \int dk k \left( \Psi^{\dagger}_{iR}(k) \Psi_{iR}(k) - \Psi^{\dagger}_{iL}(k)\Psi_{iL}(k)\right),
\end{equation}
and
\begin{eqnarray}
H_{Int} &=& \int dk_1 dk_2 dk_3 dk_4 \delta(k_1+k_3-k_2-k_4) \\ \nonumber
&\times& [c_1 \Psi^{\dagger}_{1R}(k_1) \Psi_{1R}(k_2) \Psi^{\dagger}_{1L}(k_3) \Psi_{1L}(k_4) + c_2(1\leftrightarrow 2) \\ \nonumber
&+&f_{12} (\Psi^{\dagger}_{1R}(k_1)\Psi_{1R}(k_2) \Psi^{\dagger}_{2L}(k_3)\Psi_{2L}(k_4) + 1\leftrightarrow 2) \\ \nonumber
&+&c_{12}(\Psi^{\dagger}_{1R}(k_1) \Psi_{2R}(k_2) \Psi^{\dagger}_{1L}(k_3) \Psi_{2L}(k_4) + 1\leftrightarrow 2)].
\end{eqnarray}
Here, $f_{12}$ describes the interband forward scattering, $c_{12}$ the interband Cooper coupling, and
$c_i$ the intraband Cooper coupling. Keep in mind that in this subsection we drop out all the umklapp terms (see Fig. 4) because we assume that the system is away from half-filling.

For spinless fermions, the bare values are given by (Appendix A):
\begin{equation}
c_1=c_2=f_{12}=c_{12}=U>0.
\end{equation}
Those Cooper and forward interaction channels are reminiscent of the coupling channels in a 2-patch
model in two dimensions; see Fig. 4 and Fig. 10.

\begin{figure}
\begin{center}
\includegraphics[width=9.5cm,height=9cm]{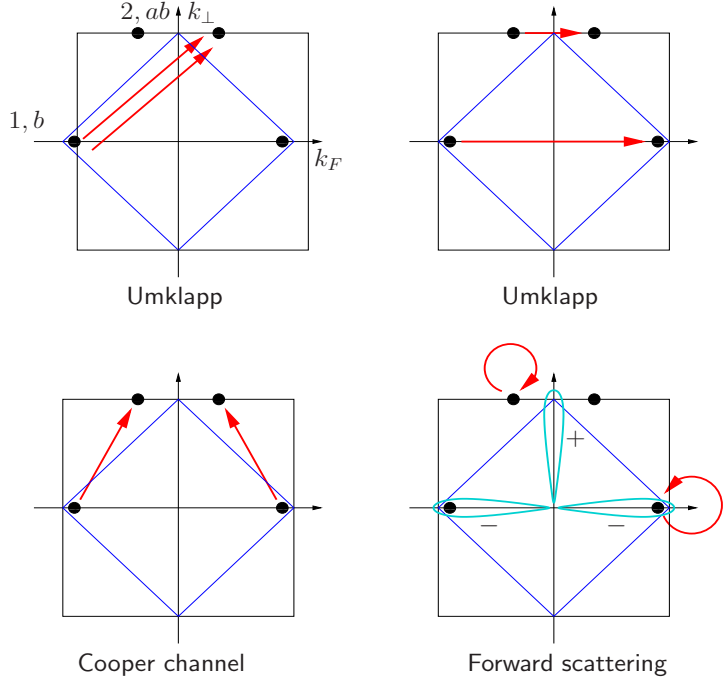}
\end{center}
\caption{Channels of interaction for the two-leg ladder in momentum space. A two-leg ladder produces
4 Fermi points in momentum space (it should be noticed that in this figure we have chosen $k_{F1}$ and $k_{F2}$ {\it arbitrarily} but we have
set $k_{\perp}=0$ for the bonding band and $k_{\perp}=\pi$ for the antibonding band.) In this basis, d-wave superconductivity emerges because the superconducting order parameter changes of sign when
going from one band to the other (or under a $\pi/2$ rotation).}
\end{figure}

To second order in the interaction $U$, the perturbative RG equations read \cite{UrsKaryn}:
\begin{eqnarray}
\frac{dc_1}{dl} &=& -\frac{1}{2\pi v_2} c_{12}^2 \\ \nonumber
\frac{dc_2}{dl} &=& -\frac{1}{2\pi v_1} c_{12}^2 \\ \nonumber
\frac{d f_{12}}{dl} &=& \frac{1}{\pi (v_1+v_2)} c_{12}^2 \\ \nonumber
\frac{d c_{12}}{dl} &=& \frac{c_{12}}{\pi}\left(\frac{2f_{12}}{v_1+v_2} -\frac{c_1}{2 v_1} -\frac{c_2}{2 v_2}\right).
\end{eqnarray}
Those RG equations can be found quite easily using the field theory rules of Eq. (A.13).
The energy scale (temperature scale) is related to $l$ through $E\sim te^{-l}$.
The plus and minus signs result from particle-hole respectively and particle-particle diagrams. For specific details, consult Appendix A.

Keeping only the terms quadratic in the coupling constants, the set of 4 differential equations turn into one differential equation for $f_{12}$. First, using all the RG equations, it is relatively easy to derive:
\begin{equation}
\frac{ d c_{12}}{dl} = c_{12}\left[A f_{12} -\frac{(v_1+v_2)U}{\pi v_1 v_2}\right],
\end{equation}
where
\begin{equation}
A = \frac{4 v_1 v_2 + (v_1+v_2)^2}{2\pi v_1 v_2(v_1+v_2)}>0.
\end{equation}
Then, resorting to the third equation, one can show that,
\begin{equation}
\frac{1}{A} \frac{d f_{12}}{dl} = \left(f_{12}-BU\right)^2 + C U^2,
\end{equation}
where
\begin{eqnarray}
B &=& \frac{2(v_1+v_2)^2}{4 v_1 v_2 + (v_1+v_2)^2}>0 \\ \nonumber
C &=& \frac{-v_1^4 + 6 v_1^3 v_2 + 6 v_1^2 v_2^2 + 6 v_1 v_2^3 -v_2^4}{\left(4 v_1 v_2 + (v_1+v_2)^2\right)^2}.
\end{eqnarray}

The solution of Eq. (22) is different for $C<0$ and $C>0$. For $C>0$, all the couplings diverge whereas for $C<0$, $c_{12}$ flows to zero and all the other couplings remain small ($\sim U$). In terms of the velocities, $C>0$ means:
\begin{equation}
1/7 < v_1/v_2< 7.
\end{equation}
One can check that the weak-coupling fixed point appearing for $C<0$ is stable at all the orders because whatever the order is, the RG equations are always multiplied at least once with $c_{12}$.

The low-energy physics can be analyzed thoroughly through standard bosonization techniques \cite{Tsvelik,Giamarchi}. The Hamiltonian takes the form \cite{UrsKaryn}:
\begin{eqnarray}
H &=& \sum_{i=1,2} \int dx \left(\frac{v_i}{2} +\frac{c_i}{4\pi}\right)\left(\partial_x\phi_i\right)^2 +
\left(\frac{v_i}{2}-\frac{c_i}{4\pi}\right)\Pi^2_i \\ \nonumber
&+&\int dx\  \frac{f_{12}}{2\pi} \left(\partial_x\phi_1 \partial_x \phi_2 -\Pi_1 \Pi_2\right) - \frac{c_{12}}{(2\pi)^2}
\cos\left(\sqrt{4\pi}(\theta_1-\theta_2)\right).
\end{eqnarray}
(Again, the lattice spacing is fixed to $1$ and the Klein factors ensuring the correct anticommutation rules for the fermions have been chosen as $\eta_{1R}\eta_{1L}\eta_{2R}\eta_{2L}=1$.) Here, $\partial_x\phi_i$ represents the charge fluctuations in each band and $\theta_i$ corresponds to the conjugate superfluid phase; we have introduced $\Pi_i=\partial_x \theta_i$. A flow to strong coupling of $c_{12}$ (quasiclassically) results in the ``pinning'' of the superfluid phase $\theta_1-\theta_2=0$ in order to minimize the energy, and in a single gapless mode, while for $c_{12}\rightarrow 0$, two gapless modes are present. 

A superconducting phase is in fact stabilized when $c_{12}$ flows to strong couplings, {\it i.e.}, for
$v_1/v_2<7$. The key point is to resort to the canonical transformations $\phi_{\pm} = (\phi_1\pm \phi_2)/\sqrt{2}$ and $\Pi_{\pm} = (\Pi_1\pm \Pi_2)/\sqrt{2}$. The mode $\theta_-$ is pinned to zero due to the prominent term $c_{12}$ whereas the symmetric mode $\phi_+$ and $\theta_+$ remain gapless. The effective Hamiltonian at low energy takes the form,
\begin{equation}
H_{+} = \int dx\ \frac{v_+}{2}\left(\frac{1}{K_+}(\partial_x\phi_+)^2 + K_+\Pi_+^2\right) + \lambda\partial_x\phi_+ \partial_x\phi_- ,
\end{equation}
We have ignored a term of the form $\Pi_+ \Pi_-$ because this term is irrelevant when $c_{12}$ flows to strong couplings. We identify:
\begin{eqnarray}
v_{\pm} &=& \sqrt{\left(\frac{v_1+v_2}{2}\right)^2 - \left(\frac{c_1+c_2\pm 
2f_{12}}{4\pi}\right)^2} \\ \nonumber
K_{\pm} &=& \sqrt{\frac{2\pi(v_1+v_2) - (c_1+c_2\pm 2f_{12})}{2\pi(v_1+v_2) + (c_1+c_2\pm 2f_{12})}}. \\ \nonumber
\lambda &=& \frac{v_1-v_2}{2} + \frac{c_1-c_2}{4\pi}.
\end{eqnarray} 
We have introduced the velocity and the Luttinger parameter of the antisymmetric bosonic mode because they will be used below\footnote{At the fixed point, as a result of $f_{12}>0$ and $c_i<0$, $K_->0$ that will favor the pinning of the superfluid phase difference $\theta_1-\theta_2$.}. 

Now, we discuss the main correlation functions in the problem. For repulsive interactions, the charge density and superconducting pairing fluctuations with the most divergent susceptibilities are\footnote{At half-filling, for spinless fermions, the ground state is a charge density wave with density alternating between the chains $(n=1/2)$. The symmetric phase $\theta_+$ becomes disordered due to the umklapp scatterings.}:
\begin{eqnarray}
O_{cdw} &=& d^{\dagger}_{1R} d_{1L} - d^{\dagger}_{2R} d_{2L} = \Psi^{\dagger}_{2R} \Psi_{1L}
+\Psi^{\dagger}_{1R} \psi_{2L} \\ \nonumber
&\propto& \exp\left(i\sqrt{2\pi}\phi_+\right)\cos\left(\sqrt{2\pi}\theta_-\right),
\end {eqnarray}
and
\begin{eqnarray}
O_{sc} &=& d_{1R} d_{2L} + d_{2R} d_{1L} = \Psi_{1R} \Psi_{1L} - \Psi_{2R} \Psi_{2L} \\ \nonumber
&\propto& \exp\left(i\sqrt{2\pi}\theta_+\right) \cos\left(\sqrt{2\pi}\theta_-\right),
\end{eqnarray}
where $d_{1L/R}$ represent the annihilation operators for the fermions in chain $i$. {\it Note that interchain superconductivity means intraband superconductivity}. Additionally, for spinless fermions, the superconducting order parameter  must be {\it odd} under parity, $O_{sc}(-x)=-O_{sc}(x)$, which is
reminiscent of p-wave like superconductivity. The {\it cdw} and {\it sc} correlation functions are given by:
\begin{eqnarray}
\langle O^{\dagger}_{cdw}(x) O_{cdw}(0)\rangle \propto x^{-\gamma} \\ \nonumber
\langle O^{\dagger}_{sc}(x) O_{sc}(0)\rangle \propto x^{-1/\gamma},
\end{eqnarray}
where \cite{UrsKaryn}:
\begin{equation}
\gamma = \frac{K_+}{1-\frac{K_+ K_-}{2 v_+ v_-}\lambda^2}.
\end{equation}
In the superconducting phase, to leading order in $t_{\perp}/t$ and in the doping $\delta=0.5-n$, one can approximate \cite{UrsKaryn}:
\begin{equation}
\gamma \approx 1+\frac{\pi^2}{8}\left(\frac{t_{\perp}}{t}\right)^2 \delta^2.
\end{equation}
It is important to observe that as a result of the finite interchain hopping $t_{\perp}\propto \lambda$
leading to $\gamma>1$, the system will develop superconducting pairing correlation functions
when $v_1/v_2<7$ (for $t_{\perp}=t$, this means $0<\delta<0.22$). 

\begin{figure}
\begin{center}
\includegraphics[width=9.5cm,height=7cm]{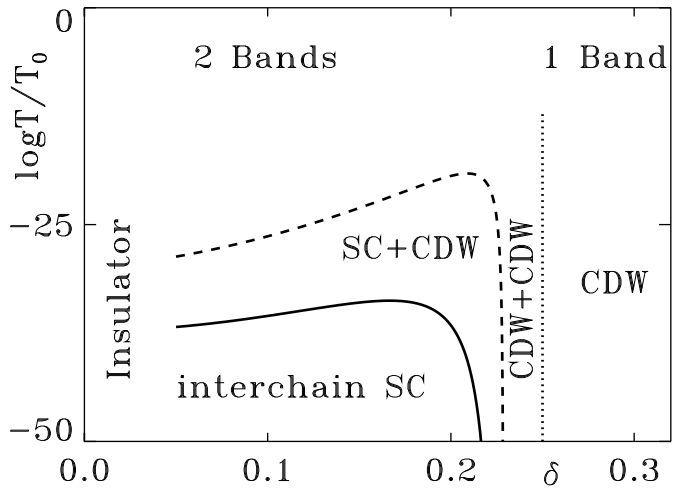}
\end{center}
\caption{Our phase diagram of the spinless ladder for $t_{\perp}/t=1$ and $U/t=0.2$ \cite{UrsKaryn}. $T_0\sim t$ represents a high-energy cutoff. The solid line shows the crossover to the
(p-wave) superconducting phase where phase coherence between the bands arises (close to half-filling, this occurs at the temperature scale $T_c\sim te^{-2\pi t/U}$) and the dashed line indicates when superconducting fluctuations appear in band $1$.}
\end{figure}

For $v_1/v_2>7$, the fixed points of the couplings are such that $c_{12}=0$, and $c_1$, $c_2$, and
$f_{12}$ are of the order of $U$. It is interesting to observe that in a short range of ratios, $7<v_1/v_2<8$, the low temperature fixed point is such that $c_1<0$ (as a result of the small velocity $v_2$) whereas $c_2>0$, implying that the bonding band 1 will still develop prominent superconducting fluctuations while the (antibonding) band 2 will exhibit charge density wave fluctuations. In this case, the superconducting correlations occur both
between the chains as well as along the chains. This phase can be seen as a precursor of the
superconducting phase occurring at $v_1/v_2<7$: the system develop phase fluctuations but no phase
coherence between the bands. Finally, for ratios $v_1/v_2>8$, the couplings $c_j>0$ and as a result
in both bands charge density wave fluctuations dominate. This corresponds to the usual metallic phase in 1D: the Luttinger phase.

Our phase diagram of the spinless ladder is shown in Fig. 5 \cite{UrsKaryn}.

This analysis can also be useful for other two-band systems such as carbon nanotubes on a superconducting substrate \cite{KSC} or quantum wires \cite{Julia}.
Now, we extend the analysis to the case of fermions with spin following Ref. \cite{Leon}, and show that the system will develop superconducting d-wave fluctuations as well as a prominent spin gap, as a reminiscence of the large $U$ analysis.

Away from half-filling the interaction channels in the band basis take the compact form:
\begin{eqnarray}
H_{Int} &=& \int dx\ \left(f_{ij}^{c} J_{Rii} J_{Ljj} - f_{ij}^s \vec{J}_{Rii}\vec{J}_{Ljj} + c_{ij}^{c} J_{Rij}
J_{Lij} - c_{ij}^s \vec{J}_{Rij}\vec{J}_{Lij}\right) \\ \nonumber
&+& \sum_{i=1,2} \int dx\ \left( c_{ii}^c J_{Rii} J_{Lii} - c_{ii}^{s} \vec{J}_{Rii} \vec{J}_{Lii}\right).
\end{eqnarray}
The only diffference with the spinless case is that now the system develops Cooper and forward scattering channels in the spin sector as well. For a precise relation between the $c_{ii}^c$ and 
$c_{ii}^s$ parameters and the usual g-ology of the Luttinger theory, consult Appendix A.
We have exploited the U(1) symmetry for charge and SU(2) symmetry for spin, introducing the charge and spin densities:
\begin{equation}
J_{pij} = \sum_s \Psi^{\dagger}_{ips} \Psi_{jps}\ \hbox{and}\ 
J^r_{pij} = \frac{1}{2} \sum_{s,s'} \Psi^{\dagger}_{ips}\sigma^r_{ss'} \Psi_{pjs'},
\end{equation}
where $p=(L,R)$ denotes the direction (chirality), $i$ represents the band, $\sigma^r$ denote the Pauli matrices, and $s$ the spin. By symmetry, one gets $c_{ij}^{c,s} =
c_{ji}^{c,s}$ and $f_{ij}^{c,s}=f_{ji}^{c,s}$. Here, the Hubbard bare values are given by (see Appendix A):
\begin{equation}
4 f_{ij}^c = f_{ij}^s = 4 c_{ij}^c = c_{ij}^s = U.
\end{equation}

In contrast to the spinless case, the (one-loop) RG equations cannot be solved analytically, such that
one needs to perform a numerical integration. On the other hand, a characteristic property of the N-leg
ladders is that the couplings flow towards {\it universal} ratios as the initial value $U/t$ is reduced; 
in particular, away from half-filling and at energies smaller than $E\sim te^{-t/U}$, one gets \cite{Leon}:
\begin{equation}
t\sim 4c_{12}^c = c_{12}^s,\ c_{11}^s/v_1=c^s_{22}/v_2\sim -1,\  \hbox{and}\ f_{12}^s\approx 0.
\end{equation}
One can bosonize the Hamiltonian; the low-energy fixed point is governed by:
\begin{eqnarray}
\tilde{H}_{Int} &=& \int dx\ -|c_{11}^s| \cos(\sqrt{8\pi}\phi_{1s}) - |c_{22}^s|\cos(\sqrt{8\pi}\phi_{2s}) 
\\ \nonumber
&-& 2c_{12}^s 
\cos(\sqrt{4\pi}\theta_{c-})\left(\cos(\sqrt{4\pi}\phi_{s-}) + \cos(\sqrt{4\pi}\phi_{s+})\right).
\end{eqnarray}
Minimizing the energy leads to the pinning of the spin modes in each band $\phi_{1s}=\phi_{2s}=0$ (or $\phi_{s+}=(\phi_{1s}+\phi_{2s})/\sqrt{2}=0$ and $\phi_{s-}=(\phi_{1s}-\phi_{2s})/\sqrt{2}=0$) and of the antisymmetric charge mode $\theta_{c-}$ similar to the spinless case\footnote{In particular, the fixed point leads to the following Luttinger parameters: $K_{s-}<1$, $K_{s+}>0$, and $K_{c-}>1$.}. Since both spin modes are pinned, this shows that the ladder has a spin gap when $U$ is small. 

The total charge mode remains gapless similar to the spinless case and one also finds that the
superconducting pairing correlation function is the most dominant one. The (singlet) pair-field operator
reads:
\begin{equation}
\Delta_j = \Psi_{jR\uparrow} \Psi_{jL\downarrow} + \Psi_{jL\uparrow} \Psi_{jR\downarrow} \propto \eta_{j\uparrow}\eta_{j\downarrow} e^{-i\sqrt{2\pi}\theta_{jc}} \cos(\sqrt{2\pi}\phi_{js});
\end{equation}
we have introduced the Klein factors $\eta_{js}$ 
(ensuring the correct anticommutation rules between two fermionic operators)
\cite{Tsvelik,Giamarchi} such that $\eta_{1\uparrow}\eta_{1\downarrow}\eta_{ 2\uparrow}\eta_{2\downarrow}=1$.

One obtains that the pairing
correlation function decays as:
\begin{equation}
\langle \Delta_i^{\dagger}(x) \Delta_j(0)\rangle \propto x^{-1/(2K_{c+})},
\end{equation}
where $K_{c+}\approx 1$ is the Luttinger parameter of the total charge sector. We check that the pairing correlations in the spin-1/2 case decay as $\propto x^{-1/2}$, whereas all the other correlation functions decay approximately as $x^{-2}$. The spin-1/2 two-leg ladder is superconducting both for $t_{\perp}>U$
and for $t_{\perp}<U$, whereas for the spinless ladder superconductivity disappears when $t_{\perp}\rightarrow 0$ or $\lambda\rightarrow 0$. 

Furthermore, for spinful electrons, one can check that the superconductivity has indeed an approximate d-wave symmetry meaning that the pair-field operator has a different sign in band 1 and in band 2, {\it i.e.}, $\langle \Delta^{\dagger}_1 \Delta_2\rangle <0$.

\subsection{D-Mott state from weak coupling}

Similar to the $t-J$ ladder, the two-leg Hubbard ladder at half-filling is an insulating spin liquid or
a Mott insulator with a spin gap. In fact, the D-Mott insulator \cite{Lin} is a concrete example of disordered d-wave superconductor: the phase coherence between the bands is already present, but the total charge mode is gapped due to the presence of extra umklapp scattering processes.

More precisely, since $k_{F1}+k_{F2}=\pi$ at half-filling, this allows for the additional two-particle umklapp processes (for an illustration, see Fig. 4):
\begin{eqnarray}
H_{Int,hf} &=& \int dx\ \Big(u_{1221}^c(I^{\dagger}_{R12} I_{L21} +h.c.) - u_{1221}^s(({\vec I}_{R12})^{\dagger}\cdot {\vec I}_{L21}+h.c.) \\ \nonumber
&+& u_{1122}^c(I_{R11}^{\dagger}I_{L22} + I_{R22}^{\dagger}I_{L11}+h.c.)\Big).
\end{eqnarray}
We have introduced the umklapp operators:
\begin{equation}
I_{pij} = \sum_{s,s'} \Psi_{pis} \epsilon_{ss'} \Psi_{pjs'},
\end{equation}
where $\epsilon = -i\sigma^y$ and $I_{pij}=I_{pji}$, and:
\begin{equation}
I^r_{pij} = \frac{1}{2} \sum_{s,s'} \Psi_{pis} (\epsilon \sigma^r)_{s s'} \Psi_{pjs'}. 
\end{equation}
The initial bare values are given by (consult Appendix A, {\it i.e.}, Eq. (A.19)):
\begin{equation}
2 u_{k \bar{k} \bar{k} k}^c = 8 u_{kk \bar{k} \bar{k}}^c = U\ \hbox{and}\ u_{k\bar{k}\bar{k} k}^s =0.
\end{equation} 
The D-Mott phase is dominated by the following set of couplings (see Fig. 6):
\begin{equation}
t\sim 4c_{12}^c = 4 f_{12}^c = c_{12}^s = - c_{ii}^{s} = 4 u_{1221}^c = 8 u_{1122}^c = u_{1221}^s,
\end{equation}
\begin{figure}
\begin{center}
\includegraphics[width=8.5cm,height=7.5cm]{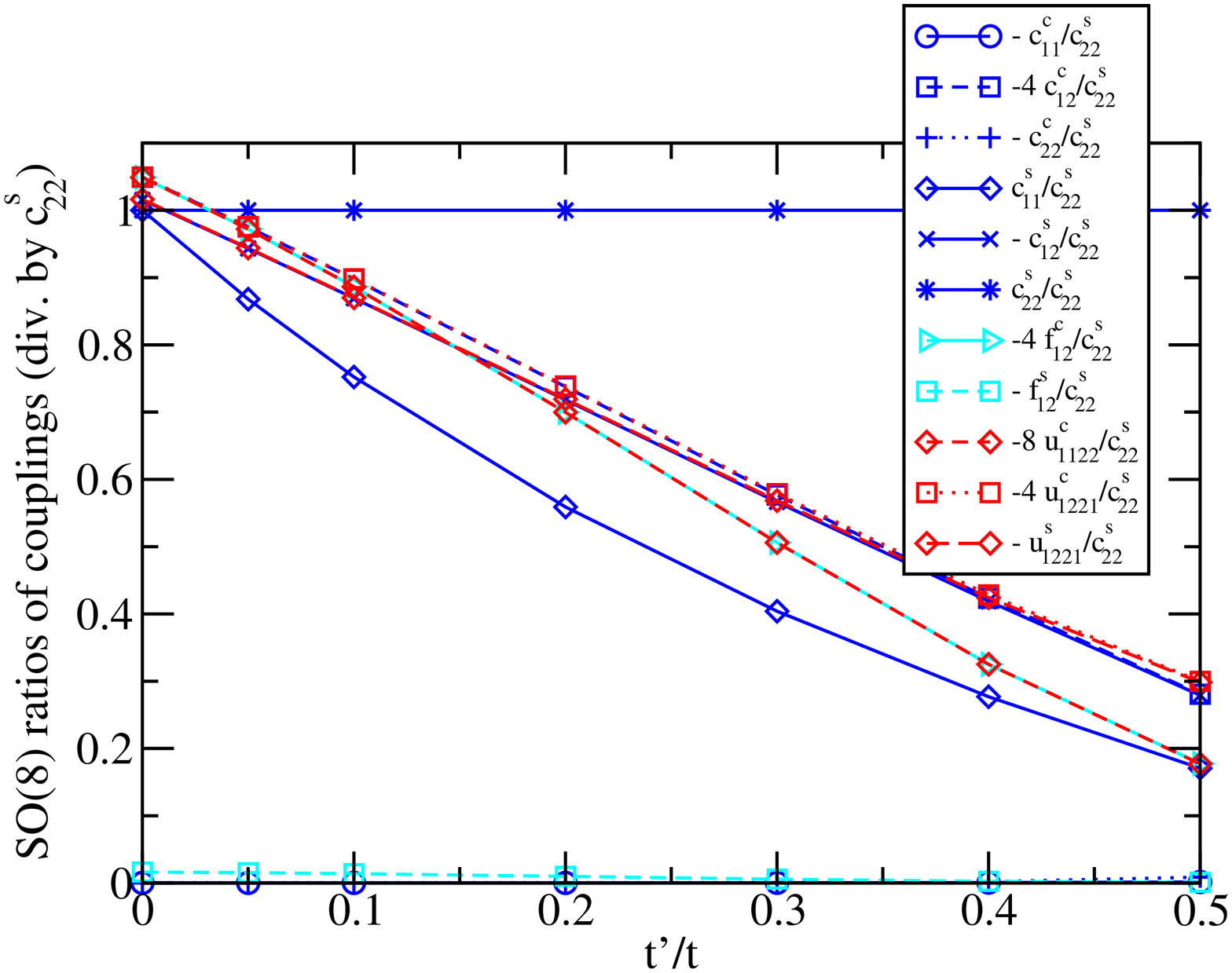}
\end{center}
\caption{Coupling ratios in the half-filled two-leg Hubbard ladder versus the next 
nearest-neighbor hopping $-t'$ following the analysis of our Ref. \cite{JohnKaryn}. For a reasonable value of $t'\sim 0.1-0.2t$, the system remains insulating and still sustains a spin gap.}
\end{figure}
and $c_{11}^c=c_{22}^c \approx f_{12}^s \approx 0$. {\it Both the Cooper
and elastic umklapp channels flow to strong couplings at half-filling; therefore, to properly describe the
low-energy fixed point, one needs an exact strong-coupling theory that can distinguish between superconductivity or insulating phase}. In one dimension, one can apply bosonization \cite{Tsvelik,Giamarchi}. The bosonized form of the umklapp interactions is:
\begin{eqnarray}
H_{Int,hf} = \int dx\ -u_{1221}^s \cos(\sqrt{4\pi}\phi_{c+})\Big(\cos(\sqrt{4\pi}\phi_{s-}) +\cos(\sqrt{4\pi}\theta_{s-}) && \\ \nonumber
+2\cos(\sqrt{4\pi}\phi_{s+})\Big) && \\ \nonumber
- 4\cos(\sqrt{4\pi}\phi_{c+})\Big(u_{1221}^c (\cos(\sqrt{4\pi} \phi_{s-})
-\cos(\sqrt{4\pi}\theta_{s-})) && \\ \nonumber
+4u_{1122}^c\cos(\sqrt{4\pi}\theta_{c-})\Big). &&
\end{eqnarray}
Since $u_{1221}^c=u_{1221}^s/4$ in the D-Mott phase, one verifies that the system
has still a spin gap: both $\phi_{s-}$ and $\phi_{s+}$ are pinned to zero. Additionally, the phase coherence between bands is already present through the pinning of the antisymmetric superfluid
mode $\theta_{c-}$. On the other hand, at half-filling, the total charge mode $\phi_{c+}$ is also pinned
that makes the system insulating: when the mode $\theta_{c-}$ is pinned, the current density only involves the total charge mode $j=v_{c+}K_{c+}\Pi_{c+}$, and half-filling when $\phi_{c+}$ is pinned this  implies $j=0$. In the weak-coupling limit, the charge gap and the spin gap are equal $(\sim te^{-v_1/U})$.

The pair field phase $\theta_{c+}$ being the conjugate field will fluctuate wildly. These quantum fluctuations will destroy the power-law 1D superconducting phase, leading to an exponential decaying
pair-field correlation function. In fact, further progress can be accomplished by re-fermionizing the
low-energy Hamiltonian \cite{Lin}: the Hamiltonian turns into an SO(8) Gross-Neveu model that has been well studied by particle field theorists. The SO(8) Gross-Neveu has a peculiar symmetry (triality \cite{Shankar2}) that allows to access the energies of various excited states. In particular, it has been shown that the energy of the lowest excited state with the quantum numbers of an electron is equal to the energy of the Cooper pair. This demonstrates pairing in the D-Mott phase. 

Finally, it is also important to underline that the charge gap and the spin gap are robust if one includes
a moderately small (frustrated) next-nearest neighbor $-t'$ hopping with $t'>0$ (relevant in high-$T_c$ cuprates); see Fig. 6 and our Ref. \cite{JohnKaryn}. In particular, for a relatively small $t'\sim 0.1-0.2t$, 
the RG flow indicates that the system still exhibits a large symmetry and only the couplings 
$c_{11}^s$ and $c_{22}^s$ progressively deviate from the other (diverging) couplings. 

\subsection{Emergent symmetries}

In Sec. 2.3, we have focused on properties of the D-Mott phase occurring in the weak-coupling regime of the two-leg Hubbard ladder; it exhibits a remarkable SO(8) symmetry \cite{Lin}. Interestingly, there exists an SO(5) subalgebra of the SO(8) group whose vector representation unifies
superconductivity and antiferromagnetism \cite{Lin}. Thus the SO(5) symmetry, proposed by Zhang \cite{ZhangS} as a phenomenological model for the cuprates, is shared by the D-Mott phase (and by the d-wave superconducting phase in the Hubbard ladder that maintains an SO(6) symmetry \cite{Schulzso6}).  Below, we discuss the SO(5) theory.

Let us consider two sites and define the fermion operators $c$ and $d$ on those
two sites. The antiferromagnetic order parameter is defined as:
\begin{equation}
m^{\alpha} =\frac{1}{2}(c^{\dagger}\sigma^{\alpha} c - d^{\dagger} \sigma^{\alpha} d), 
\end{equation}
$\sigma^{\alpha}$ represent the Pauli matrices acting on the spin space
and the `intersite' superconducting order parameter reads:
\begin{equation}
\Delta^{\dagger} = -\frac{i}{2} c^{\dagger}\sigma^y d^{\dagger} = \frac{1}{2}(-c^{\dagger}_{\uparrow} d_{\downarrow}^{\dagger} + c^{\dagger}_{\downarrow} d^{\dagger}_{\uparrow}).
\end{equation}
One can group these five components and form a vector: $(n_1,n_2,n_3,n_4,n_5)$ where $n_1=(\Delta^{\dagger}+\Delta)/2$, $n_2=m^1$, $n_3=m^2$, $n_4=m^3$, and $n_5=(\Delta^{\dagger}-\Delta)/2i$ called the {\it superspin} (it contains both superconducting and antiferromagnetic spin
components). The SO(5) theory is inspired from the Landau-Ginzburg theory; on the other hand, the SO(5) theory extends the Landau-Ginzburg approach in several ways. First, it assumes an approximately
symmetric interaction potential between the antiferromagnetism and superconducting phases (in the
underdoped regime of the cuprates). Second, it introduces a full set of dynamically variables canonically conjugate to the superspin (including the total spin, the total charge, and the so called $\pi$ operators
that perform rotations from antiferromagnetism to superconductivity and vice versa) \cite{Demler}.
Through the introduction of the dynamically conjugate variables, the SO(5) theory is capable of describing quantum disordered phases. 

Microscopic SO(5) models can be explicitly constructed from ladder or bilayer systems \cite{SZH}. Introducing the two sites $c$ and $d$ (for the ladder: $c$ belongs to a rung and $d$ to the other rung), the most general interaction Hamiltonian is:
\begin{eqnarray}
H_{2 sites} &=& U\left(n_{c\uparrow}-\frac{1}{2}\right)\left( n_{c\downarrow}-\frac{1}{2}\right) + (c\rightarrow d) + V(n_c -1)(n_d-1) \\ \nonumber
&+& J\vec{S}_d\vec{S}_c -\mu(n_c+n_d),
\end{eqnarray}
where $n_c=\sum_{\alpha} n_{c\alpha}$ and $\alpha=\ \uparrow$ or $\downarrow$. In fact, an important condition ensures the local SO(5) symmetry: $J=4(U+V)$. Lin, Balents, and Fisher have explored the possibility of SO(5) low-energy fixed points in the weak-coupling regime of the two-leg Hubbard ladder \cite{Lin}; when interactions are repulsive they found that the SO(5) system falls into the basin of attraction of the D-Mott phase.

On the other hand, the behavior of the spin-spectral function is different in the ladder system and in the SO(5) theory. The SO(5) theory predicts a sharp resonance at energy $2\mu$ and momentum $(\pi,\pi)$, the so-called $\pi$ resonance (originally introduced by Zhang \cite{ZhangS} to explain the $42meV$ neutron scattering peak in the superconducting cuprates). Lin, Balents, and Fisher have shown that, in
addition to SO(5) symmetry, the delta-function $\pi$ resonance requires a non-zero condensate density in the superconducting phase. Since condensation is not possible in one dimension, this precludes a delta-function $\pi$-resonance (however a weaker algebraic singularity is not precluded \cite{Lin}).

Regardless of the behavior in the vicinity of $\omega=2\mu$, one expects spectral weight at energies below $2\mu$ but above the spin gap.

Interestingly, the optical conductivity at hal-filling exhibits a high-frequency preformed pair continuum as well as a sharp excitonic peak below the gap as a signature of the enlarged SO(8) symmetry \cite{Lin,Konik}. 

\section{Quasi-1D solution and Truncation of the Fermi surface}

Interestingly, the weak-coupling RG approach close and at half-filling can be extended to $N$ bands with $N>2$ \cite{UKM}. The relation between the chain $i$ and the band $j$ reads:
\begin{equation}
d_{is}(x) = \sum_j \sqrt{\frac{2}{N+1}} \sin\left(\frac{\pi j i}{N+1}\right)\Psi_{js}(x).
\end{equation}
We have exploited the open boundary conditions in the direction perpendicular to the chains (legs), such that the transverse eigenfunctions are standing waves. This brings the Hamiltonian into a diagonal form in momentum space, and for a number of legs $N>2$, the dispersion relation of a given band $j$ is:
\begin{equation}
\label{bandstructure}
E_j(k) = -2t\cos k -2t_{\perp}\cos\left(\pi j/(N+1)\right).
\end{equation}

The Fermi momenta in each band $k_{Fj}$ are determined by the chemical potential $\mu_j=E_j(k_{Fj})$ and the filling $n=2(\pi N)^{-1} \sum_j k_{Fj}$. At half-filling $(\mu=0)$:
\begin{equation}
k_{Fj} = \pi - \arccos\left(\frac{t_{\perp}}{t}\cos\left(\frac{\pi j}{N+1}\right)\right).
\end{equation}
From the band structure, one can attribute a fictitious transverse momentum to each band\footnote{In fact, the transverse momenta are $\pm \pi j/(N+1)$.} $\pi j/(N+1)$. It is relevant to note that for $N$ odd, there is always a band, namely
$j=(N+1)/2$ that lies in the nodal direction at $(\pi/2,\pi/2)$ on the 2D half-filled Fermi surface whereas for $N$ even all bands lie away from the nodal points. For an illustration, see Fig. 7.  One can check that at half-filling and for $t_{\perp}=t$, all the Fermi points belong to the 2D umklapp surface.

Again, since we are only interested in the low-energy physics, we linearize the dispersion relation
at the Fermi surface, resulting in the Fermi velocities $v_j = 2t\sin(k_{Fj})$. At half-filling, one gets
the important equalities (the lowerscript $k$ referring to a given band $k$ should not be confused with the wave-vector $k$):
\begin{equation}
v_k = v_{\bar{k}} = 2\sqrt{t^2 - t_{\perp}^2 \cos(\pi k/(N+1))^2},
\end{equation}
where:
\begin{equation}
\bar{k} = N+1-k.
\end{equation}
{\it In the following, we will denote  the bands $(k,\bar{k})$ band pairs.}

It is important to note that the Fermi velocities fulfill:
\begin{equation}
v_1=v_N < v_2=v_{N-1} <...\ .
\end{equation}
The velocities are always maximum for bands in the vicinity of the nodal directions.
For $N$ not too large, these values of the velocities will lead to a hierarchy of energy scales, as shown below. Note also that at half-filling, the Fermi momenta of band pairs add up $(k_{Fk}+k_{F\bar{k}})=\pi$ allowing interband umklapp processes to take place, similar to the two-leg Hubbard ladder. 

At half-filling, the interacting part of the Hamiltonian consists of forward, Cooper, and umklapp processes
within a band and between different bands similar to the 2-leg ladder, as well as extra (non-)umklapp processes between 4 bands (or 3 bands for an odd number of chains). On the other hand, when the number of chains is quite small, say, $N=3,4,...$, one can rigorously show that the 4-band couplings (or 3-band couplings) remain {\it arbitrary small} at low energy; consult Appendix B and details in our Ref. \cite{UKM}.
This is an important characteristics of the small $N$ limit (on the other hand, the 4-band couplings will become important in the limit of large $N$ close to and at half-filling).

Integrating the RG equations\footnote{The RG equations are given in our Ref. \cite{UKM}.}, we find that the Fermi velocities $v_k$ lead to a hierarchy of energy scales, $E_k^* \sim t e^{-\alpha_k v_k/U}$ with $\alpha_k\sim 1$, where the band pairs $(k,\bar{k})$ become successively frozen out and form a D-Mott state. For example, if one starts with the 3-leg ladder where, say, $t_{\perp}/t\rightarrow \sqrt{2}$ such that $v_1=v_3\ll v_2$, one gets the following rigorous results \cite{UKM}. Firstly, the renormalization of single-band and two-band interaction channels involving the bands $1$ and $3$ is exactly the same as for the two-leg ladder and one gets the emergence of a D-Mott state at the energy scale $E_1^*\sim te^{-\alpha_1 v_1/U}$. Second, there is an umklapp term for the band $2$ since  $k_{F2}=\pi/2$, and this opens a charge gap (but no spin gap) in band $2$ at the energy scale $E_2^*\sim t e^{-\alpha_2 v_2/U}\ll E_1^*$. At half-filling, this allows us to conclude that the system has a gapless spin mode in perfect accordance with the odd-even effect for the spin ladders \cite{Dagotto}.

\begin{figure}
\begin{center}
\includegraphics[width=13.5cm,height=4cm]{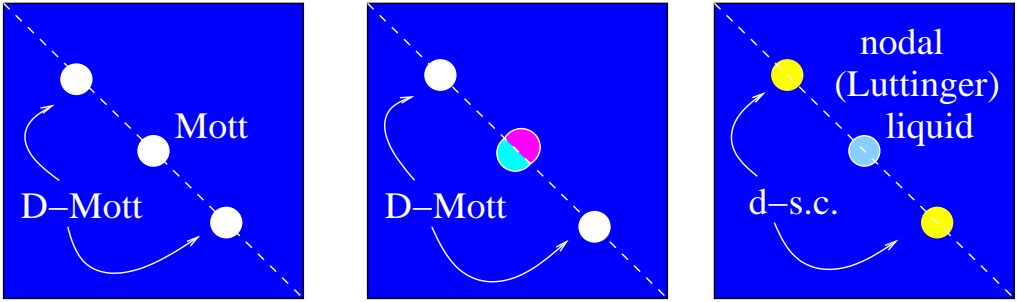}
\end{center}
\caption{The bands of the 3-leg (3-chain) Hubbard model for $t_{\perp}=t$
on the 2D umklapp surface. Band $2$ lies in the nodal direction at $(\pi/2,\pi/2)$. At half-filling, the system is insulating and is analogous to the Heisenberg chain at low energy. By slightly doping, one gets a truncated Fermi surface: the holes enter first the band with the {\it lowest} charge gap ({\it i.e.}, band 2) whereas the bands $1$ and $3$ still form an insulating spin liquid (D-Mott state). When increasing the doping further, one gets a 1D d-wave superconductor;  band $2$ then embodies the nodal (Luttinger) liquid.}
\end{figure}

When doping, the first hole will enter into band $2$ and the chemical potential jumps from 0 to 
$E_2^*$. Band $2$ forms a (Luttinger) liquid whereas bands 1 and 3 still form a D-Mott state and remain insulating\footnote{This is consistent with the Lieb-Schultz-Mattis theorem \cite{Oshikawa}; when applied to the 3-leg ladder, this theorem predicts the existence of gapless excitations at a wavevector $3\pi(1-\delta)$ away from half-filling. When holes enter only the ``odd parity'' channel (band 2 electron operator is odd under parity with respect to reflection about the middle chain)
gapless excitations will have a wavevector
$\pi(1-3\delta)$. These two wavevectors are equivalent modulo $(2\pi)$ so there is no inconsistency.}; this demonstrates the possibility of a truncated Fermi surface close to half-filling in Hubbard-type models, as a result of the band structure which favors the hierarchy of velocities $v_1=v_N<v_2=v_{N-1}<...\ $ and then a non-uniform Mott gap along the two-dimensional Fermi surface. When increasing the doping further, as soon as the chemical potential will reach $E_1^*$ then the bands 1 and 3 will form a d-wave superconductor whereas the band 2 will develop a nodal liquid (or a Luttinger liquid in one dimension). This fixed point is stable until very low energy scales because a nodal point cannot develop an Andreev process (where two paired electrons from the antinodal band $1$ or $3$ hop onto band $2$ and produce a superconducting proximity effect in band $2$) due to the d-wave symmetry \cite{KarynAndreev}. Finally, the effect of doping on the 3-band system is illustrated in Fig. 7. 

For the half-filled 4-leg ladder, the bands $1$ and $4$ form a D-Mott state with a gap $E_1^*$ and the bands $2$ and $3$ also form a D-Mott state with a smaller gap $E_2^*$. The system can be viewed
as a D-Mott state with a non-uniform Mott and spin gap along the 2D half-filled Fermi surface.
The whole system develops a spin gap at half-filling. As in 3-leg ladders, the channels decouple in pairs
so that the energy to add a hole varies around the Fermi surface. Upon doping, the holes initially enter
with the smallest energy gap. Therefore, the bands $2$ and $3$ first develop
d-wave type superconducting correlations whereas the bands $1$ and $4$ still develop a D-Mott gap. {\it This is a concrete example where Mott physics can coexist with superconductivity.}
By increasing the doping, the system will develop a full d-wave superconducting gap and finally phase coherence between the band pairs will occur when adding a small Josephson coupling between band pairs\footnote{For a large number $N$ of chains, the antiferromagnetic 4-band couplings automatically
 induce phase coherence between band pairs, {\it i.e.}, d-wave superconductivity.}. Additionally, due to the phase coherence between band pairs $i$ and $j$, one expects (here, we assume that interactions are very weak):
 \begin{equation}
 \langle \Delta^{\dagger}_i(x) \Delta_j(0)\rangle \propto x^{-1/4}.
 \end{equation}
 
 In fact, this successive opening of the Fermi surface with doping starting from the diagonals in the Brillouin zone will continue for a N-leg ladder system (for a discussion regarding the limit $N\rightarrow
 +\infty$, see Sec. 4.2).
 
  \begin{figure}
\begin{center}
\includegraphics[width=7.5cm,height=6cm]{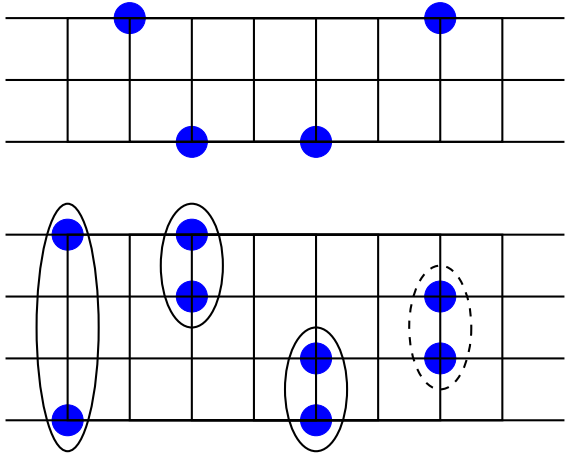}
\end{center}
\caption{When slightly doping the 3-leg ladder, the holes are situated on the outer legs (that means band $2$). For the 4-leg ladder, the highest probability to find a spin singlet is when the two holes belong to the outer legs; the lowest probability to find a singlet is on the (inner) legs $2$ and $3$; consult Eq. (\ref{prob}) and our Ref. \cite{UKM}.}
\end{figure}
 
Close to half-filling, it is also instructive to rewrite the band operators in terms of the chain operators. For
the 3-leg ladder, one can check that the band $2$ annihilation electron operator is odd under parity (with respect to reflection about the middle chain)
\begin{equation}
\Psi_{2ps} = \frac{1}{\sqrt{2}}(d_{1ps}-d_{3ps}),
\end{equation}
traducing that the unpaired holes will fill the outer legs of the ladder, as illustrated in Fig. 8. This is in
accordance with the numerical results of Refs. \cite{Maurice,Scalapino22} for the $t-J$ Hamiltonian; it should be noted that in the case of strong interactions it is necessary to use numerical techniques to analyze the problem. 

For the 4-leg
ladder close to half-filling, the superconducting order parameters of bands $2$ and $3$ are in accordance with d-wave pairing, $\langle \Delta_2 \Delta_3^{\dagger}\rangle \approx -1$, and 
\begin{eqnarray}
\label{prob}
\Delta_2 &\propto& 0.22\left(d_{1R\uparrow}d_{2L\downarrow} + d_{2R\uparrow} d_{1L\downarrow}
+R\leftrightarrow L\right) \\ \nonumber
&+& 0.22\left(d_{3R\uparrow}d_{4L\downarrow} + d_{4R\uparrow}d_{3L\downarrow} +R\leftrightarrow L\right) \\ \nonumber
&-& 0.36\left(d_{1R\uparrow}d_{4L\downarrow} + d_{4R\uparrow} d_{1L\downarrow} +R\leftrightarrow L\right) \\ \nonumber
&-& 0.14\left(d_{2R\uparrow} d_{3L\downarrow} + d_{3R\uparrow}d_{2L\downarrow} + R\leftrightarrow L\right).
\end{eqnarray}
The singlets are on the top two legs, the bottom two legs, on the legs $1$ and $4$, and with the
lowest probability on the legs $2$ and $3$, similarly as found in Ref. \cite{Scalapino3} for the $t-J$ model; consult Fig. 8. The 4-leg ladder in the strong interaction limit was also analyzed by Siller {\it et al.} \cite{Siller}. The possibility of bi-pairing and stripe phase in the 4-leg ladder has been studied in detail \cite{Chang}. Finally, the phases with $N=5,6$ legs are in agreement with numerical works \cite{Scalapino3}, emphasizing the universal behavior emerging in the Hubbard ladders.

\section{Dimensional crossovers and Theories in two dimensions}

Now, we study thoroughly how the weakly interacting $N$-leg Hubbard ladder evolves towards the two-dimensional case as $N\rightarrow +\infty$. It is relevant to note that for $t=t_{\perp}$, $v_1=v_N\sim
2\pi t/N$ (consult Eq. (52)), leading to a singular behavior for large $N$ (the quasi-1D analog of the van Hove singularities in two dimensions).  A crucial difference between the quasi-1D and 2D case are the interactions which control the low-energy physics. Our quasi-1D approach is valid as long as the energy difference between two neighboring bands is larger than the largest energy scale in the system, {\it i.e.}, $te^{-t/U}$. Using the band structure of Eq. (\ref{bandstructure}) and considering $t\sim t_{\perp}$, this implies that:
\begin{equation}
U\ll t/\ln N, t_{\perp}/\ln N.
\end{equation}
Assuming this is fulfilled, one can take $N\rightarrow \infty$ and focus on the 
crossovers. Below, we thoroughly analyze the quasi-long range antiferromagnetic correlations 
emerging for the half-filled case and the d-wave superconducting state occurring away from half-filling.
Close to half-filling, the situation is identical to the small $N$ limit: one gets a truncated Fermi surface
and the antinodal directions form an (insulating) RVB state with preformed pairs. The RG equations are a simple generalization of the 3-leg RG equations given in our Ref. \cite{UKM} and for the sake of clarity technical details are hidden in Appendix B.  We will also discuss 2D approaches in the weak-coupling and strong-coupling regimes, which also support the RVB scenario close to half-filling (pseudogap phase). We like to emphasize that the d-wave superconducting state can be understood through a Gutzwiller-type projected BCS wave-function \cite{Zhang}. We discuss the electron's Green function in the pseudogap phase by analogy with the $N$ leg ladder system \cite{UKM} and with the array of coupled ladders \cite{KTM}. We also address the crossover to the Fermi liquid regime taking place when the umkapp scattering (antiferromagnetic) processes vanish completely, making the Cooper channel irrelevant for purely repulsive interactions \cite{Shankar}. 

\subsection{Antiferromagnetic long-range correlations}

For a finite and even number of legs (bands) $N$, the ground state at half-filling is an insulating spin liquid characterized by the decoupling of band pairs $(k,\bar{k})$. On the other hand, when increasing  $N$, since $v_k-v_{k+1}\sim 1/N$, this decoupling of band pairs will be suppressed and the ``4-band'' (antiferromagnetic) interactions of Fig. 9 will start to play a dominant role. Below, we show that this results in a uniform Mott gap and in long-range antiferromagnetic correlations. 

In fact, the forward spin coupling $f_{k\bar{k}}^s$ is prominent in the (quasi-1D) antiferromagnetic phase. The sum of the couplings $R_k = c_{kk}^s + f_{k\bar{k}}^s$ then can be used to distinguish between the insulating D-Mott phase and the antiferromagnetic phase. More precisely, as shown earlier in the context of the two-leg ladder, in the insulating spin liquid (D-Mott) regime one expects that $R_k<0$ as a result of the growing of $c_{kk}^s$ (towards negative values) whereas in the antiferromagnetic phase one finds that SU(2) single-band Cooper processes remain small (but they have become attractive) $|c_{kk}^s|\ll t$. Moreover, $f_{k\bar{k}}^s$ now grows in the antiferromagnetic phase explaining that $R_k$ becomes positive (see, for example, Ref. \cite{Urs}). On the other hand, there is no broken symmetry in the quasi-1D antiferromagnetic phase and a spin liquid phase can still re-appear at half-filling at (very) low temperatures (below the antiferromagnetic phase). 

The energy scale at which the spin liquid appears is precisely defined as the energy scale $E_c(N)$ at which $R_1\approx 0$\footnote{The band pair $(1,N)$ dominates in the RG process as the result of the van Hove singularity; $v_1=v_N\sim 2\pi t/N$.}. In fact, a rigorous analysis shows that the spin liquid only survives for very small energies smaller than \cite{Urs}:
\begin{equation}
E_c(N) \sim t \exp(-a\exp(b N)),
\end{equation}
where $a\ll 1$ and $b$ is a constant of the order of unity (that depends on the ratio $t_{\perp}/t$). Note
that for the other band pairs $(k,\bar{k})$, the corresponding energy scales are smaller since the velocities $v_k$ are larger. For the spin ladder, the spin gap at large $N$ decreases as,
$J\exp(-0.68 N)$ \cite{Sudip}; while for large $U$ the spin gap decreases {\it exponentially}, for small
$U$ the decrease is {\it double-exponentially}. 

\begin{figure}
\begin{center}
\includegraphics[width=9cm,height=4.5cm]{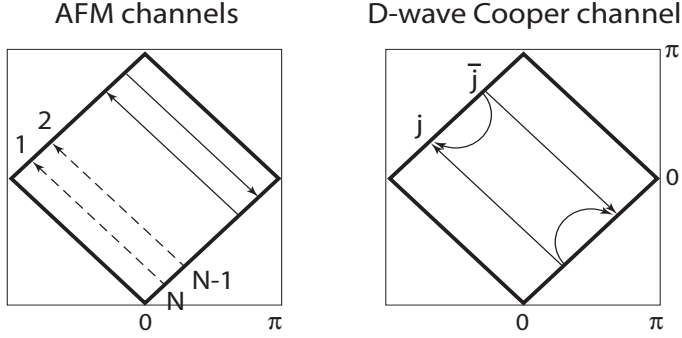}
\end{center}
\caption{4-band processes and D-wave Cooper channel in the large $N$ limit.
The square corresponds to the umklapp surface or the Fermi surface at half-filling $(t=t_{\perp})$. {\it Left}: There are two types of antiferromagnetic (AFM) processes: umklapp processes (dashed arrows) and particle-hole processes (solid arrows). {\it Right}: The AFM processes which take place within a band pair are identical to the Cooper processes.}
\end{figure}

In Appendix B.2, we study the phase for energies larger than $E_c(N)\rightarrow 0$ and show
that indeed it can be viewed as a precursor of the 2D antiferromagnetic phase.
The antiferromagnetic type ground state occurs at an energy scale
\begin{equation}
E_{AFM} \approx te^{-ct/U},
\end{equation}
where $c$ is a function of $t_{\perp}/t$; $E_{AFM}$ is enhanced for $t_{\perp}\rightarrow t$ as a result of the von Hove type singularity.
Note that the 4-band antiferromagnetic processes of Fig. 9 with
a weight $\propto N$ are responsible for the RG instability at the energy scale $E_{AFM}$. This also
favors the occurrence of a uniform Mott gap. Moreover, the spin correlation functions now obey
(for any $N$, even or odd)
\begin{equation}
\langle \vec{S}_i(x)\cdot\vec{S}_l(0)\rangle \propto (-1)^{i+l} \cos(\pi x)/x^{1/N},
\end{equation}
where the lowerscripts $i$ and $l$ represent the chains,  and the space coordinate obeys $x<1/E_c(N)$ since we are in the quasi-1D antiferromagnetic phase.

\subsection{Truncated Fermi surface and d-wave superconductivity}

By doping, the chemical potential $\mu$ couples to the total charge mode of each band pair
$\phi_{k\bar{k}c+}$ (see Appendix B.2). Since all the 4-band interactions contain the field $\phi_{k\bar{k}c+}$, they will vanish. On the other hand, the effect of doping on the 2-band interactions is the same
as when doping the D-Mott state. Then, this will lead to a spin gap and d-wave like phase coherence
between the bands $k$ and $\bar{k}$. The charge gaps close to the nodal point $(\pi/2,\pi/2)$ are
the smallest ones such that the holes enter there first. {\it We observe that the effect of doping in this quasi-1D approach is very similar to decrease the number of chains (doping suppresses the antiferromagnetic 4-band processes at low energy scales).} At very small doping, the bands close to the nodal directions will be slightly doped whereas the band pairs close to the antinodal directions will remain insulating as a result of the van Hove type singularities $v_1=v_N \sim 2\pi t/N$. It is instructive
to observe that the limit of a large number of legs $N\rightarrow +\infty$ also supports the concept of a truncated Fermi surface close to half-filling. 

Now, we show that the nodal regions are in fact unstable towards d-wave superconductivity at low energy. At a very general level, one can prove that phase coherence between band pairs around the nodal directions will take place as soon as umklapp scattering and antiferromagnetic processes in those band pairs are cutoff by finite doping effect. Remember that close to half-filling the antinodal regions still form a D-Mott state because the charge gap is bigger as a result of the van Hove type singularity $v_1=v_N\sim 2\pi t/N$. 

Using the low-energy theory of Appendix B.2, at short distances, one can check that antiferromagnetism always coincides with superconductivity within a band pair (see Fig. 9, Right):
\begin{equation}
-\vec{m}_k\cdot \vec{m}_{\bar{k}} = - \Psi^{\dagger}_{kR\downarrow}\Psi_{L\bar{k}\uparrow}\Psi^{\dagger}_{kL\uparrow}\Psi_{R\bar{k}\downarrow}\ \hbox{(+other terms)},
\end{equation}
and
\begin{equation}
\Delta^{\dagger}_k \Delta_{\bar{k}} = \Psi^{\dagger}_{kR\downarrow} \Psi^{\dagger}_{kL\uparrow}
\Psi_{\bar{k}L\uparrow}\Psi_{\bar{k}R\downarrow}\ \hbox{(+other terms)}.
\end{equation}
Now, by doping, one expects the superconducting fluctuations within the band pair $(k,\bar{k})$ in the nodal regions to become more prominent (by analogy to the 2-leg ladder). On the other hand, taking the initial values of the 2-band Cooper and forward interactions given by the antiferromagnetic phase in Eq. (\ref{afm}), one can write down an exact low-energy BCS Hamiltonian of the form \cite{Urs}:
\begin{equation}
H=H_0 +\sum_{i,k} \int dx\ V_{ik} \Delta^{\dagger}_i \Delta_k,
\end{equation}
where $V_{ik}<0$ is consistent with the symmetry
is $d_{x^2-y^2}$ and the summation is restricted to bands in the nodal region, {\it i.e.}, the umklapp scattering and antiferromagnetic processes for those bands have been cutoff at low energy by the finite chemical potential. In the large $N$-limit phase coherence between the bands $i$ and $k$ is due to a Kohn-Luttinger type attraction \cite{KohnLuttinger} mediated by the short-range antiferromagnetic fluctuations \cite{Varma,Scalapino1,Scalapino2} (In the small $N$ limit, the phase coherence must be obtained by introducing by hand a Josephson type coupling between band pairs). Since the
antiferromagnetic processes are processes involving 4 bands, this suggests that the antinodal regions should contribute to
the emergence of superconductivity in the nodal regions. In Sec. 4.4, we invoke a microscopic Andreev-like scenario for the emergence of d-wave superconductivity in the nodal regions mediated by the antinodal regions.

Finally, close to ``optimal doping'', {\it i.e.}, when there is phase coherence between all the band pairs, bosonization predicts \cite{Urs}
\begin{equation}
\langle \Delta^{\dagger}_i(x) \Delta_k(0)\rangle \propto x^{-1/N},
\end{equation}
as a result of the interband phase coherence.

\subsection{Very large doping: Reminiscence of the Landau-Fermi liquid}

For large enough dopings, all the 4-band and umklapp interactions can be neglected. Without umklapp and antiferromagnetic interactions, the weak-coupling fixed point of the large $N$ limit is reminiscent of the Fermi liquid.

More precisely, Lin {\it et al.} \cite{Lin2} have discussed the RG flow of the $N$-leg Hubbard ladder when including  Cooper and forward interactions only.  A band pair can still develop a spin gap and phase coherence; however,
there is no phase coherence between the band pairs and as a result the gap vanishes exponentially as a function of $N$. 
More precisely, the forward scattering now only have a small phase space in two dimensions \cite{Shankar} and thus
can be neglected. In the limit of $N\rightarrow +\infty$, the 2-band Cooper processes give $(4c_{ij}^c=c_{ij}^s=c_{ij})$:
\begin{equation}
\frac{dc_{il}}{dl} = -\sum_{k=1}^N \frac{1}{2v_k} c_{ik} c_{kl}.
\end{equation}
This is reminiscent of a two-dimensional Fermi liquid where one gets:
\begin{equation}
\frac{dV(\theta_1,\theta_2)}{dl} = - \frac{1}{v_F} \int d\theta V(\theta_1,\theta) V(\theta,\theta_2),
\end{equation}
and the angle $\theta$ parametrizes the 2D Fermi surface. This is a renormalization equation for a function, rather than for a constant, {\it i.e.}, one here has an example of ``functional renormalization group''.

This emphasizes that an RG instability can only occur at some points in the phase space where one gets
attractive or antiferromagnetic interactions.

It should be noted that for large doping levels, the weak-coupling fixed point of the quasi-1D theory is a Fermi liquid because the Luttinger exponents in the charge sector converge to unity when the Cooper interactions flow to zero.

\subsection{RG in two dimensions, phenomenology of the pseudogap phase, two gaps}

A major problem in two dimensions is the infinite number of interactions. On the other hand, Fermi surface patching \cite{Haldane} is a method
to study RG flows of fermionic systems in dimension larger than one. First, we will review the 2-patch model which is appropriate when the Fermi surface touches the saddle points at $(\pi,0)$ and $(0,\pi)$.
More precisely, when including a next nearest neighbor hopping $-t'$ such that $t'>0$, this modifies the band structure as:
\begin{equation}
\epsilon(\vec{k}) = -2t(\cos k_x +\cos k_y) +4t' \cos k_x \cos k_y.
\end{equation}
For $t'=0$, the Fermi surface surface at half-filling corresponds to the umklapp surface drawn in Fig. 9.
The main effect of $t'$ is to curve the Fermi surface. Therefore, when doping the system with {\it holes}, one can reach a situation where the Fermi surface touches the antinodal (or saddle) points. In the cuprates, one can estimate $t'/t\approx 1/4$ ($t'$ is supposed to vary somewhat from compound to
compound \cite{Pavarini}). For this situation, the Fermi energy intersects the van Hove singularity at a hole density $\sim 0.2$. Moreover,
the leading (van Hove type) singularity now only arises at the saddle points when adding a next nearest-neighbor hopping $-t'$ and $t'>0$. On can then simplify the RG analysis by including only the antinodal points, resulting in the {\it 2-patch} model. An elastic umklapp process, where two electrons initially near the saddle point at $(\pi,0)$ are transferred to the other saddle point $(0,\pi)$, can take place.

Although it is difficult to build a low-energy theory, one can focus on the leading divergent susceptibilities to predict the low-energy phases; one can also
use numerical techniques as shown by L\" auchli {\it et al.} \cite{Lauchli}. Here, we begin to review the results including the effect of the two-particle umklapp scattering in the two-dimensional Hubbard model close to half-filling as well as effects of the van Hove singularity \cite{Schulz,Dzya,Lederer,Furukawa}, in the 2-patch model.

In two dimensions, at the saddle points (see Fig. 10), the susceptibility for the (particle-particle) Cooper channel at $\vec{q}=\vec{0}$ shows a log-square divergence $\chi_{\vec{q}=\vec{0}}^{pp}\approx \ln^2\omega$ (instead of $\ln\omega$ in one dimension) at the energy $\omega$ due to the van Hove type singularity. For the (particle-hole) Peierls channel at $\vec{Q}=(\pi,\pi)$, there exists a crossover: $\chi_{\vec{Q}}^{ph}\propto \ln^2\omega$ when $\omega>t'$ and $\chi_{\vec{Q}}^{ph}(\omega)\propto \ln\omega$ when $\omega\ll t'$\footnote{The particle-particle Lindhard function at $\vec{q}=\vec{0}$ is unaffected by $t'$: one still gets $\epsilon(\vec{k})=\epsilon(-\vec{k})$. On the other hand, the particle-hole Lindhard function at $\vec{q}=\vec{Q}$ is affected at low energy because the $t'$ term results in $\epsilon(\vec{k}+\vec{Q}) \neq - \epsilon(\vec{k})$. Close to half-filling, this will allow the emergence of a new phase dominated by umklapp processes (consult Fig. 10) and characterized by the absence of long-range antiferromagnetism, similar to the two-leg Hubbard ladder at half-filling.}. The Peierls channels at $\vec{q}=\vec{0}$ and the Cooper channel at $\vec{q}=\vec{Q}$ also diverge log-linearly but have smaller coefficients and can be neglected. 

Similar to the 2-leg Hubbard ladder, one can identify the forward $g_2$ and $g_4$, the backward $g_1$, and the 2-particle umklapp $g_3$ processes. The relevant processes are illustrated in Fig. 10. 

\begin{figure}
\begin{center}
\includegraphics[width=9.5cm,height=8.5cm]{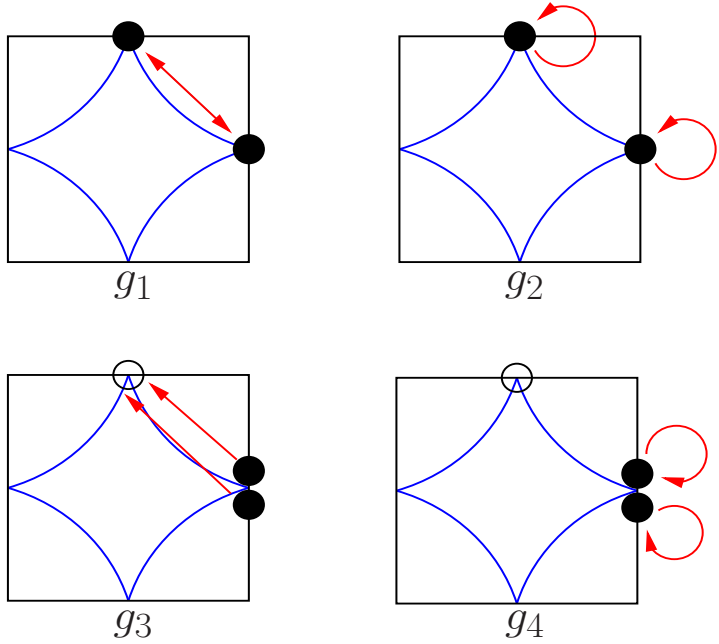}
\end{center}
\caption{Scattering processes in the 2-patch model when the Fermi surface lies at the saddle points
(for $t'/t\sim 1/4$, this approximately happens for $\delta\sim 0.2$ where $0<d_1(y)<1$): forward $g_4$ and $g_2$, backward $g_1$, and umklapp process $g_3$. In the half-filled and lightly doped case, Cooper, umklapp, and antiferromagnetic interactions are strongly coupled. Antiferromagnetic processes are shown in Fig. 9.}
\end{figure}

The RG equations take the form (we use the conventions of Refs. \cite{Lederer,Furukawa}):
\begin{eqnarray}
\dot{g}_1 &=& 2d_1(y) g_1(g_2-g_1) \\ \nonumber
\dot{g}_2 &=& d_1(y) \left(g_2^2 +g_3^2\right) \\ \nonumber
\dot{g}_3 &=& -2g_3 g_4 + 2d_1(y) g_3(2g_2-g_1) \\ \nonumber
\dot{g}_4 &=& - \left(g_3^2 +g_4^2\right).
\end{eqnarray}
Here, $\dot{g}_i = dg_i/dy$ where $y=\ln^2(\omega/E_0)\propto \chi_{\vec{q}=\vec{0}}^{pp}(\omega)$ and $E_0$ is a high-energy cutoff.
The function $d_1(y)$ describes the relative weight between Cooper and particle-hole channels:
$d_1(y)=d\chi_{\vec{Q}}^{ph}/dy$. The asymptotic forms are given by $d_1(y)\rightarrow 1$ when $y\approx 1$ and $d_1(y)\rightarrow \ln(t/t')/\sqrt{y}$ when $y\rightarrow \infty$\footnote{At small $\omega$,  one gets $\chi_{\vec{Q}}^{ph}\propto \ln\omega$ whereas $\chi_{\vec{q}=\vec{0}}^{pp}\propto \ln^2\omega$. This ensures $d_1(y)\propto 1/\sqrt{y}$.} \cite{Furukawa}.

Similar to the 2-leg ladder \cite{Lin}, the forward (inter-patch) coupling strongly renormalizes the umklapp as well as the Cooper channel.

The case $d_1=1$ arises in the limit $t'=0$ or in the strong-$U$ limit where $t'$ is irrelevant in the
Peierls channel divergence. This case has been analyzed by Schulz \cite{Schulz} and by Dzyaloshinskii
\cite{Dzya}. At half-filling, the most divergent term is the antiferromagnetic susceptibility (which has the same exponent as d-wave superconductivity but is dominant due to the next leading divergent terms). Progress in this direction has been done through the functional RG \cite{Honerkampetal}.
It is relevant to stress that in two dimensions, it is difficult to keep all the diverging channels and build
an exact low-energy theory (similar to the quasi-1D case). Nevertheless, one can use numerical approaches \cite{Lauchli}. It is also interesting to observe that the umklapp scattering $g_3\rightarrow +\infty$ strongly diverges, showing the occurrence of a Mott state at arbitrarily small $U$ in two (and one) dimensions. 

For finite doping, the flow favors the d-wave superconducting state and the antiferromagnetism is progressively weakened \cite{Schulz}. More precisely, at a critical doping that depends on the interaction strength, the growth in the antiferromagnetic channels gets cutoff at low energy and the d-wave pairing becomes the dominant instability.  The ``crossover'' from antiferromagnetism to d-wave superconductivity, for $t'=0$, has been analyzed using a N-patch model \cite{Metzner}.

The extreme limit $d_1=0$ (or $t'\rightarrow t$) has been analyzed by Dzyaloshinskii \cite{Dzya}. The RG flow takes the simple form:
\begin{eqnarray}
\dot{g}_3 &=& -2 g_3 g_4 \\ \nonumber
\dot{g}_4 &=& - \left(g_3^2 +g_4^2\right).
\end{eqnarray}
It is relevant to redefine new variables $g_{\pm}=g_4\pm g_3$ such that the RG flow takes the simple
form: $\dot{g}_{\pm}=-g_{\pm}^2$. The initial conditions are given by $g_3=g_4=U$ or $g_+=2U$ and
$g_-=0$. At low energy, we infer that $g_-(l)=0$ and $g_+(l)\rightarrow 0$. The extreme limit $d_1=0$
suggests the possibility of a weak-coupling in two dimensions close to half-filling. On the other hand,
one can easily show that this weak-coupling fixed point is unstable towards a small but finite value of
$d_1$. More precisely, $g_-$ can now access negative regions that allows a strong-coupling fixed point
governed by $g_-\rightarrow -\infty$ and $g_+\rightarrow 0$. Hence one has $g_3\rightarrow +\infty$
(and as a result $g_2\rightarrow +\infty$) and $g_4=-\infty$ at a value $y_c\sim t/U_c$\footnote{One gets the precise prefactors via a numerical integration of the RG flows.}.

Although one cannot solve for the strong-coupling fixed point using the 1-loop RG equations, one can
examine the main susceptibilities \cite{Furukawa}. It is useful to observe that the exponent for
the charge compressibility changes its sign at $d_1(y_c)\sim 0.6$. Close to half-filling, this suggests that
there exists a critical interaction $U_c$ such that for $U>U_c$ the charge compressibility is suppressed
to zero \cite{Furukawa}. Using the results above, one finds $U_c/t \propto \ln^{-2}(t/t')$. {\it Close to half-filling, this implies a transition from a superconducting phase at $U<U_c$ to a phase with a charge gap at $U>U_c$ that can be seen as a precursor of the Mott transition}. This has also been found in the $N$ patch model \cite{Honerkampetal}. It is relevant to observe that the fixed point at $U>U_c$ resembles that of the half-filled two-leg ladder which has a spin and charge gap \cite{Lauchli}. A similar phase also occurs in the 3-leg ladder away from half-filling when including the $t'$ hopping term \cite{JohnKaryn}. 

Note that in two dimensions, starting from the weak-coupling regime, this new phase for $U>U_c$ only occurs as a result of the finite value of $t'$.  In two dimensions, $t'$ is an important parameter that allows to curve the Fermi surface and therefore produces umklapp scattering only in certain directions of the Fermi surface. In contrast, in the quasi-1D band model, the band structure reveals a hierarchy of velocities, allowing for a truncated Fermi surface close to half-filling even when $t'=0$. At a general level, it is interesting to observe that new spin phases may occur as a result of  ``frustration''  \cite{Read} ({\it i.e.}, as a result of the next-nearest neighbor hopping in the weak $U$ limit).

It is not so obvious that one can describe the resulting state by a simple mean field or Hartree-Fock factorization.  In the case of the SO(N) Gross-Neveu model arising at the fixed point of the weak-coupling Hubbard ladder, a mean-field description is only exact in the large even N limit; for finite N, one has to
take into account that the gap fluctuates leading to new bound-state excitations, for example. Moreover, it is very difficult within this RG approach to clarify how the charge and spin gaps spread out laterally along the umklapp surface. Based on the experimental results of high-$T_c$ cuprates \cite{Norman,Kapitulnik} and our results on the N-leg ladder (see Sec. 2 and 3) \cite{UKM}, it sounds legitimate to argue that by increasing the electron density (remember that for $t'/t=1/4$ the Fermi surface intersects with the saddle points for $\delta=0.2$, close to optimal doping) the spin and charge gaps will propagate laterally along the umklapp surface that will let a Fermi surface consisting of 4 Fermi arcs centered at the nodal points $(\pm \pi/2,\pm \pi/2)$. This is consistent with high temperature series by Puttika {\it et al.} \cite{Puttika}. This is also consistent with the Luttinger theorem, that tells us that the area enclosed by the arcs inside the half-filled surface gives the hole density (see Sec. 4.6). A truncation of the Fermi surface has also been found within the SU(2) slave-boson approach of the $t-J$ model (consult Appendix C and Refs. \cite{Andersonetal,Ogata,Lee}).

Within this scenario, the pseudogap phase is characterized by (heavy) {\it preformed pairs} and superconductivity might emerge from the proximity effect between the Landau quasiparticles of the Fermi arcs and the preformed pairs. {\it The pseudogap is not the superconducting gap}.
Due to the vanishing of the charge compressibility in the antinodal regions, the superfluid weight will be that of the Fermi arcs alone. The RG approach in two dimensions suggests the existence of two distinct energy scales (or gaps) close to the Mott state: $T^*$ is the temperature at which umklapp scatterings produce a charge and spin gap in the antinodal regions, resulting in an (insulating) RVB state, and at a lower temperature $T_c$ the Fermi arcs will mediate d-wave superconductivity. The microscopic mechanism for the occurrence of superconductivity may be an Andreev process (proximity effect), involving the antinodal (hot) regions \cite{KarynAndreev,Geshkenbein}: two electrons from the nodal region might virtually `hop' into the antinodal region and may be converted into a Cooper pair as a result of short-range RVB correlations in the antinodal region. In the pseudogap phase, the specific heat is expected to vanish linearly as $\delta\rightarrow 0$ (only the Fermi arcs contribute), in agreement with experiment \cite{Loram}. One important consequence of this theory is that  the energy gap in the nodal regions does not scale with the pseudogap at $(\pi,0)$. In fact, this is in agreement with angle resolved photoemission spectroscopy (ARPES) measurements \cite{Dama,Tanaka}.  Other support for two gaps comes from Andreev reflection studies \cite{Andreev} and Raman scattering \cite{Raman}. 

At this point, it should be noted that the slave-boson approach applied to the $t-J$ model gives a distinct scenario for the pseudogap phase (see Appendix C). The large gap at $(\pi,0)$ is a d-wave superconducting gap and the pseudogap corresponds to a superconductor destroyed by phase fluctuations. At $T_c$, the holes (holons) condense producing a d-wave superconductor; in this theory, that relies on spin-charge separation, there is a single d-wave gap. 

A theory of $T_c$ for the underdoped cuprates has been built by analogy to the Kosterlitz-Thouless transition in two dimensions. More precisely, in 1995, Emery and Kivelson proposed a model based on phase fluctuations \cite{EK}. They pointed out that $T_c$ of the underdoped cuprates is determined by the energy scale for phase fluctuations. In their picture, the entire region below $T^*$ is characterized by phase fluctuations and at $T_c$, the phase stiffness $K_s$ jumps between zero and a finite value $K_s(T_c^-)$ given by the universal relation:
\begin{equation}
k_B T_c = \frac{\pi}{2} K_s(T_c^-) = \frac{\pi}{8} \frac{\hbar^2 \rho_s}{m}.
\end{equation}
Here, $m$ represents the (effective) mass of an electron or $2m$ embodies the mass of a Cooper pair. The precise value of $T_c$ thus depends on the superfluid density $\rho_s$. Empirically, when the concentration of holes in the CuO$_2$ planes is not too large one expects $\rho_s(T=0)\propto \delta$. In fact, this relation is automatically satisfied if the Fermi arcs only become superconducting below $T_c$ (see discussion above). This also happens in the renormalized mean-field theory of the $t-J$ model presented below. This suggests that $T_c\propto \delta$ at low doping levels, as confirmed experimentally \cite{Uemura}. On the other hand, as mentioned by Lee and Wen in 1997 \cite{superfluid}, the thermal excitation of quasiparticles near the nodes produces a linear decrease of $\rho_s$ with temperature, wich
was the earliest experimental evidence for d-wave symmetry \cite{Hardy}. The number of quasiparticles in the d-wave state varies as $T^2$ since the density of states is linear with energy. On the other hand, the amount of decrease of $\rho_s$ per quasiparticle is inversely proportional to
its energy, canceling one factor $T$: thus, $\rho_s$ decreases linearly with $T$. Following Lee and Wen \cite{superfluid} and Millis {\it et al.} \cite{Millisetal}, one predicts:
\begin{equation}
\frac{\rho_s(T)}{m} = \frac{\rho_s(T=0)}{m} - a T,
\end{equation}
and the parameter $a$ is given by (see Appendix C.4):
\begin{equation}
a = \frac{2\ln 2}{\pi}\alpha^2\frac{v_F}{v_{\perp}}.
\end{equation}
Here, $v_{\perp}$ represents the velocity of the nodal quasiparticles in the direction of the maximum gap. The ratio $v_F/v_{\perp}$ can be measured through the thermal conductivity \cite{Lee,Taillefer}.
The only assumption made is that the quasiparticles with a wave-vector $\vec{k}$ carry an electrical current $\vec{j}(\vec{k}) = -e \alpha
\vec{v}_F$, and $\alpha$ is a phenomenological Landau parameter \cite{Millisetal}, which was left out in
the original paper by Lee and Wen. Moreover, $v_F/v_{\perp}$ is known to go to a constant for small
doping $\delta$ and $\alpha^2\sim 0.5$ \cite{Lee}. This is a strong result because it states that despite the proximity of the Mott state the nodal quasiparticles carry a current which is similar to that of the tight-binding Fermi-liquid theory. 

Therefore, $\rho_s$ at $T=0$ decreases proportional to doping, yet
its rate of decrease with temperature does not vanish with doping, but in fact remains relatively constant
\cite{Randeria}. The insensitivity of the linear $T$ slope in $\rho_s$ to doping was demonstrated experimentally \cite{Lemberger}. The decrease of $\rho_s$ to zero, at a given doping level, is also sometimes considered to determine $T_c$ \cite{Lee,superfluid}. Now, coming back to the Kosterlitz-Thouless scenario, one must recognize that the $\rho_s$ which controls the transition is not $\rho_s(T=0)$ but $\rho_s(T)$ which is greatly reduced by quasiparticle excitations. By combining the two effects, the decrease of $\rho_s(T)$ becomes faster than linear and eventually becomes infinitely steep. But this happens only very close to $T_c$; therefore, the quasiparticle mechanism may give a good (better) estimate of $T_c$ \cite{superfluid}. The Lee-Wen mechanism of $T_c$ is relevant for the underdoped side of the phase diagram where it offers a natural explanation for $T_c\sim \rho_s(T=0)\sim \delta$ and holds all the way up to optimality.

\subsection{Gutzwiller projected d-wave superconducting state and mean-field theory}

Now, we analyze the strong-interaction $t-J$ model. First, we closely follow the paper by Baskaran, Zou, and Anderson \cite{BZA}, and describe a (renormalized) mean-field approach that allows to predict the existence of a superconducting dome in a quite natural way and give an idea of the pseudogap \cite{Gros,Hirsch}.

Assuming strong interactions, the $t-J$ Hamiltonian reads:
\begin{equation}
H =-t\sum_{\langle i;j\rangle,s} d^{\dagger}_{is} d_{js}  -
\sum_{i,s} \mu d^{\dagger}_{is} d_{is} 
+ J\sum_{\langle i;j\rangle} \vec{S}_i\cdot \vec{S}_j,
\end{equation}
where $J=4t^2/U$ and $S_i^+ = d^{\dagger}_{i\uparrow} d_{i\downarrow}$, ...\ , and $n_{is}=d^{\dagger}_{is} d_{is}$ is the number occupation per spin. For simplicity, we set the hopping term $t'=0$\footnote{A more realistic model for the cuprates gives a similar result, as it must be, since details of the tight-binding Hamiltonian should not matter at large $U$ \cite{Paramekanti}.}. 

To study the ground state and the excited states of the $t-J$ Hamiltonian one can use a BCS trial wavefunction $P|BCS\rangle$ \cite{Anderson} where
\begin{equation}
|BCS\rangle = \prod_{\vec{k}} (u_{\vec{k}} + v_{\vec{k}} d^{\dagger}_{\vec{k}\uparrow} d^{\dagger}_{-\vec{k}\downarrow})|Vac\rangle;
\end{equation}
$P=\prod_i (1-n_{i\uparrow} n_{i\downarrow})$ is the Gutzwiller projection operator \cite{Gutzwiller} 
that eliminates all configurations with double occupancy, as appropriate for $U\rightarrow +\infty$, $|Vac\rangle$ is the vacuum state, $u_{\vec{k}}$ and $v_{\vec{k}}$ are the variational parameters satisfying the normalization condition $|u_{\vec{k}}|^2 + |v_{\vec{k}}|^2=1$. At half-filling, the energy of the projected d-wave BCS state is very close to the ground state energy; the d-wave BCS state is in general a better trial wavefunction than the s-wave BCS wave function \cite{Trivedi}. {\it The projected BCS wavefunction is a natural generalization of the BCS state to strongly correlated systems}. This projected wavefunction is appropriate to describe three phases: a resonating valence bond insulator, a Fermi liquid metal (a state with $u_{\vec{k}}v_{\vec{k}}^*=0$), and a (d-wave) superconductor. 

In general, the Gutzwiller projection is cumbersome to implement (but properties of the trial wave function can be evaluated using Monte Carlo sampling) and a simple approximate scheme was
proposed, called the Gutzwiller approximation \cite{Zhang}. The essential step is to construct an effective
Hamiltonian:
\begin{equation}
H_{eff} = - t g_t \sum_{\langle i;j\rangle} d^{\dagger}_{is} d _{is} -
\sum_{i,s} \mu d^{\dagger}_{is} d_{is} + Jg_s \sum_{\langle i;j\rangle}\vec{S}_i \cdot \vec{S}_j,
\end{equation} 
such that the pojection operator is eliminated in favor of the reduction factor
$g_t = 2\delta/(1+\delta)$ in the kinetic term \cite{Vollhardt}. In the limit of small hole doping, $g_t\rightarrow 2\delta$; the electron number per site is $n=1-\delta$. Additionally, one gets that the projection operator enhances spin-spin correlations: $g_s=4/(1+\delta^2)$ \cite{Zhang}. The renormalization factors $g_t$ and $g_s$ are explicitly determined by the ratios of the probabilities of the corresponding physical processes ({\it e.g.}, hopping processes) in the projected state $P|BCS\rangle$ and non-projected state $|BCS\rangle$.

By analogy to the slave-boson theory \cite{Lee}, the mean-field Hamiltonian obeys:
\begin{equation}
H_{mf} = -\sum_{\langle i;j\rangle} (tg_t + \chi_{ij}) d^{\dagger}_{is} d_{js}   -
\sum_{i,s} \mu d^{\dagger}_{is} d_{is} + \sum_{\langle i;j\rangle} \Delta_{ij}\left(d^{\dagger}_{i\uparrow} d^{\dagger}_{j\downarrow} - d^{\dagger}_{i\downarrow} d^{\dagger}_{j\uparrow}\right) +h.c.,
\end{equation}
where $\chi_{ij}= \chi =(3g_sJ/4)\langle d^{\dagger}_{is} d_{js}\rangle$ for nearest neighbors and zero otherwise, and for the d-wave BCS case, the variational gap obeys
\begin{equation}
\Delta_{\vec{k}} = \Delta(\cos k_x - \cos k_y).
\end{equation}
In this section, to simplify the discussion, we assume that $\Delta$ is real.
It should be noted that the connection between antiferromagnetic fluctuations and superconductivity becomes flagrant through the mean-field approach since the Heisenberg exchange $J$ gives rise to a pairing contribution.

Here, the BCS theory is governed by the coupled equations \cite{Andersonetal,Zhang}:
\begin{equation}
\Delta_{\vec{k}} = \frac{3}{4} g_s J \sum_{\vec{k}'} \gamma_{\vec{k}-\vec{k}'} 
\frac{\Delta_{\vec{k}'}}{2E_{\vec{k}'}},
\end{equation}
and
\begin{equation}
\chi_{\vec{k}} = -\frac{3}{4} g_s J \sum_{\vec{k}'} \gamma_{\vec{k} - \vec{k}'} \frac{\xi_{\vec{k}'}}{2 E_{\vec{k}'}},
\end{equation}
where $\xi_{\vec{k}} = g_t \epsilon_{\vec{k}} -\mu- \chi_{\vec{k}}$ and (here) $\epsilon_{\vec{k}} = -t
\gamma_{\vec{k}}$ where $\gamma_{\vec{k}} = 2(\cos k_x +\cos k_y)$ is the Fourier transform of the
exchange interaction (simply the nearest neighbor result). $E_{\vec{k}}=  \sqrt{\xi_{\vec{k}}^2 + \Delta_{\vec{k}}^2}$ has the same form as in the BCS theory. 

\begin{figure}
\begin{center}
\includegraphics[width=5cm,height=7cm]{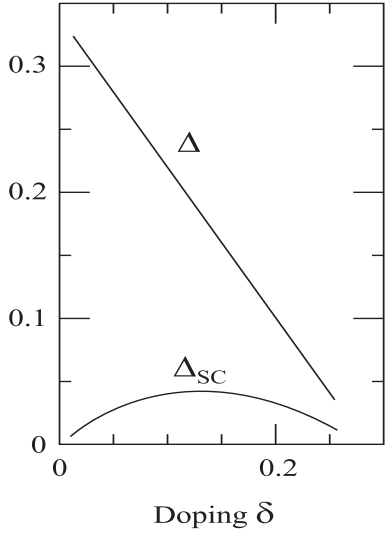}
\end{center}
\caption{The d-wave gap $\Delta$ and superconducting order parameter $\Delta_{SC}$ $(\sim T_c)$, from the renormalized mean-field theory for $J/t=0.2$ \cite{Zhang}, as a function of hole doping $\delta$ (close to half-filling the d-wave gap $\Delta$ converges to $J$; the gap is in unit of $3g_s J/4$).}
\vskip 0.4cm
\end{figure}

These gap equations can be solved \cite{Zhang}. An explicit derivation of the solutions is given in Ref. \cite{Zhang}.
Interestingly, the d-wave (spin) gap $\Delta$ falls almost linearly with the hole doping
from a number of order $J$, and vanishes around $\delta=0.3$ for $J/t=0.2$. This quantity may represent the pseudogap, which is known to vary experimentally in this way. A more realistic band structure
for the cuprates (including the next-nearest neighbor hopping $t'$) gives a similar result \cite{Paramekanti}.
The renormalized mean-field theory supports the RVB approach, {\it i.e.}, the appearance of a large energy
scale representing the spin gap. The spin gap was first observed by Nuclear Magnetic Resonance (NMR) at the end of the eighties \cite{Alloul}. Its significance was only slowly recognized by the mid-nineties.

$\Delta$ represents the d-wave gap or the ``pseudogap''. On the other hand, one can define a superconducting order parameter $\Delta_{SC}$ which obeys \cite{Andersonetal,Paramekanti} 
\begin{equation}
\label{SC}
\Delta_{SC}^2 = \langle BCS|P (d_{i\uparrow}^{\dagger} d_{j\downarrow}^{\dagger} d_{i+l\uparrow} d_{j+l\downarrow}) P|BCS\rangle,
\end{equation}
 for large distance $l$; the real superconducting order parameter is the off diagonal long range order eigenvalue of the density matrix. Using the {\it projected} BCS state, this
averaged value is renormalized by a factor $g_t^2$, implying that the superconducting order parameter vanishes linearly at $\delta=0$. {\it In fact, $\Delta_{SC}\propto \delta$ at low doping is a general property of projected superconducting wavefunctions}. The local fixed number constraint imposed by $P$ at $\delta=0$ leads to large quantum phase fluctuations that fatally destroy the superconducting order at very low doping levels. The state at $\delta=0$ is an insulator with a vanishing Drude weight. Remarkably, $\Delta_{SC}$ bears a striking resemblance to the variation of $T_c$ with doping, and was suggested to be a measure a $T_c$ within this approach. Eq. (\ref{SC}) suggests that $T_c$ (only) characterizes the emergence of the off-diagonal long-range order. However, as pointed out in Ref. \cite{Zhang}, below $T_c$, the superconducting order parameter $\Delta_{SC}$ could also be defined as:
\begin{equation}
\Delta_{SC} = \langle BCS|P (d_{i\uparrow} d_{j\downarrow}) P|BCS\rangle,
\end{equation}
which is supposed to renormalize with $g_t$ (when taking the averaged value using the projected
BCS state). This definition would suggest a two-gap scenario where the superconducting gap is distinct from the pseudogap (in relation to Sec. 4.4). The physical amplitude of the superconducting gap $\Delta_{SC}$ is shown in Fig. 11 \cite{Zhang} and grows linearly with $\delta$ at low dopings. This is
relevant to note that those predictions regarding the d-wave nature of the superconducting gap, made in 1988, came before the experimental confirmations in 1993-94 \cite{phase1,phase2,Hardy}. The d-wave pairing symmetry was also predicted by the ``spin fluctuation theory'' (see introduction), based on a more orthodox structure \cite{Scalapino2,Monthoux}. On the other hand, the renormalized mean-field theory is different from the spin-fluctuation theory since the latter is based on a Fermi liquid theory and is unable to deal with the unusual properties close to a Mott insulator.

Variational Monte-Carlo methods for projected d-wave states support these results \cite{Paramekanti}. 
For more details and references, see the recent review by Ogata and Fukuyama \cite{Ogata}. They also provide a quantitative description on the nodal quasiparticles (``nodons'') in the BCS state. The Gutzwiller projected d-wave superconducting ground state displays sharp nodal quasiparticle excitations and is in good agreement with experiment (below $T_c$) \cite{Johnson}. These low-lying excitations are expected to dominate the low temperature thermodynamics, transport, and response functions in the superconducting state. 

Within this scheme, the nodal quasiparticles have a coherent spectral weight $Z$ that goes to zero as $g_t$ but whose Fermi velocity is very weakly doping dependent\footnote{This stems from the fact that $\chi$ does not depend much on doping \cite{Lee}.} and remains non-zero as $\delta\rightarrow 0$ \cite{Andersonetal,Paramekanti}. $Z\sim \delta$ also implies that the real part of the self-energy $\Re e\Sigma$ obeys $|\partial \Re e\Sigma/\partial \omega|\sim 1/\delta$ and $|\partial \Re e\Sigma/\partial k|\sim 1/\delta$ ($v_F\neq 0$). The effects of the renormalization $g_t$ on the superfluid density $\rho_s$ and on the Drude weight \cite{Sawatzky} is a consequence of the RVB based theories. In fact, the renormalization $g_t$ applies to any term in the Hamiltonian which is a one-electron energy. In particular, the Green's function of the quasiparticles in the superconducting state
contains a sharp coherence peak at the quasiparticle energy on top of a very broad incoherent spectrum
and ARPES experiments have confirmed that the amplitude of that peak is proportional to $2\delta$ \cite{Feng,Ding}.

To summarize briefly, the Gutzwiller projected d-wave wavefunction confirm the occurrence of d-wave superconductivity in the strong-coupling $t-J$ model for $0<\delta<\delta_c$ where $\delta_c\sim 0.3$ for $J/t=0.2$;  results in the superconducting state are in very good agreement with experimental results (for a more detailed comparison between theory and experiment, consult Ref. \cite{Andersonetal}). 

\subsection{Pseudogap phase, generalized Luttinger sum rule and electron propagator}

However, the situation from the renormalized mean-field theory is less clear in the pseudogap phase. Phase fluctuations have to be considered thoroughly.

Moreover, there is another difficulty: the proliferation of nearby alternative states of different symmetry. One important issue is the evolution to the antiferromagnetic state at very low doping. On general principles mesoscopically inhomogeneous states (stripes) are likely to be stable at low doping on some scale. Charge and spin stripe order have been seen experimentally in a few special cuprate compounds, specifically La$_{2-\delta}$Ba$_{\delta}$CuO$_4$ and La$_{16-\delta}$Nd$_{0.4}$Sr$_{\delta}$CuO$_4$. The highest stripe ordering temperatures occur at $\delta=1/8$, where $T_c$ is strongly depressed. It has been shown experimentally that the dominant impact of the (spin) stripe ordering is to electronically decouple the CuO$_2$ planes \cite{Tranquada}. The data on anisotropic transport and magnetization properties of La$_{2-\delta}$Ba$_{\delta}$CuO$_4$ at $\delta=1/8$ report a crossover to a state of 2D fluctuating superconductivity, which eventually reaches a 2D superconducting state below a Kosterlitz-Thouless transition. Interestingly, it appears that the stripe order in  La$_{2-\delta}$Ba$_{\delta}$CuO$_4$ frustrates the three-dimensional phase order, but is fully compatible with 2D superconductivity. The layer decoupling mechanism has been theoretically addressed in Ref. \cite{Berg}. Additionally, the nature of stripes in a generalized $t-J$ model has been recently investigated in Ref. \cite{Yangstripe}. 

A second class of competing states has its origin in the SU(2) gauge symmetry emerging at half-filling, as explained by Affleck {\it et al.} \cite{Afflecketal}. In the undoped state, with exactly one electron per site, the presence of an up spin is equivalent to the absence of a down spin and vice-versa, allowing independent SU(2) rotations at each site. This implies that an undoped RVB state or any Mott state can be represented by an enormous number of wave functions before projection; other candidates are the staggered flux state \cite{Affleck} and the d-density wave \cite{Chakravarty}. Wen and Lee \cite{WL} have proposed that SU(2) rotations, connecting fluctuations of staggered flux states and d-wave superconductivity, may play a crucial role in explaining the pseudogap phase (see Appendix C). Very recently, Kaul {\it et al.} have discussed the possibility of an algebraic charge liquid (in this context, the algebraic correlations reside in the charge sector) which seems to arise between the antiferromagnet and the superconductor \cite{Kaul}.
 
Another approach to capture the pseudogap region is to start with the $N$-leg ladder close to half-filling
\cite{UKM} or similarly a doped spin liquid  comprising an array of two-leg ladders close to half-filling, in the weak-coupling regime \cite{KTM}.

First, using the exact SO(8) low-energy theory of the weak-coupling regime (see Sec. 2.3), one can compute the electron Green's function around each of the four Fermi points of the two-band model. The single-particle Green's function at half-filling $(\mu=0)$ is governed by the coherent part \cite{Konik}:
 \begin{equation}
 G_a(k,\omega) = \frac{Z_a(\omega+ E_a(k))}{\omega^2- E_a^2(k)-\Delta^2} +G_{inc},
 \end{equation}
 where $E_a(k)$ is the bare dispersion in the corresponding bonding or antibonding band close to
 the Fermi wave-vector $k_{Fa}$  (close to half-filling, the two bands
 are characterized by the same Fermi velocity), resulting in
 \begin{equation}
 E_a(k) = v_{a}(k-k_{Fa}).
 \end{equation}
The quasiparticle weight $Z_a\sim 1$ and $\Delta$ is the D-Mott single-particle gap \cite{Konik}. While the coherent part has a form similar to the diagonal Green's function in BCS theory, there is no off-diagonal component of the Green's function in this case since the system is in a Mott insulating state at half-filling.

It should also be noted that $G_a(k,0)$ changes sign from positive to negative through
 a {\it zero} as $k$ crosses $k_{Fa}$: this occurs when $E_a(k)=0$. This shows that the area characterized by $\Re e G_a(k,0)>0$ is unchanged by the interactions. This satisfies the Luttinger sum rule (LSR) \cite{LSR} for the 2-band system:
 \begin{equation}
 2\sum_{a=1,2} \int_{\Re e G_a(k,0)>0} dk =  4\sum_a k_{Fa} = 4\pi n.
 \end{equation}
 The Luttinger sum rule, of course, works for a single-chain Luttinger liquid \cite{KarynGreen}.
 The key point about the LSR as pointed out in the famous textbook by Abrikosov, Gorkov, and Dzyaloshinski \cite{AGD} and more recently by Tsvelik \cite{tsv}, is that the sign change from positive to 
 negative values of $\Re e G(k,\omega=0)$ is not restricted to an infinity in $\Re e G(k,0)$ as in the case of a Fermi surface of a Landau-Fermi liquid. It can occur through a zero in 
 $\Re e G(k,\omega=0)$ as in the BCS theory of superconductivity or as in the 2-leg Hubbard ladder \cite{Konik}.  
  
In fact, we can naturally extend this analysis to the case of a $N$-leg ladder close to half-filling which unambiguously demonstrates a possible truncation of the Fermi surface; consult Sec. 3 and Sec. 4.2.  
When doping, the holes enter the band pairs successively from the diagonals of the Fermi surface. For a small hole doping $\delta$, the insulating antinodal directions yield exactly the same
fixed point as at half-filling. This implies that the electron Green's function in the antinodal regions is still governed by a self-energy of the form \cite{Konik}:
\begin{equation}
\label{self}
\Sigma_{a}(k,\omega) = \frac{\Delta_{a\bar{a}}^2}{\omega+E_a(k)},
\end{equation}
and the gap $\Delta_{a\bar{a}}$ is related to the energy scale $E_a^*$ defined in Sec. 3, corresponding
to the energy scale at which a band pair $(a,\bar{a})$ becomes successively frozen out and forms a D-Mott state. It is relevant to mention that this form of self-energy also emerges in a doped spin liquid consisting of 2-leg Hubbard ladders coupled by a long-range interladder hopping \cite{KTM}.

\begin{figure}
\begin{center}
\includegraphics[width=6.2cm,height=5.5cm]{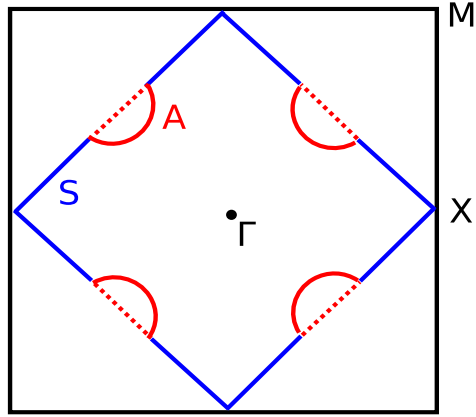}
\end{center}
\caption{A sketch of the electronic structure in the pseudogap phase by analogy with the  
$N$-leg ladder weak-coupling investigation. In the region $S$ the Fermi surface shown as ``heavy lines'' is gapped due to umklapp scattering processes. From the quasi-1D analysis with $t_{\perp}\sim t$, one can rigorously prove that the D-Mott gap propagates from the saddle points along the 2D umklapp surface. The generalized LSR for the $N$-leg ladder shows that the area enclosed by the Fermi arcs (region A) and the umklapp surface (dashed lines) should be proportional to the hole density.}
\end{figure}

Now, it is interesting to discuss the LSR through this large $N$-limit. At half-filling, the
LSR for the $N$-leg ladder system is automatically satisfied:
 \begin{equation}
 2\sum_{a=1,2,...,N} \int_{\Re e G_a(k,0)>0} dk =  4\sum_a k_{Fa} = 2\pi n N.
 \end{equation}
The Luttinger contour of zeroes in $\Re e G_a(k,0)$ corresponding to $E_a(k)=0$ coincides with the 2D umklapp surface or with the antiferromagnetic Brillouin zone, assuming that $t_{\perp}=t$. 
Close to half-filling, the bands located around the antinodal directions remain insulating and as a result
have a self-energy of the form of Eq. (\ref{self}). The remaining Fermi surface then can be viewed as four arcs (by analogy to the 2D situation): applying the general formulation of the Luttinger theorem requires that the area enclosed by the arcs inside the umklapp surface gives the hole density, as illustrated in Fig. 12. Even though the Luttinger theorem is applicable in various cases and is general, it can also break down in specific two-orbital Mott insulator systems \cite{RoschLut}. 

By analogy to the ladder systems \cite{UKM} and to the array of coupled ladders \cite{KTM}, taking the parameters from the renormalized mean-field theory in Sec. 4.5, Yang {\it et al.} \cite{zhangnew} have suggested a similar single-particle Green's function in the pseudogap phase of the underdoped cuprates for the large $U$ limit:
\begin{equation}
G^{RVB}(\vec{k},\omega) = \frac{g_t}{\omega-\xi(\vec{k}) - |\Delta_{\vec{k}}|^2/(\omega+\xi_0(\vec{k}))} + G_{inc},
\end{equation}
where \cite{zhangnew}
\begin{eqnarray}
\xi_0(\vec{k}) &=& -2t(\delta)(\cos k_x +\cos k_y) \\ \nonumber
\xi(\vec{k}) &=& \xi_0(\vec{k}) -4t'(\delta)\cos k_x \cos k_y - 2t''(\delta)(\cos 2k_x + \cos 2k_y) -\mu \\ \nonumber
\Delta_{\vec{k}} &=& \Delta(\delta)(\cos k_x - \cos k_y).
\end{eqnarray}
(We have noted $\omega=\omega+i\delta$).
The factor $\xi(\vec{k})-\xi_0(\vec{k})$ has been adapted from the $t_{\perp}(k_{\perp})$ term emerging in the electron Green's function for an array of coupled ladders \cite{KTM}. From the renormalized 
mean-field theory of Sec. 4.5 \cite{zhangnew}:
\begin{eqnarray}
t(\delta) &=& g_t t +\frac{3}{8} g_s J \chi \\ \nonumber
t'(\delta) &=& -g_t t' \\ \nonumber
t''(\delta) &=& g_t t''.
\end{eqnarray}
where $t'>0$ within our conventions; see Eq. (68). Here $\chi$ is a dimensionless parameter
(we have slightly changed the notations compared to Sec. 4.5). We also introduce a 3$^{rd}$ nearest neighbor hopping $t''$. The chemical potential $\mu$ is determined from the LSR on the electron density. It should be noted that the only difference with the weak-coupling Green's function from the quasi-1D analysis resides in the factor $g_t$ of the renormalized mean-field theory.

In the limit of zero doping, $\delta\rightarrow 0$, then $g_t\rightarrow 0$ and the quasiparticle dispersion
reduces to the spin dispersion and the quasiparticles has a vanishing weight in the single particle
Green's function. At small but finite $\delta$, similar to the D-Mott phase, the zero frequency Green's function $G^{RVB}(\vec{k},0)$ that enters the LSR has lines of zeroes when $\xi_0(\vec{k})=0=-2t(\delta)
(\cos k_x +\cos k_y)$. The Luttinger contour of zeroes in $G^{RVB}(\vec{k},0)$ then consists of straight lines connecting the saddle points $(\pm \pi,0)$ and $(0,\pm \pi)$ and perfectly coincides with the 2D umklapp surface (see Fig. 12). This Luttinger contour is also in accordance with the functional RG calculations \cite{Honerkampetal} (supplied by a numerical treatment at the fixed point \cite{Lauchli}) on the weak-coupling 2D $t-t'-U$ Hubbard model of Sec. 4.4. 

\begin{figure}
\begin{center}
\includegraphics[width=11.2cm,height=8cm]{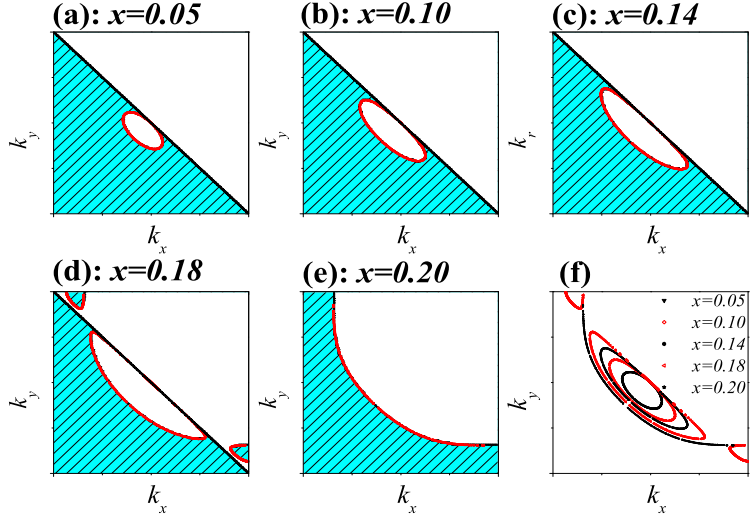}
\end{center}
\vskip -0.5cm
\caption{From Yang {\it et al.} \cite{zhangnew}: Contours on which $G(\vec{k},0)$ changes sign at various hole concentrations $\delta$ are shown in (a)-(e). A summary is shown in (f). The shaded area corresponding to $\Re e G(\vec{k},0)>0$
satisfying the LSR. In the normal pseudogap phase, the line connecting the saddle points $(\pi,0)$ and $(0,\pi)$ is the Luttinger surface of zeroes and the pockets in the thick line represent the infinities of $\Re e G(\vec{k},0)$. The chosen parameters are: $\chi=0.338$, $J/t=1/3$, $t'/t=0.3$, and $t''/t=0.2$.}
\end{figure}

A second feature that emerges from $G^{RVB}(\vec{k},\omega)$ is the appearance of
hole pockets (rather than arcs) at finite hole doping. The hole pockets define Fermi surfaces where $G^{RVB}(\vec{k},\omega)$ changes sign through {\it infinities} and contain a total area equal to the hole density. The LSR is satisfied since the area with $\Re e G^{RVB}(\vec{k},0)>0$ is bounded by the Luttinger surface which
contains exactly 1 electron per site, minus four hole pockets which have a total area related to the hole density. The Fermi surface of the four hole pockets is defined as:
\begin{equation}
\xi(\vec{k}) + |\Delta_{\vec{k}}|^2/\xi_0(\vec{k}) = 0.
\end{equation}
The evolution of the hole pockets as a function of doping is shown in Fig. 13. The chemical potential $\mu$ was adjusted at each $\delta$ to reproduce the correct area for the hole pockets. One can see that the
hole pocket evolves gradually into a more normal surface as $\delta$ increases. It is also instructive to observe that within this ansatz of $G^{RVB}$ the spectral weight of the quasiparticle pole varies strongly around the pocket \cite{zhangnew}. In fact, the quasiparticle weight in $G^{RVB}$ have very small values on the back sides of the hole pockets, which can account for the failure to observe these parts from photoemission experiments. However, it should be noted that Fermi pockets have been recently inferred from quantum oscillation datas \cite{ProustN}. Another explanation based
on the algebraic charge liquid can be found in Ref. \cite{Kaul}.
Many experimental features, {\it e.g.}, the slow variation of the nodal Fermi velocity $v_F$ and the
Drude weight scaling with $\delta$ are reproduced by the propagator of Eq. (88).
The ansatz for $G^{RVB}$ agrees with ARPES and resolves an appearent disagreement with the LSR \cite{zhangnew}.

The Green's function $G^{RVB}$ also produces zeroes in the nodal regions along the 2D umklapp surface: in fact, the zeroes only disappear at the nodal points since the d-wave RVB gap vanishes at those points. The umklapp surface along which the energy gap opens up lies above the chemical potential, causing particle-hole asymmetry in the quasiparticle spectra in the pseudogap state. The spectra are symmetric only
along the nodal directions where the gap vanishes, and the asymmetry increases away from the nodal directions in agreement with ARPES measurements. For a recent discussion, see Ref. \cite{Yang}.

Below $T_c$, following the quasi-1D analysis of Sec. 4.2, the Fermi arcs now become superconducting. Therefore, the self-energy in the superconducting phase must be modified to convert the infinities of the single particle Green's function at the Fermi surface of the four hole pockets onto zeroes, except along the Brillouin zone diagonals where the superconducting gap $\Delta_{SC}(\vec{k})$ vanishes.  

In the publication of Ref. \cite{zhangnew}, Yang {\it et al.} have proposed the following ansatz for the d-wave superconducting state (pushing forward the ``2 gap'' idea):
\begin{equation}
\Sigma_{SC}(\vec{k},\omega) =\Sigma_{RVB}(\vec{k},\omega) + \frac{|\Delta_{SC}(\vec{k})|^2}{\omega+\xi(\vec{k}) +\Sigma_{RVB}(\vec{k},-\omega)},
\end{equation}
and $\Sigma_{RVB}(\vec{k},\omega)$ is the RVB spin liquid self-energy in Eq. (87):
\begin{equation}
\Sigma_{RVB}(\vec{k},\omega) = |\Delta_{\vec{k}}|^2/(\omega+\xi_0(\vec{k})).
\end{equation}
Consult also Ref. \cite{Bascones}. Following Yang {\it et al.}, the coherent part of the single-particle Green's function in the d-wave superconducting state then reads:
\begin{equation}
G_{coh}^s (\vec{k},\omega) = \frac{g_t}{\omega-\xi(\vec{k}) - \Sigma_{SC}(\vec{k},\omega)}.
\end{equation}
This form of Green's function has been found by solving the coupled equations which connect the regular $(G_{coh}^s = -i\langle d_{\uparrow} d_{\uparrow}^{\dagger}\rangle)$ and the anomalous $(F_s=-i\langle d_{\uparrow} d_{\downarrow}\rangle)$ Green's functions (following 
Abrikosov, Gorkov, Dzyaloshinski, page 295 \cite{AGD}). In the superconducting state, the off-diagonal
component of the Green's function becomes non-zero. Including the pseudogap in the normal phase
and ignoring the effect of the Gutzwiller projector, then one gets \cite{zhangnew}:
\begin{eqnarray}
(\omega - \xi(\vec{k}) - \Sigma_{RVB}(\vec{k},\omega))G_{coh}^s - i\Delta_{SC} F^{\dagger}_s(\vec{k},\omega) &=& 1 \\ \nonumber
(\omega+\xi(\vec{k}) + \Sigma_{RVB}(\vec{k},-\omega))F_s^{\dagger}(\vec{k},\omega) + i \Delta_{SC}^{\dagger}G_{coh}^s(\vec{k},\omega) &=& 0.
\end{eqnarray}
The superconducting gap function has a d-wave form and is related to the anomalous Green's function
in a standard way:
\begin{equation}
\Delta_{SC}(\vec{k}) = \int d\vec{k}' d\omega g(\vec{k}-\vec{k}')F(\vec{k}',\omega),
\end{equation}
where $g(\vec{k}-\vec{k}')$ is the d-wave pairing interaction. Within the renormalized mean-field theory,
any single-particle operator must be renormalized by $g_t$. This justifies Eq. (93) (which includes the self-energy of the pseudogap phase).

First, since $\Sigma_{RVB}(\vec{k},0)\rightarrow +\infty$  on the surface where $\xi_0(\vec{k})=0$, $G_{coh}^s$ continues to have a Luttinger surface of zeroes in the superconducting state on the umklapp surface. However, there is an additional set of Luttinger surfaces defined by $\xi(\vec{k}) +\Sigma_{RVB}(\vec{k},0)=0$.  But these are just the Fermi surface of the four hole pockets in the normal phase which have now been converted to a Luttinger surface of zeroes in the superconducting state. The $G_{coh}^s$ satisfies the LSR. We also remark that along the Brillouin zone diagonals both $\Delta_{\vec{k}}$ and $\Delta_{SC}(\vec{k})$ vanish, and exactly along these directions there is only a quasiparticle pole
which crosses the Fermi energy at a Fermi wave-vector given by $\xi(\vec{k}_F)=0$. 

In a more recent publication, Yang {\it et al.} \cite{Yangnew} have observed that in the absence of particle-hole symmetry in the normal phase, it is more rigorous (justified) to use the proper {\it quasiparticle} basis in the normal phase such that:
\begin{equation}
G_{coh}^s = \sum_{\alpha=\pm} \frac{W_{\vec{k}}^{\alpha}}{\omega-E_{\vec{k}}^{\alpha} - \frac{\Delta_{SC}^2(\vec{k})}{\omega+E_{\vec{k}}^{\alpha}}},
\end{equation}
where $E_{\vec{k}}^{\pm} = \bar{\xi}_{\vec{k}}\pm \sqrt{\bar{\xi}^2_{\vec{k}}+\epsilon^2_{\vec{k}}}$, $\epsilon^2_{\vec{k}}=\xi(\vec{k})\xi_0(\vec{k})+|\Delta_{\vec{k}}|^2$, $\bar{\xi}_{\vec{k}} = (\xi(\vec{k})-\xi_0(\vec{k}))/2$, and $(W_{\vec{k}}^{\pm})^{-1} = 1+ |\Delta_{\vec{k}}|^2/(E_{\vec{k}}^{\pm} +\xi_0(\vec{k}))^2$. In the normal phase, the hole pockets satisfy the equations $E_{\vec{k}}^{\pm}=0$ that is equivalent to $\epsilon^2_{\vec{k}}=0$ or Eq. (91). In the superconducting phase, the four hole pockets are
naturally converted into Luttinger surfaces of zeroes. In fact, this propagator in the superconducting phase gives a better agreement with the experimental results and in particular with Scanning Tunneling Microscopy (STM) results of Kohsaka {\it et al.} \cite{Kohsaka}. Especially, the asymmetry in the tunneling density of states then becomes weaker due to the inherent particle-hole symmetry of superconductivity.

At this point, it is relevant to compare the phenomenological forms for $G^{RVB}$ and $G_{coh}^s$
to other proposals. It is very close to the form derived by Ng \cite{Ng} based on spin-charge separation but with an added ``not-too-rigorous'' attraction between spinon and holon (see discussion in Appendix C) which leads to quasiparticle poles and a self-energy similar to Eq. (97). Earlier, Norman {\it et al.} \cite{norman} introduced d-wave pairs plus lifetime broadening in the pseudogap phase.
The effect of classical phase fluctuations on the quasiparticle spectra has been addressed by Franz and Millis \cite{FranzMillis}. A recent comparison between a number of theoretical proposals for the arcs has
been drawn in Ref. \cite{Normanarcs}.

 \subsection{Overdoped cuprates and Breakdown of Landau-Fermi liquid}
 
 Now, we discuss the crossover to the overdoped regime. From the $N$-leg ladder analysis of Sec. 4.2 and 4.3, one can conclude that the Fermi liquid is stable at low energy when antiferromagnetic processes vanish completely, as a result of the circular Fermi surface that forms when overdoping the system. In this sense, the onset of superconductivity with doping should coincide in the normal phase with the onset of antiferromagnetic (umklapp) scattering. 

Recently, Abdel-Jawad {\it et al.} \cite{Abdel} have reported a correlation between charge transport and superconductivity in heavily overdoped high-$T_c$ superconducting cuprates. For a review, consult Ref. \cite{TailleferN}. They found that the onset
of superconductivity with doping coincides with the appearance in the normal phase of strong anisotropic quasiparticle scattering. The application of a magnetic field to suppress the superconductivity revealed that the anisotropic term in the inplane $(ab)$ transport scattering rate is {\it linear} in temperature violating the quadratic dependence characteristic of a Landau-Fermi liquid. 

Through a careful analysis, the authors report a scattering rate of the form:
\begin{equation}
\Gamma = \Gamma_0 + aT^2 + bT \cos^2(2\phi),
\end{equation}
where $\phi$ is the azimuthal angle. The isotropic part is standard for metals: the sum of impurity scattering
$(\Gamma_0)$ and electron-electron scattering $(aT^2)$. The last term, the anomalous one, is of particular
interest. The linear $T$ dependence of the in-plane resistivity stems from this term. In Tl-2201 samples \cite{Abdel} where doping $\delta=0.25$, the resistivity is not perfectly linear, but best described by:
$\rho=\rho_0+AT^2+BT$ \cite{McKenzie,Proust}. The existence of a large Fermi surface in the overdoped regime has been seen through the observation of quantum oscillations \cite{Vignolle}. The key finding of Ref. \cite{Abdel} is that the anomalous scattering is 
deeply anisotropic. It goes to zero as $\phi=\pi/4$ and is maximum at $\phi=\pi$. This angle dependence
suggests that the anomalous scattering has the d-wave symmetry. It should be noted that the linear
$T$ term disappears at $\delta=0.3$ \cite{McKenzie,Proust}; it coincides roughly with the onset of superconductivity at $\delta=0.27$.

Earlier theoretical investigations of the 2D square lattice via the functional RG method found that d-wave pairing in the overdoped region of the phase diagram was driven by the appearance of a strongly anisotropic
scattering vertex in the particle-particle and particle-hole channels 
\cite{Honerkampetal,Zanchi,Halboth}. Further investigations revealed that the self-energy is also anisotropic
\cite{Zanchi2,Honerkamp,Katanin,Rohe}.

 \begin{figure}
\begin{center}
\includegraphics[width=12.2cm,height=7.5cm]{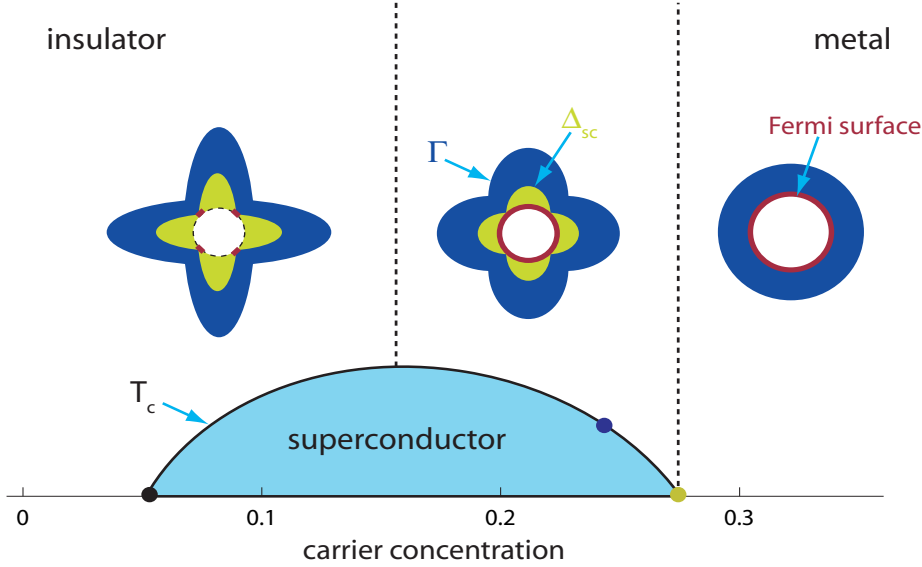}
\end{center}
\caption{Phase diagram inspired from Taillefer \cite{TailleferN}. Three regions can be distinguished from the behavior in the normal phase. For very large dopings, one gets a fully cylinrical Fermi surface, no superconductivity, and an isotropic scattering rate $(\delta>0.27)$. The overdoped region $(0.15<\delta<0.27)$ marks the simultaneous appearance of anisotropic d-wave superconductivity and anisotropic scattering mediated by antiferromagnetic processes at 
$(\pi;\pi)$. In the underdoped region, when the van Hove singularity touches the saddle points then the Fermi surface will become truncated in the antinodal directions where the scattering is maximum; the antinodal regions will form an insulating RVB region for $U>U_c$ where $U_c/t\propto \ln^{-2}(t/t')$ (see Sec. 4.4). By increasing the electron density, the charge and spin gaps propagate along the 2D umklapp surface and satisfy the generalized LSR (see Sec. 3 and 4).}
\end{figure}

Recently, Ossadnik {\it et al.} \cite{Ossadnik} have presented an extensive study of the doping and temperature dependence of the quasiparticle scattering rate with the pairing instability suppressed. Within
the RG approach, it is difficult to introduce a magnetic field, therefore the d-wave pairing instability has been
suppresed by an isotropic elastic scattering term added in the free part of the action. 

The self-energy at the Fermi surface is given the by the expression:
\begin{eqnarray}
\Sigma_{\Lambda=0} &=& \int_{\Lambda_0}^0 d\Lambda\  \theta(|\xi(\vec{k}_1)|-\Lambda)\delta(|\xi(\vec{k}_2)|-\Lambda)\theta(\Lambda-|\xi(\vec{k}_3)|) \\ \nonumber
&\times& V_{\Lambda}^2 G_0(k_1) G_0(k_2) G_0(k_3),
\end{eqnarray}
and summation and integration over internal momenta and Matsubara frequencies is implied. The two-loop
diagrams contributing to the self-energy are shown in Fig. 6 of Ref. \cite{Honerkamp}.
Here, $k_i = (\vec{k}_i,\omega_i)$, $\xi(\vec{k})=\epsilon(\vec{k})-\mu$ where $\epsilon(\vec{k})$ is given
in Eq. (68), and $G_0$ is the free propagator. The first propagator contains a sharp infrared cutoff $\chi_{\Lambda}=\theta(|\xi(\vec{k}_1)|-\Lambda)$, the second has support at, and the third below the cutoff
$\Lambda$. The initial condition gives $V_{\infty}(k_1,k_2,k_3)=U$ and the frequency dependence of all
the vertices has been neglected. To compute the self-energy, one needs to evaluate the 4-point vertex
$V_{\Lambda}(\vec{k}_1,\vec{k}_2,\vec{k}_3)$ at the energy scale $\Lambda$. As we focus on 
scattering rates at the Fermi surface or $\Im m\Sigma_{\Lambda=0}(\vec{k}_F,\omega)$, for a frequency 
$\lim_{\omega\rightarrow 0} \omega+i\delta$ with $\omega=0$, one
gets an imaginary part $\propto \delta(\xi(\vec{k}_3)-\xi(\vec{k}_2) - \xi(\vec{k}_1))$, reflecting energy
conservation \cite{Honerkamp}. 

It turns out that neglecting the flow of the 4-point vertices, or setting $V_{\Lambda}=U$, is equivalent to
a second order perturbative calculation of the scattering rate, which gives a $T^2$ behavior away from
the van Hove singularities. All deviations from the Landau theory scaling form then must be attributed
to the renormalization of the 4-point vertices. Now, let us imagine that at energy $\Lambda$, we focus on
the 4-point vertex $V_{\Lambda}(\vec{k}_1,\vec{k}_2,\vec{k}_3)$, and we fix the outgoing wave-vector
$\vec{k}_3$ close to $(\pi,0)$; Ossadnik {\it et al.} have shown that the strongest scattering process occurs at $(\pi,\pi)$. They also predict that these scattering vertices are roughly $\propto \Lambda^{-1/2}$ for small values of the cutoff $\Lambda\sim T$ \cite{Ossadnik}. Moreover, for very small
cutoff, the square root divergence is suppressed due to the presence of the isotropic scattering rate added in the free part of the action.

The factor $V_{\Lambda}^2$ in the self-energy calculation then results in $\Im m\Sigma \propto T$ \cite{Honerkamp}. It should be noted that this result is not due to 
a proximity to the van Hove singularity at the saddle point of the band structure: in this case, this lies well below the Fermi energy. In contrast, the breakdown of the Landau-Fermi liquid theory here emerges from the increase of the 4-point vertex with decreasing the energy scale. It should also be noted that this increase in the 4-point vertex is not restricted to the Cooper channel; examination of the RG flows shows that several channels in the 4-point vertex grow simultaneously, {\it e.g.}, the particle-particle umklapp process. On the other hand, it is quite surprising that this increase of the 4-point vertex gives exactly the same profile of $\Im m\Sigma \propto T$ as when the van Hove singularity lies at the (antinodal) saddle points \cite{Honerkamp,Nick}. 

We argue that the phenomenon associated with the increase of the 4-point vertex $V_{\Lambda}$ when decreasing the energy scale (thus producing the simultaneous reinforcement of several channels) can be seen as a precursory effect of the RVB behavior taking place at lower doping levels; see Fig. 14. 

To summarize briefly, Ossadnik {\it et al.} \cite{Ossadnik} have found that the scattering rates have a minimum in the nodal direction $(\phi=\pi/4)$ and they increase towards the antinodal direction. Moreover,  the coefficient $b$ in the scattering rate is found to increase when decreasing hole doping, whereas the coefficient $a$ of the Landau-Fermi liquid does not change much with doping. This is in agreement with the experimental results of Abdel-Jawad {\it et al.} \cite{Abdel}.

\section{Cold Atomic Fermions and Hubbard model in optical lattices}

The experimental realization of degenerate Fermi atomic samples (see Appendix D.1) have stimulated a new wave of investigations of quantum many-body systems \cite{Fermions1,Fermions2,Fermions3,Fermions4}.
Experimental advances include the observation of (molecular) fermion-pair condensates \cite{Fermion1}. Recently, both repulsive and attractive Hubbard models have been successfully implemented in optical lattices with fermions (using $^{40}$K atoms) \cite{fermion,Esslinger}. Fingerprints of the Mott state have been observed for repulsive interactions \cite{Esslinger,Blochnew}. This suggests that d-wave superfluidity on a square lattice might be realized in the future; s-wave superfluidity in an optical lattice has  already been reached (using $^6$Li) \cite{Ketterle}. Superfluidity in optical lattices has been discussed by Hofstetter {\it et al.} \cite{Walter}.

\subsection{Light and Atomic Parameters}

First, we find it instructive to discuss the parameters of those atomic Hubbard models, resulting from the trapping of atoms in a periodic potential creating by interfering laser beams (for a review, see Ref. \cite{Blochtrap}).

Consider an ensemble of fermionic atoms illuminated by several orthogonal laser fields (tuned far from
atomic resonance). These fields produce a periodic potential for atomic motion in two or three dimensions of the form (for more details, consult Appendix D.2):
\begin{equation}
V(\vec{x}) = V_0 \sum_{i=1}^{2(3)} \cos^2(k x_i),
\end{equation}
where $k$ is the wave-vector of the light. The depth of the lattice in each direction is determined by the intensity of the corresponding pair of laser beams which is easily controlled in an experiment. The potential depth $V_0$ is usually expressed in terms of the {\it recoil energy} $E_r= \hbar^2 k^2/2m$.
It is interesting to observe that for a YAG laser with $\lambda=2\pi/k=1.06\mu$m and $^6$Li atoms, $E_r\sim 1.4\mu$K. When venturing in the community of cold atoms, condensed matter physicists who are used to express energy scales in Kelvins (or electron-volts) need to remember that in units of frequency:
$1\mu \hbox{K}\sim 20.8$kHz.

The Bloch functions depend on the quasi-momentum and on the band index. It is
instructive to focus on the corresponding eigenenergies for different lattice depths $V_0/E_r$. 
Already for moderate lattice depths of a few recoils, one finds that the separation between the lowest lying bands is much larger than their extend \cite{Jaksch}. One can then focus on the lowest Bloch band and define Wannier wavefunctions that are optimally localized at the site $\vec{x}_i$. We are interested in a situation in which there is roughly {\it one} atom per lattice site; such atomic densities generally correspond to free-space Fermi energies on the order of $E_r$ (note that the kinetic energy and Fermi energy of atoms in the optical lattice are not the recoil energy but rather the bandwidth of the Bloch band under consideration, which strongly depends on the laser intensity $V_0$). Then, the atoms can tunnel from one site to another.

To realize the fermionic Hubbard model, one requires that two kinds of atoms are present; they can differ
by angular momentum or by generalized spin $(s=\ \uparrow,\downarrow)$. For sufficiently low temperatures, the atoms are confined to the lowest Bloch band, and the system can be described by the
Hubbard Hamiltonian given in Eq. (1). The parameter $t$ (depicted in Fig. 15A) corresponds to the tunneling matrix element
between adjacent sites\footnote{For a description of the Wannier wave functions and their corresponding tunneling amplitudes, consult Ref. \cite{Bloch}. The behavior of the hopping parameter $t$ can be qualitatively understood from a Wentzel-Kramers-Brillouin type argument. The value of the Hubbard parameter is obtained by taking the
Wannier state localized around a bottom well as the Gaussian ground state in the localized oscillator
potential. In the quantum optics community, the hopping term is often called $J$.}: 
\begin{equation}
t \approx E_r (4/\sqrt{\pi}) \xi^3 \exp(-2\xi^2),
\end{equation}
where $\xi$ is precisely given by $\xi=(V_0/E_r)^{1/4}$, and the parameter \cite{Bloch}
\begin{equation}
U = E_r a_s k \sqrt{8/\pi} \xi^3,
\end{equation}
characterizes the strength of the onsite interaction. The sign of the scattering length $a_s$ determines the nature of atomic interactions and can be controlled (tuned) experimentally: negative $a_s$ corresponds to attraction between atoms $(U<0)$ and positive $a_s$ corresponds to repulsion $(U>0)$. 

For example, in the case of $^6$Li both cases can be realized depending on the particular electronic states that are being trapped \cite{Fermions2,Fermions3}. The hopping amplitude $t$ and the on-site interaction $U$, calculated for the lowest band, are plotted as a function of $V_0/E_r$ in Fig. 15B (for a cubic lattice). 

\begin{figure}
\begin{center}
\includegraphics[width=10cm,height=8.8cm]{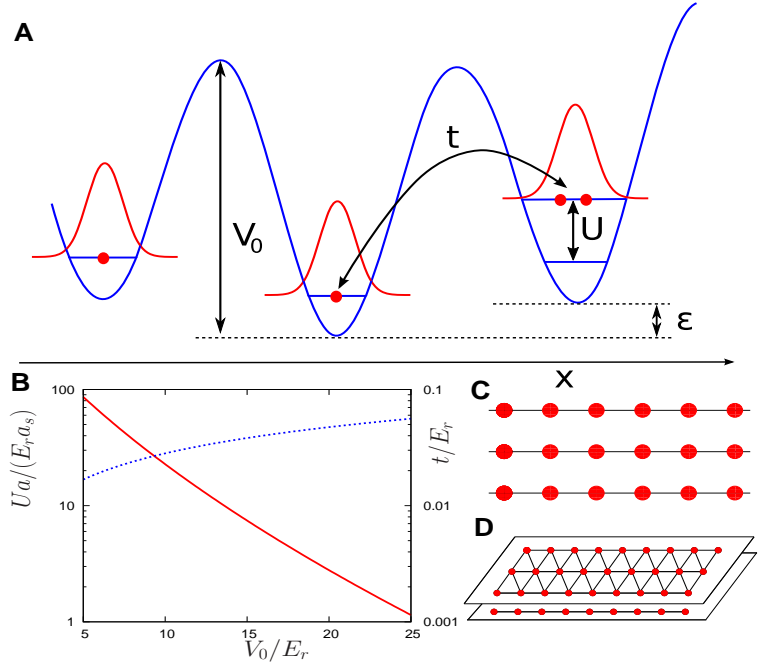}
\end{center}
\caption{A) Implementation of the atomic Hubbard model. B) The hopping strength $t$ and the Hubbard interaction $U$ can be tuned via the lattice $V_0$. C) and D) Different lattice geometries can be engineered by resorting to various pairs of lasers and by controlling the laser intensities. We have introduced the lattice spacing $a=\lambda/2$.}
\end{figure}

The realization of the Hubbard model with ultracold fermionic atoms requires 3 prerequisites: i) neglect
the second band, ii) neglect other interactions besides the Hubbard interaction $U$, and iii) replace
the interatomic potential by a pseudopotential approximation. Assumption i) is justified as long as the second band is not populated (less than two fermions per site, and $V_0$ not too small; typically, 
$V_0\gg 2.3 E_r$ such that the two bands do not overlap). We also assume that the on-site interaction $U$ is smaller than the separation between bands $\delta_S$. The assumption ii) is justified if other interactions are negligible. The most dangerous coupling turns out to be a `density-assisted' hopping
between two nearest-neighbor sites \cite{Georges}. The pseudopotential approximation consists to approximate the interaction potential by $V_{int}^{s,-s}(\vec{x}-\vec{x}')=g\delta(\vec{x}-\vec{x}')$ where $g=4\pi \hbar^2 a_s/m$. The Hubbard interaction $U$ is obtained by rewriting the interaction Hamiltonian in the basis of Wannier functions. Therefore, the assumption iii) is justified
when the typical distance between two atoms in a lattice well (which is given by the extension of the Wannier wavefunction $l$) is must be larger than the scattering length $l\gg a_s$. It is relevant to observe that for deep lattices $V_0\gg E_r$ this assumption coincides with the requirement that\footnote{One can use $l\sim a(E_r/V_0)^{1/4}$, $U/E_r\sim (a_s/a)(V_0/E_r)^{3/4}$, and $\delta_S\sim (E_r V_0)^{1/2}$; we have re-introduced the lattice spacing $a$. We have used the model of an harmonic oscillator with a typical frequency $\omega\sim \sqrt{E_r V_0}$; from the ground state of the harmonic oscillator, one finds $l\sim (E_r V_0)^{-1/4}$ ($\lambda\sim 1/k \sim 1/\sqrt{E_r}$).} $U\ll \delta_S$ and, for large $V_0/E_r$, boils down to:
\begin{equation}
\frac{a_s}{a} \leq \left(\frac{V_0}{E_r}\right)^{-1/4}.
\end{equation}
For a deep lattice, the scattering length should not be increased too much.

It is important to emphasize that optical lattices allow the realization of a variety of lattice geometries. The
lattice site positions $\vec{x}_i$ determine the lattice geometry. For instance, the arrangement of three (two) pairs of orthogonal laser beams leads to a simple cubic (square) lattice. Since the laser setup is very versatile different lattice geometries can be achieved quite easily. For example, three leaser beams propagating at angles $2\pi/3$ with respect to each other in the $xy$ plane (and all of them being polarized in the $z$ direction) allow to realize a 2D triangular lattice; an additional pair of lasers in the $z$ direction can be used to create localized lattice sites (see Fig. 15D). The motion of the atoms can be
restricted to one spatial dimension by large laser intensities. 

The size of an optical lattice is relatively small in comparison to its periodicity $a$ with typically a few hundred lattice sites in each dimension only. Also, this lattice is out into a {\it slowly} varying harmonic potential (trap) which translates into a space dependent site offset or
a local chemical potential $\epsilon(\vec{x})$ (see Fig. 15A). This can lead to the simultaneous coexistence of spatially separated regions where alternating superfluid and Mott phases are present \cite{Jaksch}. 

Below, we consider that the harmonic potential is smooth enough such that a large number of sites are subject to an almost flat potential. Those sites are described by the
same filling factor which can be controlled experimentally, by either increasing the
total number of atoms, by reducing $t$ via an increase of the lattice depth or via an increase
of the overall harmonic confinement. 

A next important question is the temperature of the many-body atomic fermion system. It has been shown that one can determine the temperature of a {\it non-interacting} 50/50 spin mixture of fermions by measuring the number of doubly occupied sites. When $U=0$, for zero temperature and deep optical lattices, one expects all lattice sites to be occupied by a spin-up and a spin-down equally. For finite temperatures, atoms can be thermally excited to higher lying lattice sites at the border of the system, thus reducing the number of doubly occupied sites. By converting doubly-occupied sites into molecules, it has been possible to determine the number of doubly occupied sites versus singly occupied sites and obtain an estimate of the temperature for the system \cite{kohl}. Nowadays, one can reach temperatures $\sim 0.1$T$_F$ where $T_F$ is the Fermi energy ($k_B=1$) of a non-interacting Fermi gas in an harmonic trap. 

When loading the atoms in the optical lattice, the bandwidth of the Bloch band under consideration 
becomes much smaller than the recoil energy $E_r$. When the lattice is gradually turned on, adiabatic
cooling is expected to take place (the entropy of the system cannot flow and the degree of degeneracy
is conserved). This leads to a (smaller) temperature of the lattice, say, $T_{lat}$. Here, the temperature here is not determined by some external bath but just by the entropy. The density of states is enhanced considerably as the band shrinks (since the one-particle states all fit in a smaller energy window); therefore, one expects that $T_{lat}$ is reduced compared to the initial temperature in the trap. Interactions can significantly modify these effects; see below. The Mott insulating regime can be reached assuming that $U\gg (t,T_{lat},\mu)$.  Recently, it has been confirmed experimentally that for $U\sim 24t$ the double occupancy is strongly reduced to values systematically below 2$\%$ (for small atom numbers) \cite{Esslinger}. This constitutes a fingerprint for the suppression of fluctuations in the occupation number. Schneider {\it et al.} have recently developed a new method to measure the compressibility of the many-body system \cite{Blochnew}. The entrance into the Mott regime is signaled by a minimum in the global compressibility. 

Concerning the antiferromagnetic ground state, a recent discussion on the N\' eel temperature $T_N$ can be found in Ref. \cite{Huse}. It is relevant to observe that according to quantum Monte Carlo simulations of the Hubbard model on a cube \cite{MC}, for a given  nearest-neighbor hopping matrix element $t$, the highest $T_N\sim t/3$ occurs for $U\approx 8t$, while for a given $U$ the maximal $T_N\sim U/20$ occurs at $t\approx 0.15U$. Thus, to increase $T_N$ one wants to move to large $t$ (which implies a weaker optical lattice, or smaller $V_0$), and to larger $U$ (which means a larger scattering length $a_s$). On the other hand, this moves the system away from the regime where it is well-approximated by the single-band Hubbard model; see Eq. (103).  Mathy and Huse \cite{Huse} have focused on the precise conditions which maximize the N\' eel temperature in the Mott insulating regime. 

It should be noted that numerical simulations \cite{MC} predict $T_N/J\sim 0.957$ on the cubic lattice (quantum fluctuations reduce $T_N/J$ from its mean-field value). 
Hence, $T_N/t$ or $J/t$ becomes small (as $t/U$) in the Mott insulating phase. Assuming $t\sim 0.1U$ then temperatures $T_{lat}$ on the scale of $\sim 10^{-2}E_r$ must be reached. Experimentally, it is crucial to reach such energy scales to prove the disappearance of the huge spin degeneracy below $T_N$ and at the same time to observe d-wave superfluidity since $T_c$ is smaller than $J$.

In the experiment \cite{Blochnew}, initial temperatures in the trap are in the range $0.15\leq T/T_F\leq 0.2$. At these temperatures, the entropy per particle is larger than $2\ln 2$. When switching on the lattice, a Mott insulator with unit filling and $\ln 2$ entropy per particle form in the center of the trap
even for $T/T_F=0.15$, for which the average entropy per particle is above $2\ln 2$. This is possible only due to the inhomogeneity of the system, as most entropy is carried by the metallic shells at the edges (where the entropy per particle can be very large). Therefore, the temperature $T_{lat}\sim t$ remains larger than $T_N$ (but much smaller than $U$). The next experimental challenge now consists to access temperature scales $T_{lat}$ of the order of the Anderson's superexchange $\sim J=4t^2/U$. Experimentally, one may first cool down the gas (in the absence of the optical lattice) down to a temperature where the entropy per particle takes a low enough initial value (smaller than $\ln 2$); interaction effects in the optical lattice then lead to adiabatic cooling mechanisms which could help to reach the N\' eel phase as $U/t$ is increased \cite{Achim}. A detailed discussion on this issue can be found in Refs. \cite{Georges,Georgeslight} (but the effect of the trapping potential has not been discussed). It seems very useful to think in terms of entropy... .

Nevertheless, it is relevant to mention that the existence of superexchange antiferromagnetic correlations has been shown  in an array of {\it isolated} double wells \cite{BlochS}. Also, the regime $T_{lat}\gg J$ is appropriate to observe the so-called spin-incoherent Luttinger liquid behavior of 1D fermion systems \cite{GKL,G}. 

\subsection{Plaquette models with ultracold fermions and d-wave superfluidity}

In fact, it has been known for some time that the minimal 2D unit that can sustain d-wave pairing physics is a single square plaquette \cite{ST}. 

Similar to the ladder system, the single plaquette model has been another relevant starting point of various approaches to the Hubbard model on a square  lattice 
\cite{Altman,Kivelson}. Interestingly, even for purely repulsive interactions, there is a range of parameters where two holes tend to bind together on a single plaquette rather than to delocalize among
different plaquettes. Here, we review the main results of the plaquette system along the lines of Ref. \cite{ST}. Those plaquette systems can be realized with atoms loaded
in 2D optical superlattices \cite{Rey}.

First, let us consider a plaquette of four sites with {\it four} electrons. As soon as one switches on the Hubbard interaction $U\neq 0$, it has been shown that the resulting ground state is a singlet with 
d-wave symmetry. For large $U$, the gap to the excited states is $\sim J$ (the splitting between the lowest lying singlet and triplet state: ``the spin gap''), whereas at small $U$ the (spin)  gap varies as $U^2$ \cite{CK}. In fact, all eigenstates of a 4-site Hubbard model can be calculated analytically by taking advantage of the symmetries of the model and the results have been published by R. Schumann \cite{Schumann}. Assuming that the inter-plaquette hopping $t'$  is small enough (see Fig. 16), then the gap can be treated as big. When the plaquette contains {\it two} electrons, the ground state has also been shown to be a singlet as a result of antiferromagnetic correlations.

\begin{figure}
\begin{center}
\includegraphics[width=8cm,height=7.8cm]{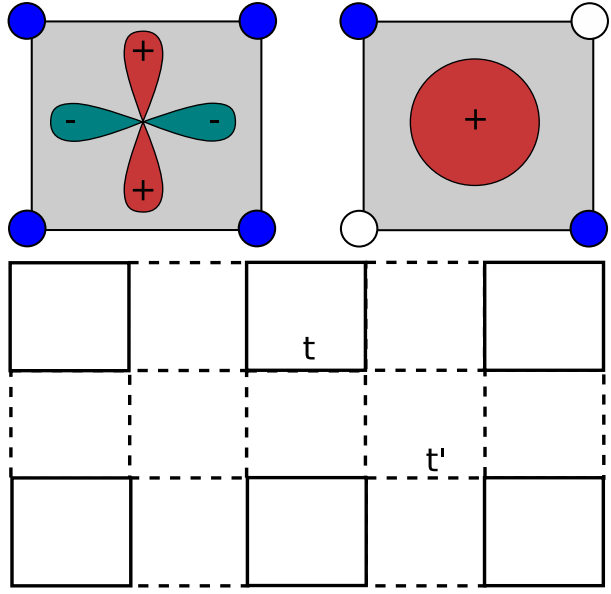}
\end{center}
\caption{A plaquette is the minimal system that exhibits d-wave symmetry. When loaded with 4 fermions
the ground state is d-wave symmetric whereas with 2 electrons the ground state exhibits s-wave
symmetry. Consequently those two states are connected through the d-wave pair creation operator. An atomic checkerboard lattice may be realized by coupling the plaquette through an hopping amplitude $t'$.}
\end{figure}

Now, let us discuss the emergence of d-wave pairing. For non-zero $U$, this can be seen by writing the wave function with two electrons (on 4 sites) as \cite{ST}:
\begin{equation}
|\Psi_2\rangle = K\left(d^{\dagger}_{2\downarrow}d^{\dagger}_{1\uparrow} + d^{\dagger}_{4\downarrow} d^{\dagger}_{1\uparrow}+ ...\right) |Vac\rangle,
\end{equation}
where $K$ is a normalization factor\footnote{The signs are in agreement with the  wavefunction at $U=0$: $|\Psi_2\rangle \propto (d_{1\downarrow}^{\dagger}+d^{\dagger}_{2\downarrow}+d_{3\downarrow}^{\dagger}+d_{4\downarrow}^{\dagger})(d^{\dagger}_{1\uparrow}+d^{\dagger}_{2\uparrow}+d_{3\uparrow}^{\dagger}+d_{4\uparrow}^{\dagger})|Vac\rangle$. If a nonzero $U$ is added to the Hamiltonian, all amplitudes remain positive, although they no longer have the same amplitude.}. The $(1,2)$ amplitude has the same sign as the $90^o$ rotated $(1,4)$
amplitude.  One can thus write:
\begin{equation}
|\Psi_2\rangle = \Delta^{\dagger}_s |Vac\rangle,
\end{equation}
where $\Delta^{\dagger}_s$ is an operator that creates an s-wave pair. On the other hand, when one starts from the Mott state the ground state is not the vacuum but rather the N\' eel state with exactly one particle per site $|\Psi_4\rangle$. The interesting point is that this same two-particle ground state 
$|\Psi_2\rangle$ can also be created by a $d_{x^2-y^2}$-wave operator removing particles from the Mott insulating state. For the
model system with four sites, $\langle \Psi_2| \Delta_d|\Psi_4\rangle$ is large whereas $\langle \Psi_2|\Delta_s|\Psi_4\rangle=0$.

More precisely, for a repulsive $U$ the largest real space
amplitudes in the 4-particle ground state wavefunction are for the ``N\' eel'' configurations:
\begin{equation}
|\Psi_N^1\rangle = d^{\dagger}_{4\downarrow} d^{\dagger}_{2\downarrow} d^{\dagger}_{3\uparrow} d^{\dagger}_{1\uparrow}|Vac\rangle,
\end{equation}
and the spin reversed state
\begin{equation}
|\Psi^2_N\rangle = d^{\dagger}_{3\downarrow} d^{\dagger}_{1\downarrow} d^{\dagger}_{4\uparrow} d^{\dagger}_{2\uparrow}|Vac\rangle.
\end{equation}
The emergence of d-wave pairing in the plaquette can be seen as follows. We examine the relative
phase for an electron pair on sites $(1,2)$ and the $90^o$ rotated pair on sites $(1,4)$. Annihilating the
appropriate electrons, one gets
\begin{equation}
d_{3\uparrow} d_{4\downarrow} |\Psi^1_N\rangle = (d_{3\uparrow}d_{4\downarrow})\left(d^{\dagger}_{4\downarrow} d^{\dagger}_{2\downarrow} d^{\dagger}_{3\uparrow} d^{\dagger}_{1\uparrow}\right)|Vac\rangle = - d_{2\downarrow}^{\dagger} d^{\dagger}_{1\uparrow} |Vac\rangle,
\end{equation}
and:
\begin{equation}
d_{3\uparrow} d_{2\downarrow} |\Psi^1_N\rangle = (d_{3\uparrow}d_{2\downarrow})\left(d^{\dagger}_{4\downarrow} d^{\dagger}_{2\downarrow} d^{\dagger}_{3\uparrow} d^{\dagger}_{1\uparrow}\right)|Vac\rangle = + d_{4\downarrow}^{\dagger} d^{\dagger}_{1\uparrow} |Vac\rangle.
\end{equation}
Thus, to have a non-zero overlap against the state $|\Psi_2\rangle$, one must use 
\begin{equation}
\Delta_d = (d_{3\uparrow} d_{2\downarrow} - d_{3\uparrow} d_{4\downarrow}+...),
\end{equation}
on the four-electron Mott insulating state; because of the minus sign, this is a $d_{x^2-y^2}$-wave operator.
To check that $\Delta_d$ is a singlet one can operate on the linear combination $|\Psi^1_N\rangle +|\Psi^2_N\rangle$; in the 4-particle ground state, $|\Psi^1_N\rangle$ and $|\Psi^2_N\rangle$ enter with a relative plus sign. The linear combination which gives the same relative signs as in $|\Psi_2\rangle$ is
obtained using (the d-wave operator) \cite{ST}
\begin{equation}
\left[(d_{3\uparrow}d_{2\downarrow} - d_{3\downarrow}d_{2\uparrow}) - (d_{3\uparrow}d_{4\downarrow}-d_{3\downarrow} d_{4\uparrow})+...\right] (|\Psi^1_N\rangle+|\Psi_N^2\rangle).
\end{equation}
{\it To summarize, the hole-pair creation operator connecting the $|\Psi_4\rangle$ and $|\Psi_2\rangle$ state must have a d-wave symmetry.}

When the hopping between plaquettes is zero, $t'=0$, thus the ground state of the model is known exactly for all $U/t$ and (also all band-fillings or dopings). In fact, the ground state of the single plaquette
problem can be obtained either by brute force \cite{Schumann} and Bethe Ansatz \cite{CK,LiebWu} (four site plaquette is a 1D ring with four sites). At exactly half-filling, the ground state is an insulator for $U/t>0$ (similar to the 2D Hubbard model and the ladder systems). 

Additionally, the insulator has a d-wave symmetry and is called the D-Mott plaquette state \cite{Ki}. It should be noted that this D-Mott plaquette state is slightly different from the D-Mott state for the two-leg Hubbard ladder system: here, the plaquette state is odd under $\pi/2$ rotations about the plaquette center.

\begin{figure}
\begin{center}
\includegraphics[width=11.1cm,height=5.5cm]{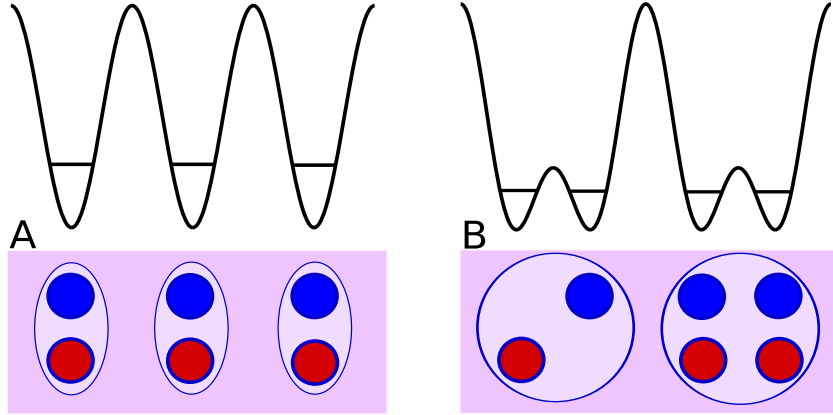}
\end{center}
\caption{In principle, different lattice geometries may be realized through superlattices \cite{Paredes}. (A) Ladder system. (B) Checkerboard lattice around $3/8$ filling.}
\end{figure}

Now, let us look at the hole-doped case, with 2 holes in two isolated plaquettes. Similar to the ladder,
the holes can bind together within the same plaquette or separate as single holes in each plaquette, depending on the binding energy:
\begin{equation}
\Delta_b = 2E_g(3) - E_g(4) - E_g(2),
\end{equation}
being positive or negative, respectively. Here $E_g(N_a)$ is the single plaquette ground state energy
when loaded with $N_a$ atoms. A positive $\Delta_b$ signifies an effective attraction between doped holes in the sense that for two doped holes, it is energetically preferable to place both on one cluster than to place one on each of two clusters. For the isolated square, it has been shown that $\Delta_b$ is positive for $0<U<U_c\sim 4.58t$ \cite{Schumann}. Then, two adjacent plaquettes can be coupled through a weak tunneling $t'$ to form a super-plaquette. As long as $0<t' < \Delta_b$, the states with 4 particles in one plaquette and 2 particles in the other are lower in energy and the states with one hole in each plaquette are energetically suppressed. Thus, starting with the half-filled case and doping with holes, may result in d-wave superfluidity since $\langle \Psi_2| \Delta_d|\Psi_4\rangle$ is large \cite{Rey}.  

In fact, a quite ideal situation to
observe d-wave superfluidity and avoid occupation states of a plaquette with 3 particles is to tune
$U\sim 2t$ ($\Delta_b$ is maximum) and to consider situations close to the $3/8$ filling, where 
the number of 4 particle states and the number of 2 particle states are (almost) equal. 

More precisely, by treating the $|\Psi_2\rangle$ and $|\Psi_4\rangle$ states as the pseudo-spin components $|\uparrow\rangle$ and $|\downarrow\rangle$ of an effective spin system, in the limit where $\Delta_b\gg t'_s$
where $t'_s$ is the hopping amplitude for each specie, then one can build an effective XXZ Hamiltonian (along the lines of the ladder case of Sec. 2.1):
\begin{equation}
H_{eff} = \sum_{\vec{R};\vec{R}'} \left(-J_{\perp}(\sigma^x_{\vec{R}}\sigma^x_{\vec{R}'}+\sigma^y_{\vec{R}}\sigma^y_{\vec{R}'})\right)+J^z \sigma^z_{\vec{R}}\sigma^z_{\vec{R}'},
\end{equation}
where the summation involves two neighboring plaquettes and \cite{Rey}:
\begin{equation}
J_{\perp} = g^2 \frac{t'_{\uparrow}t'_{\downarrow}}{\Delta_b},\hskip 0.5cm J^z = (t'^2_{\uparrow}+t'^2_{\downarrow})\frac{g_z^2}{2\Delta_b};
\end{equation}
In fact, the parameters $g(U/t)$ and $g_z(U/t)$ have been properly derived in Ref. \cite{Rey}; see Appendix D.3. At $3/8$ filling, the behavior of the function $g_z(U/t)$ is quite complicated. On the other hand, for $t_{\uparrow}' \sim t_{\downarrow}'=t'$, the system will be in the d-wave superfluid phase as long as $g\gg g_z$: this requires $0\ll U<2.7t$. Away from the 3/8 filling, the propagation of the hole pairs (plaquette states) in the system will be always favored (dilute limit), producing $d-wave$ superfluidity for $U<U_c\sim 4.58 t$. 

In the optical lattice, it should be noted that the critical temperature to reach the d-wave superfluid state is very small since
$t'\ll\Delta_b\sim 0.04t$ for $U\sim 2t$ \cite{Rey}. However, adiabatic cooling may however come to the rescue when the lattice is gradually turned on (the system uses constant entropy trajectories). An estimate of the entropy of the ordered state near $T_c$ has led to the prediction that one needs $T/T_F\sim 0.01$ to observe the d-wave (plaquette) superfluid \cite{Rey}.

\subsection{Noise Correlations and Entanglement within a BCS pair}

ARPES meaurements in condensed-matter systems allow to directly access the electron spectral function. This technique has played a key role in revealing the highly unconventional nature of single-particle excitations in cuprate superconductors \cite{Dama}. In the experimental study of ultra-cold atomic systems one can also resort to spectroscopic techniques, quite similar in spirit to what is done
in condensed-matter physics. This is the case, for example, when the observable we want to access is
a local observable such as the local density or the local spin density. Light (possibly polarized) directly couples to those, and light scattering is the tool of choice in the context of cold atomic systems. Recently, 
Stewart {\it et al.} have applied photoemission to cold atomic systems via an rf photon that ejects an atom
from the strongly interacting system \cite{Stewart}. 
Additionally, Bragg scattering can be used to measure the density-density dynamical correlation function \cite{Kurn}.  In optical lattices, the lattice spacing is set by the wavelength of the laser, hence lasers in the same range of wavelength can be used to sample the momentum-dependence of various observables, with momentum transfers possibly spanning the full extent of the Brillouin zone.

On the other hand, innovative noise correlations in cold atomic systems can be used to detect different quantum phases \cite{Altman2,Blochp,Greiner}. This technique makes use of the fact that the fluctuations in the momentum distribution after release from the trap contains information of the initial correlated quantum state. 

More specifically, what may be measured is the momentum space correlation function of the (fermionic) atoms in the trap:
\begin{equation}
 G_{s s'}(\vec{k},\vec{k}') = \langle n_{s\vec{k}} n_{s'\vec{k}'} \rangle - \langle n_{s\vec{k}} \rangle \langle n_{s' \vec{k}'}\rangle.
 \end{equation}
 Such correlation techniques in expanding atom clouds have begun to be successfully applied in recent experiments, revealing the quantum statistics of bosonic or fermionic atoms optical lattices (see Ref. \cite{Bloch} and references therein).
 
 Firstly, let us imagine that atoms are released from a single macroscopic trap.  In a typical experimental setup, the trapping potential is turned off suddenly, and the atoms are free to evolve {\it independently} (the atom velocity is assumed to be constant).
 This is valid if the collision cross-section is quite small; such conditions can be achieved experimentally.
 If interactions can be neglected during the time-of-flight, the average density distribution is related to the
 in-trap quantum state via \cite{Bloch,Altman2}:
 \begin{equation}
 \label{dens}
 \langle n_s(\vec{r}(t))\rangle_{tof} \approx \langle n_s(\vec{k}(\vec{r}))\rangle_{trap}.
  \end{equation}
  After a long time of flight the density distribution becomes proportional to the momentum distribution
  in the initial trapped state $\langle n(\vec{r}(t))\rangle \sim (m/ht) \langle n(\vec{k}(\vec{r}))\rangle$; the wave-vector $\vec{k}(\vec{r}) = m\vec{r}/(\hbar t)$ defines a correspondence between position in the cloud and momentum in the trap. For long time-of-flight times, the initial size of the atom cloud in the
  trap can be neglected. In fact, in each experimental image, a {\it single} realization of the density is observed, not the expectation value. On the other hand, Eq. ({\ref{dens}) is still meaningful because each bin $s$ in the image represents a substantial number of atoms $N_s$, while the atomic noise scales
  as $O(\sqrt{N_s})$. In fact, since $N_s$ is not macroscopic the density fluctuations are
  visible. They are characterized by the correlation function:
  \begin{equation}
  G_{s s'}(\vec{r},\vec{r'}) = \langle n_{s}(\vec{r}(t)) n_{s'}(\vec{r'}(t)) \rangle - \langle n_s(\vec{r}(t))\rangle\langle n_{s'}(\vec{r'}(t))\rangle.
  \end{equation}
  In analogy with Eq. (\ref{dens}), this can be related to ground state momentum correlations:
  \begin{equation}
  G_{s s'}\sim \langle n_s(\vec{k}(\vec{r}))n_{s'}(\vec{k}(\vec{r'}))\rangle - \langle n_s(\vec{k}(\vec{r}))\rangle \langle n_{s'}(\vec{k}(\vec{r'}))\rangle.
  \end{equation}
 Of course, there is a proportionality constant that we ignore, for simplicity. Now, let us consider a superfluid state of fermionic atoms. Using the BCS wave-function of Eq. (75), one easily obtains that
 $\langle n(\vec{r})(t)\rangle_{tof} = 2|v_{\vec{k}(\vec{r})}|^2$ \cite{Altman}. It is interesting to observe that the latter at $T=0$ is qualitatively indistinguishable from a Fermi distribution at $T=T_c$ \cite{Tinkham}. 
In particular, $|v_{\vec{k}}|^2$ at $T=0$ approaches unity well below the Fermi surface and zero well above. This traduces that the change in a `metal' on cooling from $T_c$ to $T=0$ should not be
described by the occupation numbers of one-electron momentum eigenstates.

The essential difference between these states in fact lies in the two-particle correlations. For every atom with wave-vector $\vec{k}$ in the BCS ground state, there is another one at exactly $-\vec{k}$. This implies prominent correlations between density fluctuations on diametrically opposite points in the expanding cloud:
 \begin{equation}
  G_{\uparrow\downarrow}(\vec{r},\vec{r'}) = 2|u_{\vec{k}(\vec{r})}|^2 |v_{\vec{k}(\vec{r})}|^2 \tilde{\delta}(\vec{r}+\vec{r'}),
 \end{equation}
 where $\tilde{\delta}(\vec{r}+\vec{r'})$ is a sharply peaked function of $\vec{r}+\vec{r'}$. This sharp peak
 is a direct analogue of the {\it first order} coherence peak in bosonic systems. It should be noted that
 the product $|u_{\vec{k}} v_{\vec{k}}|$ is proportional to the BCS gap, $\Delta_{SC}$.
 
 Now, let us consider atoms that are initially in an optical lattice. Then \cite{Altman2}
  \begin{equation}
 \langle n_s(\vec{r}(t))\rangle \approx \langle n_s (\vec{k}(\vec{r}))\rangle \propto \sum_{i;j} 
 e^{i(\vec{R}_i-\vec{R}_j).\vec{k}(\vec{r})}\langle d^{\dagger}_{is} d_{js} \rangle,
 \end{equation}
 where the wave-vector 
 $\vec{k}(\vec{r})$ now defines a correspondence between position in the cloud and quasi-momentum
 in the lattice. For example, in a superfluid state of bosons, where $\langle d^{\dagger}_{i} d_{j} \rangle
 =|\Psi|^2$ (where $\Psi$ is the superfluid order parameter), $n(\vec{r}(t))$ exhibits Bragg peaks at $\vec{k}(\vec{r})$ corresponding to reciprocal lattice
 vectors $\vec{P}$.  In the Mott state, on the other hand, $\langle d^{\dagger}_{i} d_j \rangle \sim \delta_{ij}$, and there is no interference pattern in $n(\vec{r}(t))$. However, it is interesting to observe that the D-Mott state
 for the plaquette system yields unusual interference peaks \cite{Sarma}, which could be observed experimentally in a time-of-flight experiment. 
 
 The density-density correlation function after a time-of-flight takes the form:
 \begin{eqnarray}
  G_{s s'}(\vec{r},\vec{r'}) &=& \sum_{i i' j j'} e^{i \vec{R}_{ii'}.\vec{k}(\vec{r})+i\vec{R}_{j j'}.\vec{k}(\vec{r'})}
  \langle d^{\dagger}_{is} d^{\dagger}_{js'} d_{j' s'} d_{i' s}\rangle \\ \nonumber
  &+&\delta(\vec{r}-\vec{r'})\langle n(\vec{r}(t))\rangle - \langle n(\vec{r})\rangle \langle n(\vec{r'})\rangle.
  \end{eqnarray}
  In a BCS state,  $G_{\uparrow \downarrow}(\vec{r},\vec{r'})$ contains terms proportional to $|\Delta_{SC}|^2$, and a superfluid with $|\Delta_{SC}| \neq 0$ must exhibit interference fringes at $\vec{k}+\vec{k}'= n\vec{P}$, where $n$ is an integer. The dominant contribution is for $i=j$ and $i'=j'$. 
  
In the case of the Hubbard model with repulsive interactions, one must take into account the proximity to the Mott state. In the pseudogap phase, one expects that the phase fluctuations of the pseudogap order parameter will destroy the coherence peak in $G_s(\vec{r},\vec{r'})$.  In the d-wave superconducting phase, using the renormalized mean-field theory, one predicts that the coherence peak in $G_{\uparrow \downarrow}$ will be proportional to $|\Delta_{SC}|^2$; see Eq. (81). The d-wave nature of the state will be signaled by a modulation of the peaks with an overall enveloppe showing d-wave nodal directions along $k_x=\pm k_y$ and $k'_x = \pm k'_y$. In the d-wave BCS state of the checkerboard system, $G_{\uparrow\downarrow}(\vec{k},\vec{k}')$ exhibits interference fringes at $\vec{k}+\vec{k}'=n\vec{K}$ where $\vec{K}$ is the reciprocal lattice vector of the plaquette array (which is half of the reciprocal lattice vector of the underlying lattice) \cite{Rey}. 
  
Noise correlations can also be used to detect superfluidity in ladder type systems.   For example, for the spinless Hubbard ladder in the p-wave superconducting state, using Eq. (30), we predict that $G(q,q') \propto |q+q'|^{1/\gamma -1}$; there is a singularity at $q'=-q$ since $\gamma>1$ even though phase fluctuations cannot be completely avoided in 1D. For electrons with spin, the singularity is more pronounced $G_{\uparrow\downarrow}\propto |q+q'|^{\eta-1}$ and $\eta\rightarrow 1/2$ (Eq. (11)). For the 1D case,  $q=k-k_F$ and $q'=k'+k_F$ denote wave-vectors (momenta) relative to the respective Fermi points. For a comparison, it should be noted that for a single Hubbard chain with attractive interactions \cite{Ashwin} $G_{\uparrow\downarrow}(q,q') \propto |q+q'|^{1/K_c -1}$, where $K_c$ is the Luttinger parameter in the charge sector (the system exhibits only a divergence for attractive interactions $K_c>1$). In contrast, a charge density wave or a Mott state would always exhibit an anticorrelation in $G_{\uparrow\uparrow}$ at $q'=q$.

It is useful to note that the noise correlations in Eq. (118) allow to confirm that within the BCS
state ``entanglement'' (singlet formation) only occurs between each Bell pair $(\vec{k},s)$ and $(-\vec{k},-s)$.
The amount of entanglement entropy is:
\begin{equation}
S = -z_{\vec{k}}\ln {z}_{\vec{k}} - (1- z_{\vec{k}})\ln(1-z_{\vec{k}}),
\end{equation}
where $z_{\vec{k}} = |u_{\vec{k}}|^2=1-|v_{\vec{k}}|^2$ in the BCS theory ($|v_{\vec{k}}|^2$ represents
the probability of a pair $(\vec{k},\uparrow;-\vec{k},\downarrow)$ being occupied). In a Fermi liquid phase, since $u_{\vec{k}}v_{\vec{k}}^*=0$, then we recover that there is no entanglement entropy within a pair of electrons $(\vec{k},s)$ and $(-\vec{k},-s)$. It is relevant to note the correspondence between entanglement and the superconducting energy gap via the product $u_{\vec{k}} v_{\vec{k}}^*$. 

\subsection{BCS-BEC crossover and pseudogap}

At this point, it is suggestive to discuss the ``BCS-BEC'' crossover taking place in harmonic traps, and address possible connections with high-$T_c$ cuprates concerning the existence of a pseudogap phase.  A discussion on the BCS-BEC crossover can also be found in Ref. \cite{Bloch}. What makes those gases so important is their remarkable tunability and controllability. Using a Feshbach resonance, one can tune the
{\it attractive} two-body interaction from weak to strong, and thereby make a smooth crossover from a BCS superfluid to a Bose-Einstein condensation (BEC) \cite{Leggett,and}. The field of BCS-BEC crossover is built around early observations by Eagles \cite{and} that the BCS ground state is much more general than was originally thought. If one increases the strength of the attraction and self-consistently solves for the fermionic
chemical potential $\mu$, this wave-function will correspond to a more BEC-like form of superfluidity. In the weak-coupling regime, $\mu=E_F$ where $E_F$ is the Fermi energy and BCS theory results. However at sufficiently strong coupling, $\mu$ begins to decrease, crossing zero and then ultimately becoming negative in the BEC regime. For negative $\mu$, the Fermi surface is gone
and the material is bosonic. Knowing the ground state what is the nature of superfluidity at finite $T$? 
Nozi\` eres and Schmitt-Rink were the first to address non-zero temperature \cite{NSR}.

In fact, one can make relevant observations \cite{Levin}. 

As we go from BCS to BEC, pairs will form above $T_c$ {\it without} condensation. In the normal state, the system takes advantage of the pairing attractive interaction. Only in the extreme BCS limit do pairs form exactly at $T_c$.

The fundamental entities in these superfluids are fermions. Pairs of fermions form a boson. Similar
to the pseudogap phase of high-$T_c$ cuprates, one measures these bosonic pairs indirectly via the fermionic gap parameter $\Delta(T)$. 

There will be {\it two} types of excitations in these BCS-BEC crossover systems. Only in strict BCS
theory are the excitations of solely fermionic character, and only in the strict BEC limit will the excitations
be solely of bosonic character. In the intermediate (``unitary'') regime the excitations consist of a mix of both fermions and bosons below a crossover temperature denoted $T^*$ (by analogy to the cuprates) while the superconducting transition takes place at $T_c<T^*$. 

\begin{figure}
\begin{center}
\includegraphics[width=12.7cm,height=3.4cm]{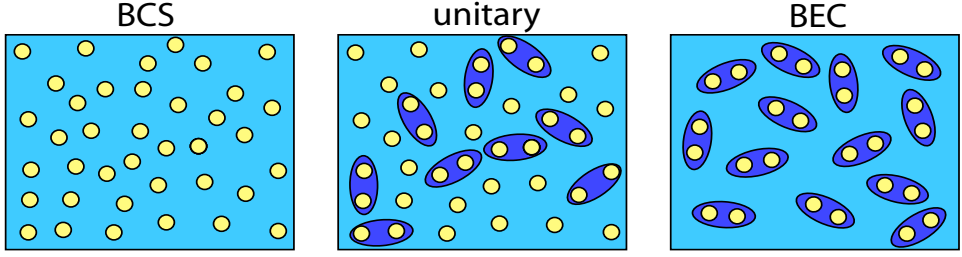}
\end{center}
\caption{Sketch of excitations in the BCS, unitary, and BEC regimes. The yellow disks represent fermionic excitations. Pair excitations become progressively dominant as the system evolves from the BCS to BEC regime.}
\end{figure}

An approach to address the BCS-BEC crossover is through the BCS Leggett T-matrix theory \cite{Levin}. 
One considers coupled equations between the particles (with dressed propagator $G$) and the pairs with propagator $t(P)$. The pairs and the fermions are coupled: at the mean-field level, the attractive BCS interaction term $U$ between fermions produces a coupling between the pairs and the fermions. At the mean-field level,  a pair can be converted into two fermions (Andreev type scattering). Pair-pair interactions
are only treated in a mean-field averaging procedure. Below $T_c$, then there are two contributions to the full T-matrix, from the non-condensed pairs with non-zero momenta or pseudogap (pg) and from the condensate (sc), $t=t_{pg}+t_{sc}$ where $t_{sc}(Q)=-\Delta_{SC}^2\delta(Q)/T$ and we denote $Q=(\vec{q},i\omega_n)$ and write $\sum_Q = T\sum_{i\omega_n} \sum_{\vec{q}}$, where $\omega_n$ is the Matsubara frequency. The electron self-energy then reads:
\begin{equation}
\Sigma(K) = \Sigma_{sc}(K) +\Sigma_{pg}(K) = \sum_P t(P)G_{0}(P-K),
\end{equation}
where $G_{0}$ is the free electron propagator: $G_0^{-1}=i\omega_n -\xi_{\vec{k}}$ and $\xi_{\vec{k}}=\epsilon_{\vec{k}} - \mu$ where $\epsilon_{\vec{k}} = \hbar^2 k^2/(2m)$. The label {\it pg} serves to label the ``pseudogap'' and the corresponding non condensed pair propagator. It follows:
\begin{equation}
\Sigma_{sc}(\vec{k},\omega) = \frac{\Delta_{SC}^2}{\omega+\epsilon_{\vec{k}} -\mu}.
\end{equation}
This is the usual BCS self-energy. Now, one needs to include the effect of the non-condensed pairs. An approximation below $T_c$ consists to replace \cite{Levin}:
\begin{equation}
\Sigma_{pg}(K) \approx -G_0(-K)\Delta_{pg}^2.
\end{equation}
This is valid below $T_c$ since there is a vanishing chemical potential for the pairs, which means that $t_{pg}(Q)$ diverges at $Q=0$ when $T<T_c$. It is reasonable to assume that noncondensed bosons in equilibrium with a condensate must necessarily have a zero chemical potential.
One gets $\Delta_{pg}^2 = -\sum_{P\neq 0} t_{pg}(P)$ (this is the pseudogap equation), and the pair propagator obeys
\begin{equation}
t_{pg}(P) = \frac{U}{1+U\chi(P)} \hskip 1cm P\neq 0,
\end{equation}
where $\chi(P)$ is the pair susceptibility defined as \cite{Levin}:
\begin{equation}
\chi(P) = \frac{1}{2}\left(\chi_{\uparrow\downarrow}(P) + \chi_{\downarrow\uparrow}(P)\right).
\end{equation}
Levin {\it et al.} assume that the latter is given by the product of one dressed and one bare Green's function,
\begin{equation}
\chi_{\uparrow\downarrow}(P) = \sum_K G_{0\uparrow}(P-K)G_{\downarrow}(K).
\end{equation}
(There are more than one possible choices in the normal phase.) Within this definition, one obtains consistent answers between T-matrix based approaches and the BCS-Leggett ground state equations. We have the usual BCS-like form for the electron self-energy $\Sigma(K)=
\frac{\Delta^2}{\omega+\epsilon_{\vec{k}} -\mu}$ below $T_c$ with 
\begin{equation}
\Delta^2(T) = \Delta^2_{pg}(T) + \Delta_{sc}^2(T).
\end{equation}
Since these finite momentum pairs do not have phase coherence, their contributions lead to a pseudogap in the fermion spectrum. In this context, the gaps add in quadrature as a result of the (valid) approximation in Eq. (125).

 It should be noted that this form of self-energy is distinct from that derived by Yang {\it et al.} \cite{zhangnew} for the underdoped cuprates in Eq. (97); see Sec. 4.6. 

Knowing the electron self-energy, one finds $G$ and $t_{pg}$. In fact, Levin {\it et al.} argue that $\Delta_{pg}^2(T)=Z^{-1}\sum_{q\neq 0} b(\Omega_q,T)$,
where $Z^{-1}$ is a residue, $b(x)$ is the Bose-distribution, and $\Omega_q=q^2/2M^*$ is the pair dispersion where $M^*$ is the ``effective'' pair mass \cite{Levina}. It should be noted that within this approach, the collective modes do not couple
to the pair excitations: this leads to a $q^2/2M^*$ form of the pair dispersion. In this sense, all inter-boson effects are treated in a mean-field sense and (only) enter to renormalize the effective pair mass $M^*$. The role of these modes at unitarity is currently unresolved. 

At $T=0$, $\Delta_{pg}$ vanishes since all pairs condense.

The BEC condition $\mu_{pair}=0$ gives the usual BCS gap equation $\Delta\sim\Delta_{sc}$:
\begin{equation}
1+ U\chi(0) = 0 = 1+ U\sum_k \frac{1-2f(E_k)}{2\sqrt{(\epsilon_k-\mu)^2+\Delta^2}}.
\end{equation} 
The number of fermions determines the chemical potential $n=2\sum_K G(K)$.
Above $T_c$, one must set $\Delta_{SC}=0$ and $\mu_{pair}\neq 0$. By introducing the Bose-Einstein distribution function $b(x)$ for non-condensed pairs, one predicts
\begin{equation}
\Delta_{pg}^{2} = Z^{-1}\sum_{q\neq 0} b(\Omega_q-\mu_{pair},T),
\end{equation}
Moreover, in the normal state, one gets $t_{pg}^{-1}=U^{-1}+\chi(0)=Z\mu_{pair}$. In the pseudogap phase, then one can determine $\mu_{pair}$, $\Delta_{pg}$, and $\mu$ by adding the
density equation $n= 2\sum_{K} G(K)$ \cite{Levin3}. It should be noted that the pseudogap physics in cold atomic systems has been observed, {\it e.g.}, in Ref. \cite{Thomas}.

It is certainly interesting to note the intersection of two different fields: high-$T_c$ superconductivity and superfluidity in atomic traps. In both cases, the pseudogap in the normal state consists of noncondensed fermion pairs. On the other hand, in high-$T_c$ cuprates the pseudogap is also related to Mott physics
which reflects the anomalously small zero temperature superfluidity; this aspect of the problem is not present in cold atomic systems with attractive interactions. Moreover, it is relevant to underline that for one particle per site on a bipartite lattice, the Hubbard model with $U>0$ maps onto the same model with $U<0$  under a particle-hole transformation; therefore, the spin density wave (weak coupling) regime corresponds to the BCS one and the Heisenberg (strong coupling) regime to the BEC one.

\section{Conclusion and Discussion}

First, we emphasize that superconductivity close to the Mott state appears in a variety of single-band Hubbard systems with repulsive interactions such as ladders, the checkerboard Hubbard model, and the homogeneous Hubbard model in two dimensions. The antiferromagnetic fluctuations at a large 
wave-vector $(\pi,\pi)$ unambiguously favor a d-wave type singlet pairing (see Eq. (2)). 

It is maybe important to underline that the concept of an electron-electron interaction mediated by magnetic spin fluctuations is not specific to the cuprates \cite{reviewS}. A similar process is believed to help facilitate p-wave spin-triplet pairing $(L=1, S=1)$ in superfluid $^3$He, and to lead to other pairing states in organic superconductors as well as heavy fermion systems \cite{Monthoux2,Lohneysen}. 

Now, we summarize the main ideas of this review and address a few questions.

{\it Theoretical Proof of d-wave superconductivity:} 
For quasi-1D systems, one can rigorously show that the (4-band) antiferromagnetic processes of Fig. 9 are responsible for the Cooper instability close to half-filling (even though phase fluctuations cannot be avoided even at zero temperature due to the reduced dimensionality). For the 2D Hubbard model in the strong coupling limit, {\it i.e.}, for the $t-J$ model, at a mean-field level, one can always decouple the Heisenberg exchange interaction as a BCS pairing channel, as shown in Sec. 4.5. The
renormalized mean-field theory plus variational Monte-Carlo methods applied to the $t-J$ model provide a good understanding of the d-wave BCS state for $0 < \delta < \delta_c$ where $\delta_c \sim 0.3$ and $J/t \sim 0.2$; results in the superconducting state are in agreement with experiments. It should be emphasized that a single square plaquette (the minimal 2D unit) already sustains d-wave pairing.

{\it Introducing Gutzwiller-type projected BCS wavefunctions:}
The discovery of high-$T_c$ superconductivity has certainly shown the way of using Gutzwiller-type projected BCS wavefunctions as trial wavefunctions to study complex strongly correlated electron systems. In general, the projected BCS wavefunction is appropriate to describe a resonating valence bond insulator, a Fermi liquid phase and the superconducting region. 
Strong correlation close to the Mott state is the driving force behind the phase diagram of high-$T_c$ cuprates and the resulting BCS state can be captured through a projected BCS wavefunction. Due to the proximity of the Mott state, the
situation is well distinguishable from the conventional BCS superconductors which are controlled
by the electron-phonon coupling \cite{Carbotte}. In particular, using the standard Gutzwiller approximation, one gets that all the quantities containing a single-particle operator are renormalized
with a factor $g_t$ vanishing as $\delta$ at small doping levels.

{\it $T_c$:}
Concerning the typical value of the critical temperature $T_c$ in the 2D Hubbard model for weak interactions, since the Cooper interaction is a `marginal' interaction in the RG sense, this results
in a $T_c$ that varies as $\exp-v_F/U$, and therefore is very small. For very large interactions, the
renormalized mean-field theory predicts that close to optimal doping $T_c \sim g_t J/2$ where $J$
is the Heisenberg superexchange (we have used the fact that the variational d-wave gap decreases linearly with $\delta$). A typical value for the cuprates is $J\sim 0.13$eV\footnote{The relatively large $T_c$ in the cuprates certainly results from this large value of $J$; for a comparison, note that
the Debye temperature for Cu is only $\sim 315$K.}; therefore, around optimal doping where $\delta\sim 0.15$ and $g_t\sim 0.26$, gives $T_c\sim 169$K. Dynamical Cluster approaches provide the same typical values for $T_c$ \cite{Andre,Jarrell}. Cuprates close to optimal
doping are governed by a $T_c$ from $90$K (for YBCO) to $138$K (for a Hg copper oxide superconductor). The connection between the spin gap and $J$ has been discussed experimentally, {\it e.g.}, in Ref. \cite{Amit}. At low doping levels, following the argument by Lee and Wen \cite{superfluid} for the destruction of the d-wave superconductivity by the quasiparticle excitations, one gets that $T_c$ follows the superfluid density at $T=0$, $\rho_s(T=0)$. From the renormalized mean-field, then one gets $\rho_s\propto \delta$ at low doping levels, reproducing the Uemura {\it et al.} plot, which draws a precise relationship between $T_c$ and carrier density. Recently, it has been suggested that significant enhancement of $T_c$ is possible in cuprate heterostructures \cite{Altmansu} under realistic conditions providing a possible explanation for measurements on LSCO bilayers \cite{Yulli}.

{\it Single-Band Model:} 
At this point, it is relevant to discuss the validity of the single-band Hubbard to describe cuprates.
At stoichiometry the $S=1/2$ spins of the Cu$^{2+}$ ions in the Mott insulator  order classically below
a N\' eel temperature $T_N\sim 300$K; $T_N$ is small as a result of the small interlayer coupling. 
The highest occupied orbital is is the $d_{x^2-y^2}$ orbital.
On the other hand, the copper is surrounded by six oxygens in a octahedral environment. The Cu orbital is singly occupied while the $p$ orbitals are doubly occupied. The $d$ and the $p$ orbital form
a strong covalent bond resulting in a large hopping integral $t_{pd}$. Using the hole representation, the oxygen is empty of holes and lies at an energy $E_p$ which is higher than $E_d$. In the parent compound, the virtual hopping 
process $t_{pd}$ then induces an effective exchange interaction: $J=t_{pd}^4/(E_p-E_d)^3$ where the effective hopping amplitude $t\sim t_{pd}^2/(E_p-E_d)$ and $E_p-E_d$ plays the role of the Hubbard $U$ in the one-band model. Experimentally, an energy gap of $2$eV is observed from optical conductivity. By doping the Cu-O planes, a doped hole resonates on the four oxygens surrounding a Cu and the spin of the doped hole combines with the spin on the Cu to form a Zhang-Rice singlet \cite{singlet}; the Zhang-Rice singlet can hop from site to site with an hopping amplitude $\sim t$, justifying  the single-band Hubbard model. This effectively produces sites with a Cu$^{3+}$ valence. A tight-binding model for Cu can be built thoroughly; see Ref. \cite{JohnKaryn,Pavarini}.

{\it Pseudogap phase:} Even though the d-wave superconducting state of the cuprates can be relatively well-described through a projected BCS wavefunction, the pseudogap phase still remains a ``delicate'' issue from the theory side. Several theories predict the possibility of an RVB-type state, {\it i.e.}, the existence of a pseudogap phase close to half-filling with preformed pairs and a truncated Fermi surface. 
More precisely, the ladders are nice prototype systems allowing to show the possibility of a truncated Fermi surface close to the Mott state \cite{UKM} where antinodal regions form the celebrated D-Mott state, which is characterized by a spin and a charge gap (see Sec. 3 and 4.2). The ladder systems also provide a quantitative expression for the electron's Green function in the pseudogap phase and allow to prove the (generalized) Luttinger Sum Rule.

In two dimensions, the RG approach in the 2-patch model, when the Fermi surface lies at the saddle points $(0,\pi)$ and $(\pi,0)$ (for $t'/t\sim 1/4$ this happens for $\delta\sim 0.2$), allows to predict a novel phase close to half-filling in the $t-t'-U$ model in which the antinodal directions become incompressible as a result of elastic umklapp processes. The opening of a spin gap has also been confirmed through numerical calculations at the low-energy fixed point \cite{Lauchli}. This incompressible spin-gapped region yields some resemblance with the D-Mott spin liquid region of the ladder system. By increasing the electron density, from the quasi-1D analysis for $t_{\perp}\approx t$, one predicts the propagation of this spin-gapped Mott region along the umklapp surface; this weak-coupling RG approach also suggests the appearance of Fermi arcs in the pseudogap phase. In the strong-coupling limit, the SU(2) slave-boson approach developed by Wen and Lee \cite{WL} also tends to predict the existence of Fermi arcs along the nodal directions, through a spinon-holon recombination. However, several approximations are used to study the effect of gauge fluctuations on the electron Green's function. Finally,  by combining the results from the ladder systems \cite{UKM} and those from the renormalized mean-field theory \cite{Zhang}, Yang {\it et al.} \cite{zhangnew,Yangnew} have proposed an ansatz for the single-particle Green's function in the pseudogap and in the superconducting phase, which  shows very good agreement with ARPES and STM measurements. A comparison with other existing theories for the electron propagator in the pseudogap phase can be found in Ref. \cite{Normanarcs}.

{\it Nodal-Antinodal Dichotomy and 2 gaps versus 1 gap:} It is relevant to observe that the RG approach in the weak-coupling limit and the slave-boson approach in the strong-coupling limit suggest two distinct scenarios for the occurrence of the pseudogap in the antinodal regions. The RG approach
in the $t-t'-U$ model shows the occurrence of an {\it incompressible} phase for $U>U_c$ where
$U_c/t\propto \ln^{-2}(t/t')$. As a reminiscence of the ladder systems, this pseudogap phase then can be seen as a precursor of the Mott transition. Below $T_c$, for the underdoped regime, superconductivity proliferates in the Fermi arc regions only as a result of the antiferromagnetic fluctuations (preformed pairs) in the antinodal regions; this scenario suggests that superconductivity in the nodal regions appear through an Andreev type process involving the hot (antinodal) regions. The quasi-1D and the weak-coupling RG approach in two dimensions predict the appearance of two energy scales and two distinct 
gaps close to half-filling: $T^*$ embodies the energy scale at which umklapp scatterings play an important role in the antinodal regions opening a charge and spin gap. $T_c$ represents the superconducting energy scale at which the Fermi arcs become superconducting. One important consequence is that the energy gap in the nodal regions does not scale with the pseudogap at $(\pi,0)$.
The two gap scenario is supported experimentally, by ARPES measurements, Andreev reflection studies, and Raman scattering. This is contrast with the Bose condensation scenario which only involves
a single gap, {\it i.e.}, the d-wave pseudogap. A direct evidence for competition between the pseudogap and superconductivity has been recently reported using ARPES \cite{Kondo}.

{\it Fermi liquid for severely overdoped systems:}
The Fermi liquid phase is relatively simple to capture. From RG arguments, it appears when antiferromagnetic fluctuations and umklapp scatterings are negligible due to the relatively large doping. The Cooper processes then become marginally irrelevant. 
In the large $U$ limit, the antiferromagnetic fluctuations are governed by the Heisenberg exchange
coupling $J$. The emergence of the Fermi liquid phase can be understood from the competition between the exchange energy $J$ and the kinetic energy which is of order $t$ per hole or $\delta t$
per unit area. When $\delta t\gg J$, we expect the kinetic energy to win and we expect a Fermi-liquid
metal with a weak residual antiferromagnetic correlation. This fact is also transparent in the slave-boson
theory where spin and charge bind when the d-wave gap is zero. One expects that the crossover to
the Fermi liquid phase should occur when $\delta t\sim J$ or $\delta\sim J/t\sim 1/3$ (in agreement with experiment at $T\rightarrow 0$). 

{\it Breakdown of Landau-Fermi liquid in the overdoped regime:} When $\delta t<J$,
in general, the outcome is less clear because the system would like to maintain the antiferromagnetic
correlation while allowing the holes to move as freely as possible. In particular, the evolution of (or to) the antiferromagnetic state at low dopings $\delta\ll 1/3$ remains a subtle question due to the proliferation of alternative states (d-density wave state, stripe phases, for example) even though the RVB theory emerging from the ladder analysis gives a reasonable (and quantitative) idea of the pseudogap phase. It is interesting to observe that Abdel-Jawad {\it et al.} \cite{Abdel} have reported a flagrant correlation between charge transport and superconductivity in overdoped high-$T_c$ cuprates. Recently, Ossadnik {\it et al.} \cite{Ossadnik} have shown that d-wave pairing in the overdoped region of the phase diagram is strongly related to the appearance of a strongly anisotropic scattering vertex both in the particle-particle and particle-hole channels.  Through a functional RG analysis, they have found a scattering rate with an isotropic Fermi-liquid like part and a strongly anomalous part that varies linearly with $T$ and presents a d-wave character. This scattering rate with d-wave symmetry emerges as a result of antiferromagnetic processes connecting the antinodal points and must be attributed to the increase of the 4-point vertex when decreasing the energy.
This phenomenon can be perceived as a precursory effect of the pseudogap region (which should emerge as soon as the Fermi surface touches the saddles points producing a van Hove singularity and an incompressible spin-gapped antinodal region as soon as $U>U_c$.)

{\it Other ingredients?} There are many reasons to believe that the single-band model contains most (but maybe not all) of the ingredients necessary for understanding high-temperature superconductivity in the cuprates. This does not preclude that additional phenomena, such as interactions between electrons and the lattice (phonons) and inhomogeneities at the mesoscopic level, must also be included to have a quantitative description of the phase diagram. In particular, recently it has been proven that the stripe order in  La$_{2-\delta}$Ba$_{\delta}$CuO$_4$ frustrates the three-dimensional phase order, but is fully compatible with 2D superconductivity \cite{Tranquada,Berg}.The strict interplanar decoupling arises because the planar superconductivity contains a periodic array of lines of $\pi$-phase shift which rotate
of $\pi/2$ up the $c$-axis together with the spin and charge stripe ordering in the low-temperature
tetragonal phase.

{\it Why is the high-$T_c$ problem hard, but important?} The superconducting region close to
the Mott state may already be unconventional and is an interesting question on its own. The weak-coupling RG approach in the quasi-1D limit and for the $t-t'-U$ model in two dimensions suggests a d-wave superconducting phase (with two distinct gaps) where only the Fermi arcs participate in the superfluid weight, whereas the antinodal regions remain incompressible. By increasing the doping further, the superconducting region then proliferates along the antinodal directions. The coexistence between Mott and superconducting regions in momentum space is an interesting idea that has to be pushed further.  The high-$T_c$ problem is hard because one must take into account the proximity to the Mott state that produces prominent phase fluctuations: the Gutzwiller projection is difficult to implement beyond the renormalize mean-field theory and the slave-boson method suffers from several approximations in the treatment of gauge fluctuations. Understanding the anomalous
pseudogap phase in the underdoped cuprates remains a central challenge. Rather than deal directly with the very difficult strong correlation, one can relax the condition and approach the problem assuming a weak to moderately strong onsite interaction. In fact, in the quasi-1D regime and in the 2-patch model, the RG approach gives quantitative (and encouraging) results close to the Mott state. From an experimental point of view, the situation is also encouraging: it now appears possible to probe the normal phase(s) at zero temperature by applying a large magnetic field \cite{ProustN}; this represents a powerful new tool to study the normal regime and especially the underlying physics in the antinodal regions.

{\it Similitudes with $Fe$-pnictide superconductors?}
The recent discovery of superconductivity in a family of Fe-based oxypnictides with quite large transition temperatures has led to tremendous activity aimed at identifying the pairing mechanism
in these materials. The high transition temperatures and the electronic structure of the Fe-pnictide
superconductors suggest that the pairing interaction is of electronic origin. Additionally, the involvment of antiferromagnetic correlations in the superconductivity also seems to emerge. On the other hand, a few
distinctions already appear between the two classes of unconventional superconductors: first, the Fe-based oxypnictides are described by a complex multi-band system and second, in the parent
iron based compounds, antiferromagnetism seems to emerge from a collective spin-density-wave order instability of an itinerant system (and not from local moment physics).

{\it Ultracold atoms as Quantum Simulators:} Ultracold atoms
in an optical lattice allow in principle to engineer the Hubbard model in a controlled way and to detect d-wave superfluidity via novel tools such as noise correlation measurements \cite{Altman}. In this sense, this would represent a quantum simulator as Feynman first envisaged it. Spectroscopic tools have been successfully implemented and it is now possible to apply photoemission or Bragg scattering type experiments to study cold atomic fermions. In fact, there are many promising roads for research with ultra-cold fermionic atoms. In particular, optical lattices have opened the possibility of studying
strongly correlated regimes, such as high-temperature superfluidity and its interplay with Mott localization. They also offer the possibility of studying these systems in regimes which are not usually reachable in condensed-matter physics ({\it e.g.}, under time-dependent perturbations bringing the system out of equilibrium), and to do this within a highly controllable and clean setting. At a very general level, optical lattices allow to engineer the many-body wave-function of large quantum systems by manipulating atoms individually or globally. On the other hand, to probe the N\' eel ground state of the half-filled Hubbard model  and the d-wave superfluid state when doping, this requires to reach initial temperatures in the trap such that the entropy per particle is smaller than $\ln 2$ (we have set $k_B=1$). It would also be worthwhile to understand more deeply the physics of atomic fermions at the unitary limit and the pseudogap phase in these systems. Finally, it is also relevant to underline that optical lattices also offer the possibility of exploring exotic forms of superfluidity out of equilibrium \cite{Rasch}.

{\bf Acknowledgments:} We thank our collaborators E. Dagotto, N. Furukawa, C. Gros, C. Honerkamp, J. Hopkinson, R. Konik, A. L\" auchli, U. Ledermann,  M. Ossadnik, M. Salmhofer, M. Sigrist, A. Tsvelik, M. Troyer, K. Y. Yang, and F. C. Zhang.  K. L. H. is also grateful to the National Science Foundation (NSF) for financial support related to this research, under the contract DMR-0803200, and to the members of CIFAR for stimulating discussions.

\appendix
\section{Renormalization Group}

Renormalization Group (RG) techniques have been developed in particle physics to investigate the
high-energy behavior. In condensed matter systems, RG techniques are rather applied to
study the low-energy condensed matter physics.

The Kadanoff-Wilson type high-energy mode elimination (see {\it e.g.} the review by Shankar \cite{Shankar}) has been
successfully applied to describe the physics of a single Hubbard band. The kinetic Hamiltonian $H_0$
describes Dirac fermions with the velocity $v=2t\sin(k_F)$ and the interaction part $H_{Int}$ takes the form:
\begin{equation}
H_{Int} = \Big(\prod_i \int_{-\Lambda}^{\Lambda} \frac{dk_i}{2\pi}\Big) \delta(k_1+k_2-k_3-k_4){h}_{Int},
\end{equation}
where
\begin{eqnarray}
{h}_{Int} &=& g_1 \Psi^{\dagger}_{Rs}(k_1)\Psi^{\dagger}_{Ls'}(k_2)\Psi_{Rs'}(k_3)\Psi_{Ls}(k_4)
\\ \nonumber
&+& g_2 \Psi^{\dagger}_{Rs}(k_1)\Psi^{\dagger}_{Ls'}(k_2) \Psi_{Ls'}(k_3)\Psi_{Rs}(k_4)
\\ \nonumber
&+& g_4 \big(\Psi^{\dagger}_{Rs}(k_1)\Psi^{\dagger}_{R\bar{s}}(k_2)\Psi_{R\bar{s}}(k_3)\Psi_{Rs}(k_4)
+R\leftrightarrow L\big),
\end{eqnarray}
where we have introduced $\bar{s}=\downarrow$ if $s=\uparrow$ and vice-versa. Here, we have neglected the umklapp scattering assuming that the system is not half-filled. The $g_4$ term includes
completely chiral interactions. This term does not affect the low-energy properties, but only renormalizes
the Fermi velocity (this is transparent when using bosonization); therefore in this work we systematically drop such terms. The $g_2$ term corresponds to forward scattering whereas $g_1$ describes backward scattering (or Cooper scattering). 

The Kadanoff-Wilson scheme is as follows. Let us denote $S$ the action corresponding to $H=H_0+H_{Int}$ and to the cutoff $\Lambda$, and the action corresponding to the kinetic term (Dirac fermions) reads:
\begin{equation}
S_0 = \sum_{p=R/L,s} \int_{-\infty}^{+\infty} \frac{d\omega}{2\pi} \int_{-\Lambda}^{+\Lambda} \frac{dk}{2\pi} \Psi^{\dagger}_{ps}(-i\omega \pm v k) \Psi_{ps},
\end{equation}
where $\pm$ corresponds to right and left movers respectively. The action $S'$ of the ``energy-reduced'' system is given by
\begin{equation}
e^{-S'} = \int D\bar{\Psi} e^{-S},
\end{equation}
where the (Grassmann) integration $D\bar{\Psi}$ is carried out over fields with $\Lambda/b<|k|<\Lambda$ and $b>1$; the wave-vector $k$ is measured from $k_F$ or $-k_F$. The renormalization group transformation is completed by a rescaling procedure which returns the cutoff to itÕs original value ($\tau=it$ is the imaginary time):
\begin{equation}
x \rightarrow bx; \tau \rightarrow b\tau; \Psi \rightarrow b^{-1/2}\Psi,
\end{equation}
such that the action $S_0' = S_0(\Lambda/b)=S_0(\Lambda)$ is invariant under the RG procedure. Now, let us decompose explicitly $\Psi = \bar{\Psi} +\Psi'$, where $\Psi'$ is the field to keep:
\begin{equation}
\int D\bar{\Psi} e^{-S} = e^{-S_0' - S_{I0}} \int D\bar{\Psi} e^{-\bar{S}_0 -\sum_{j=1}^4 S_{Ij}},
\end{equation}
where $S_{Ij}$ denotes the interacting part containing $j$ times the field $\bar{\Psi}$.  More precisely,
we have decomposed the noninteracting action as $S_0(\Psi) = S_0(\Psi') + S_0(\bar{\Psi}) = S_0' + \bar{S}_0$. The interacting part is a bit more involved and one has:
\begin{equation}
S_I = S_{I0} + \sum_{j=1}^4 S_{Ij}.
\end{equation}
The integration over $\bar{\Psi}$ is not straightforward and one has to perform a perturbative expansion.
The important contribution stems from the second order:
\begin{equation}
S' \approx S'_0 + S_{I0} -1/2\langle S_{I2}^2\rangle_{\bar{0}},
\end{equation}
where $\langle ...\rangle_{\bar{0}}$ indicates averaging over $\bar{\Psi}$. The calculation is done
diagramatically. Taking into account particle-particle and particle-hole
diagrams (that differ from an overall minus sign), in the action $S'$, the couplings $g_1$ and $g_2$
are renormalized as:
\begin{equation}
\delta g_1 = - \frac{\ln b}{\pi v} g_1^2\  \hbox{and}\ \delta g_2 = - \frac{\ln b}{2\pi v} g_1^2.
\end{equation}
Iterating the RG process $\Lambda\rightarrow \Lambda/b$, $\Lambda/b\rightarrow \Lambda/b^2$,..., 
then results in the RG equations for the couplings,
\begin{equation}
\frac{d g_1(l)}{dl} = -\frac{1}{\pi v} g_1^2(l)\ \hbox{and}\ \frac{dg_2(l)}{dl} = -\frac{1}{2\pi v} g_1^2(l),
\end{equation}
where $l=\ln b$. This shows that $g_1-2g_2=\ \hbox{const.}$ and that
\begin{equation}
g_1(l) = \frac{g_1(0)}{1+g_1(0)l/(\pi v)}.
\end{equation}
For repulsive interactions, $g_1(0)=g_2(0)=U>0$, then $g_1$ vanishes in the low-energy limit and $g_2=U/2$ in agreement with the Luttinger liquid. In the text, 
for a given {\it band}, say $1$, we use the notations $c_{11}^c=-(g_1/2-g_2)>0$ and $c_{11}^s=2 g_1$.

For the N-leg ladder case, the Operator Product Expansion (OPE) allows to derive the one-loop RG equations in a quite simple way. The RG equations for the situation close to and at half-filling are given in our Ref. \cite{UKM}. This approach systematically takes into account the U(1) symmetry
for charge and SU(2) symmetry for spin. The charge and spin currents have been introduced in Sec. 2.3.
A product of normal-ordered currents obeys the following short-distance expansion for $z\rightarrow 0$ (Wick-Theorem):
\begin{eqnarray}
J_R(z) J_R(0) &=& :\Psi^{\dagger}_{Rs}(z) \Psi_{Rs}(z): :\Psi^{\dagger}_{Rs'} (0) \Psi_{Rs'}(0):
\\ \nonumber
&=& : \Psi^{\dagger}_{Rs}(z) \Psi_{Rs}(z)\Psi^{\dagger}_{Rs'} (0) \Psi_{Rs'}(0): -\frac{2}{(2\pi z)^2}
\\ \nonumber
&-& \frac{i}{2\pi z} :\Psi^{\dagger}_{Rs}(z)\Psi_{Rs}(0): +\frac{i}{2\pi z} :\Psi^{\dagger}_{Rs}(0)\Psi_{Rs}(z):;
\end{eqnarray}
here, we have denoted $z=x+i v \tau$ for right movers (for left movers $z\rightarrow z^*$). Now, when including explicitly
the different bands, one gets the (non-umklapp) currents (for convenience, we redefine: $z_i=v_i\tau - ix$):
\begin{eqnarray}
J_{Rij} J_{Rlm} &\sim& \frac{\delta_{jl}}{2\pi z_j}J_{Rim} - \frac{\delta_{im}}{2\pi z_i}J_{Rlj} \\ \nonumber
J_{Rij}^a J_{Rlm}^b &\sim& \frac{\delta^{ab}}{4}\left(\frac{\delta_{jl}}{2\pi z_j}J_{Rim} - \frac{\delta_{im}}{2\pi z_i} J_{Rlj}\right)
+\frac{i\epsilon^{abc}}{2}\left(\frac{\delta_{jl}}{2\pi z_j} J_{Rim}^c +\frac{\delta_{im}}{2\pi z_i} J_{Rij}^c\right) \\ \nonumber
J_{Rij}^a J_{Rlm} &\sim& \frac{\delta_{jl}}{2\pi z_j} J_{Rim}^a - \frac{\delta_{im}}{2\pi z_i} J_{Rij}^a,
\end{eqnarray}
where we have introduced the totally antisymmetric tensor $\epsilon^{abc}$. These rules can be easily extended by including
umklapp scatterings. Since all the couplings are marginal, the RG equations take the general form:
\begin{equation}
\frac{d g_i}{d l} = A_{ijk} g_j g_k.
\end{equation}
For the (spinless) two-leg ladder, it is straightforward to show that the interband
Cooper interaction $c_{12}$ will renormalize all the other coupling channels. 

It is also instructive to fix the Hubbard initial values of the various couplings; to avoid any mistake, one
can make use of certain rules, such as:
\begin{equation}
J_{Rij} J_{Llm} - 4{\vec J}_{Rij}\cdot {\vec J}_{Llm} = 2(\Psi^{\dagger}_{Ris} \Psi_{Rjs} \Psi^{\dagger}_{Ll\bar{s}} \Psi_{L m\bar{s}} - \Psi^{\dagger}_{Ris} \Psi_{R j \bar{s}} \Psi^{\dagger}_{Ll\bar{s}} \Psi_{Lms}).
\end{equation}
In the case of spinful fermions, this results in:
\begin{equation}
 2 f_{k\bar{k}}^c = 2 c_{kk}^c = 2 c_{k\bar{k}}^c  = \frac{3U}{2(N+1)},
\end{equation}
where $(k=1,\bar{k}=N)$, $(k=2,\bar{k}=N-1)$,...\  , and for $i \neq k,\bar{k}$ one gets:
\begin{equation}
f_{ik}^c = c_{ik}^c = \frac{U}{2(N+1)}.
\end{equation}
For the SU(2) spin interactions one gets $f_{ik}^s = 4 f_{ik}^c$ and $c_{ik}^s = 4 c_{ik}^c$ for all $i$
and $k$. To determine the bare values of the umklapp processes, one can resort to:
\begin{equation}
I_{Rij}^{\dagger} I_{Llm} = \Psi^{\dagger}_{Rjs} \Psi_{Lls} \Psi^{\dagger}_{Ri\bar{s}} \Psi_{Lm\bar{s}}
+ \Psi^{\dagger}_{Rjs} \Psi_{Lms} \Psi^{\dagger}_{Ri\bar{s}} \Psi_{Ll\bar{s}}.
\end{equation}
One also infers that:
\begin{equation}
u_{k\bar{k} \bar{k} k}^c = 4 u_{kk \bar{k}\bar{k}}^c = \frac{3U}{2(N+1)},
\end{equation}
whereas $u_{k\bar{k}\bar{k} k}^s=0$ at the bare level. 

\section{4-band interactions and quasi-1D antiferromagnetism}

\subsection{Definitions}

Here, let us assume that the number of legs $N$ is even. 

The 4-band interactions combine processes 
between two different band pairs $(k,\bar{k})$ and $(l,\bar{l})$ (Fig. 9):
\begin{equation}
H_{4B} = \sum_{l=1}^{N/2-1} \sum_{k=l+1}^{N-l} \int dx\ h_{lk}^{4B},
\end{equation}
where,
\begin{eqnarray}
h_{lk}^{4B} &=& c_{lk\bar{k}\bar{l}}^c \left(J_{Rlk} J_{L\bar{k}\bar{l}} + J_{R\bar{l}\bar{k}} J_{Lkl} +h.c.\right) \\ \nonumber
&-& c_{lk\bar{k}\bar{l}}^s\left(\vec{J}_{Rlk}\cdot \vec{J}_{L\bar{l}\bar{k}} + \vec{J}_{R\bar{l}\bar{k}}\cdot
\vec{J}_{Lkl} +h.c.\right) \\ \nonumber
&+& u_{lk\bar{k}\bar{l}}^c\left(I^{\dagger}_{Rlk} I_{L\bar{k}\bar{l}} + I^{\dagger}_{R\bar{k}\bar{l}}I_{Llk}+h.c.\right) \\ \nonumber
&+& u_{l\bar{l}k\bar{k}}^c\left(I^{\dagger}_{Rl\bar{l}} I_{Lk\bar{k}} + I^{\dagger}_{Rk\bar{k}} I_{Ll\bar{l}} +h.c.\right) \\ \nonumber
&-& u_{lk\bar{k}\bar{l}}^s\left((\vec{I}_{Rlk})^{\dagger}\cdot\vec{I}_{L\bar{k}\bar{l}} + 
(\vec{I}_{R\bar{k}\bar{l}})^{\dagger}\cdot\vec{I}_{Llk}+h.c.\right).
\end{eqnarray}
Since $v_{k}=v_{\bar{k}}$ this also implies:
\begin{eqnarray}
c_{lk\bar{k}\bar{l}}^{c,s} &=& c_{l\bar{k}k\bar{l}}^{c,s} \\ \nonumber
u_{lk\bar{k}\bar{l}}^{c,s} &=& u_{l\bar{k}k\bar{l}}^{c,s}.
\end{eqnarray} 
The bare values of the four-band interactions:
\begin{equation}
2 c_{lk\bar{k}\bar{l}}^c = u_{lk\bar{k}\bar{l}}^c = u_{l\bar{l} k \bar{k}}^c = \frac{U}{N+1},\ c_{lk\bar{k} \bar{l}}^s=4 c_{l k\bar{k} \bar{l}}^c,\ u_{lk\bar{k}\bar{l}}^s=0.
\end{equation}
The 4-band couplings describe antiferromagnetic processes, {\it i.e.}, terms such as $\Psi^{\dagger}_{lRs}\Psi_{\bar{l}Ls'}$ involves a longitudinal momentum transfer of $\pi$. 

The RG equations are given explicitly in our Ref. \cite{UKM}. The key point is that, 
for a small number of chains, say $N=4$, the 4-band interactions stay in a fixed ratio \cite{UKM}:
\begin{equation}
\frac{c_{1234}^c}{c_{14}^c} \propto \left(\frac{U}{c_{14}^c}\right)^{5/12} \rightarrow 0.
\end{equation}
At low energy, the bands $1$ and $4$ are therefore decoupled from the bands $2$ and
$3$. The low-energy Hamiltonian is thus the sum of two-leg ladder Hamiltonians $H=H_{14}+H_{23}$.
All charge and spin excitations are therefore gapped (similar to the Heisenberg 4-leg spin ladder).

For $N$ odd, there is an umklapp term for the band $l=(N+1)/2$,
\begin{equation}
\int dx\ u_{ll}^c (I^{\dagger}_{Rll} I_{Lll} + h.c.),
\end{equation}
leading to a Mott gap in this band as well. 

\subsection{Quasi-1D antiferromagnetism}

Now, we study the quasi-1D antiferromagnetic phase. The RG equations for the 4-band couplings in the (simplified) limit of large $N$ are given in Ref. \cite{Urs}. We have checked that the 4-band couplings converge to the precise values \cite{Urs}
\begin{equation}
\label{flow}
t \sim 3g_{lk}=4 c_{lk\bar{k}\bar{l}}^c = 3 c_{lk\bar{k}\bar{l}}^s = 4u_{lk\bar{k}\bar{l}}^c = -3u_{lk\bar{k}\bar{l}}^s.
\end{equation}
Additionally, one can check that $u_{l\bar{l}k\bar{k}}^c$
remains small. For large $N$, the 2-band and single-band interactions are dominantly renormalized by the 4-band interactions. The asymptotic ratios of the 2-band and single-band interactions can be calculated by inserting the ratios (\ref{flow}) in the RG equations for these
couplings \cite{UKM}. One then finds that the couplings of the band pairs $(k,\bar{k})$ flow and approach (novel) fixed
ratios \cite{Urs}:
\begin{equation}
\label{afm}
t\sim 3g_{k\bar{k}}=4f_{k\bar{k}}^c = 3 f_{k\bar{k}}^s = 4 c_{k\bar{k}}^c = 3c_{k\bar{k}}^s = 4u_{k\bar{k}\bar{k}k}^c =
8 u_{kk\bar{k}\bar{k}}^c = 3 u_{k\bar{k}\bar{k}k}^s.
\end{equation}
The main difference with the D-Mott phase is that now $f_{k\bar{k}}^s$ flows to strong coupling and that $c_{kk}^s$ remain small (but attractive). For bands $k$ and $l$ that are close together
on the Fermi surface, $k\rightarrow l$, we check that the 4-band couplings become the same as the
corresponding 2-band couplings. In the limit $N\rightarrow \infty$, the RG equations simplify considerably and an {\it analytic} solution is in fact possible. In particular, 
the energy scale at which the
system enters in the antiferromagnetic phase and opens a {\it uniform} Mott gap can be 
determined using that \cite{Urs}:
\begin{equation}
\label{slk}
\frac{d}{dl} s_{lk} = \sum_{i\neq l,k}^{N/2} \frac{1}{v_i}s_{li}s_{ki},
\end{equation}
where
\begin{equation}
s_{lk} = \frac{c_{lk\bar{k}\bar{l}}^s}{4} - \frac{u_{lk\bar{k}\bar{l}}^s}{4} +c_{lk\bar{k}\bar{l}}^c + u_{lk\bar{k}\bar{l}}^c.
\end{equation}
For the extreme case $t_{\perp}\rightarrow t$, the scale of divergence is given by:
\begin{equation}
l_c = \frac{\pi t}{U}\frac{1}{\ln(2N/\pi)},
\end{equation}
where $U<t/\ln N$ for the validity of our calculations. For $t_{\perp}/t=0$, we find:
\begin{equation}
l_c=\frac{2t}{U}.
\end{equation}
The logarithmic correction for $t_{\perp}\rightarrow t$ stems from the van Hove
type singularity $(v_1=v_N\sim t/N)$. This allows us to check that a quasi-1D antiferromagnetic type ground state occurs at the energy scale
\begin{equation}
E_{AFM} \approx te^{-l_c} = te^{-ct/U},
\end{equation}
where $c$ is a function of $t_{\perp}/t$. Note that the 4-band antiferromagnetic processes with
a weight $\propto N$ are responsible for the RG instability at the energy scale $E_{AFM}$; see Eq. (\ref{slk}). This also favors the occurrence of a uniform Mott gap.

Remarkably, with the ratios of the couplings given above, one can check that the low-energy Hamiltonian for energy $E_c(N)<E<E_{AFM}$ takes the form
\begin{equation}
H = H_0-\frac{1}{2}\sum_{i,j} \int dx\ g_{ij} \vec{m}_i\cdot \vec{m}_j,
\end{equation}
where $H_0$ is the kinetic term and the staggered magnetizations obey:
\begin{equation}
m_k^r  = \Psi^{\dagger}_{kRs}\sigma_{ss'}^r \Psi_{\bar{k}Ls'} +h.c.
\end{equation}
In fact, the 2-band couplings $f_{k\bar{k}}^{c,s}$ and $u_{kk\bar{k}\bar{k}}^c$ give the contributions $\vec{m}_k\cdot \vec{m}_k$ and the 2-band couplings $c_{k\bar{k}}^{c,s}$ and $u_{k\bar{k}\bar{k}k}^{c,s}$ lead to
the products $\vec{m}_k\cdot \vec{m}_{\bar{k}}$. The 4-band couplings provide the other couplings in 
$\vec{m}_i\cdot \vec{m}_j$. For large $N$, one can check that all couplings take the same value $g_{ij}=g>0$. {\it This represents an effective spin Hamiltonian for the half-filled weakly-interacting 
Hubbard Hamiltonian.}

As a result, all the band pairs $(k,\bar{k})$ are interacting with each other. In particular, for $N$ odd, the (nodal) band $j=(N+1)/2$ is now interacting with all the other band pairs and there is no fundamental
difference between odd and even $N$. This contrasts the ladder-case at energies smaller than $E_c(N)$, where only interactions within the band pairs $(k,\bar{k})$ are present\footnote{In particular, the 4-band processes become suppressed due to the occurrence of the spin gap in the band pair $(k,\bar{k})$ or as a result of $K_{k\bar{k}s+}<1$ since $c_{kk}^s$ changes of sign. Here, $K_{k\bar{k}s+}$ denotes the Luttinger parameter associated with the symmetric spin mode within a band pair $(k,\bar{k})$.}; this leads to an odd-even effect in the spin sector that is reminiscent of the spin ladder. 

One can also resort to bosonization to analyze the low-energy properties \cite{Urs}. Interestingly, the charge sector is reminiscent of the two-leg ladder system, {\it i.e.}, for a band pair $(k,\bar{k})$, the total charge mode $\phi_{k\bar{k}c+}$ is pinned to zero whereas in the antisymmetric charge sector
$\theta_{k\bar{k}c-}\approx 0$; see Sec. 2.3. In contrast, in the spin sector, both the fields $\phi_{k\bar{k}s\pm}$ and the dual fields $\theta_{k\bar{k}s\pm}$ appear in a cosine, resulting in a competition between
different phases. However, since the Luttinger parameters in the spin sectors (for a given band pair $(k,\bar{k})$) now obey $K_{k\bar{k}s+}>1$ and $K_{k\bar{k}s-}<1$, this favors the pinning of the fields $\phi_{k\bar{k}s-}$ and $\theta_{k\bar{k}s+}$. This is the main difference with the D-Mott phase where one rather gets $K_{k\bar{k}s+}<1$ and $K_{k\bar{k}s-}<1$. 

To be more precise, only the differences $\theta_{k\bar{k}s+} - \theta_{l\bar{l}s+}$ appear in the
4-band Hamiltonian and as a result the total magnon mode, given by
\begin{equation}
\theta_T = \sqrt{\frac{2}{N}} \sum_{k=1}^{N/2} \theta_{k\bar{k}s+}\ \hbox{and}\ \phi_T = \sqrt{\frac{2}{N}}
\sum_{k=1}^{N/2} \phi_{k\bar{k}s+},
\end{equation}
remains gapless \cite{Urs}. The real-space spin operator at the site $x$ in chain $i$ reads:
\begin{equation}
S_i^r(x) = \frac{1}{2} \sum_k \gamma_{ik}\gamma_{i\bar{k}}\left(\Psi^{\dagger}_{kRs}\sigma^r_{ss'}
\Psi_{\bar{k}Ls'}e^{i(k_{Fk}+k_{F\bar{k}})x} + R\leftrightarrow L\right),
\end{equation}
where $\gamma_{jm}=\sqrt{2/(N+1)}\sin(\pi j m/(N+1))$ (consult Eq. (49)) and at half-filling $k_{Fk}+k_{F\bar{k}}=\pi$ for the band pair $(k,\bar{k})$.
Using the expression of $S_i^r(x)$ in terms of the different bosonic fields, one gets:
\begin{equation}
\langle \vec{S}_i(x)\cdot\vec{S}_l(0)\rangle \propto (-1)^{i+l} \cos(\pi x)/x^{1/N},
\end{equation}
and $x<1/E_c(N)$ since we are in the quasi-1D antiferromagnetic phase.

It is relevant to remember that the quasi-1D antiferromagnetic phase has the same charge properties
as the D-Mott state. By doping, the chemical potential $\mu$ couples to the total charge mode of each
band pair $\phi_{k\bar{k}c+}$. Since all the 4-band interactions contain the charge field $\phi_{k\bar{k}c+}$, it implies that they will be suppressed at low very energy scales. This shows that a (quasi-) long-range order is difficult (impossible) to stabilize away from half-filling, {\it i.e.}, in the presence of mobile holes. On the other hand, the effect of doping on the 2-band interactions is the same as when doping the D-Mott state. 

Therefore, close to half-filling, we still expect the occurrence of a truncated Fermi surface where the holes enter first the nodal directions and the antinodal directions form a D-Mott spin liquid state.

\section{Slave-boson approach for the $t-J$ model}

Here, we briefly review the slave-boson approach to the $t-J$ model. For more details, consult Refs. \cite{Lee,Lee2}. The slave-boson approach suggests that antiferromagnetic fluctuations (singlet pairs)
are prominent in the pseudogap phase below $T^*$. This explains the reduction of the magnetic susceptibility and the reduction of the c-axis conductivity. On the other hand, within this approach that
relies on spin-charge separation, superconductivity in the underdoped regime occurs when the holons (holes) condense \cite{Anderson}. In quasi-2D systems, the Bose-Einstein condensation temperature is roughly proportional to the areal density. Above $T_c$, preformed pairs exist, but the hole motion is incoherent resulting in ``gauge'' fluctuations. This approach predicts a single gap scenario but suffers from several approximations in the pseudogap phase.

\subsection{Slave-Boson scheme and Bose condensation}

The effect of the strong Coulomb interaction is represented by the fact that the electron operators
$d_{is}$ and $d_{is}^{\dagger}$ are the projected ones in which double occupancy is forbidden. This
means:
\begin{equation}
\sum_s d^{\dagger}_{is} d_{is} \leq 1.
\end{equation}
A method for treating this constraint is the slave-boson approach \cite{Barnes,Coleman}. 
This method was successfully developped for the Kondo problem.
The electron operator is then decomposed as:
\begin{equation}
d^{\dagger}_{is} = f^{\dagger}_{is}b_{i} +\epsilon_{ss'} f_{is'} a^{\dagger}_i,
\end{equation}
where $\epsilon_{\uparrow\downarrow}=-\epsilon_{\downarrow\uparrow}=1$ is the antisymmetric tensor, $f^{\dagger}_{is}$ is the fermion operator whereas $b_i$, $d^{\dagger}_{is}$ are the slave-boson operators. More precisely, $a^{\dagger}_i$ refers to a doubly-occupied site whereas $b^{\dagger}_i$ corresponds to the vacant state. Within this formulation, the constraint becomes:
\begin{equation}
\sum_{s=\uparrow,\downarrow} f^{\dagger}_{is} f_{is} + b^{\dagger}_i b_i + d^{\dagger}_i d_i  =1.
\end{equation}
In the limit of large onsite interaction, one can ignore doubly-occupied sites and as a result the
projected electron operator is written as:
\begin{equation}
d^{\dagger}_{is} = f^{\dagger}_{is} b_i,
\end{equation}
with the prerequisite:
\begin{equation}
\sum_{s=\uparrow,\downarrow} f^{\dagger}_{is} f_{is} + b^{\dagger}_i b_i =1.
\end{equation}
This constraint can be implemented through a Lagrange multiplier $\lambda_i$. The idea is to write
down the electron operator as the product of boson (charge or holon) and fermion which carries the spin index.

The Heisenberg term 
$\vec{S}_i\cdot\vec{S}_j-n_i n_j/4$ where $n_i=(1-b^{\dagger}_i b_i)$ can be re-written in terms of the 
spin operator $f_{is}$ if we neglect the nearest-neighbor hole-hole interaction $b^{\dagger}_i b_i b^{\dagger}_j b_j$ (this is reasonable if one considers cases close to half-filling with a small number of vacant sites). One can then apply a Hubbard-Stratonovich transformation to
decouple the exchange term both in the particle-particle and in the particle-hole channel. The partition
function then becomes:
\begin{equation}
Z = \int Df Df^{\dagger} D b Db^{*} D\lambda D\chi D \Delta \exp\left(-\int_0^{\beta} d\tau L\right),
\end{equation}
(formally $f$ becomes a (Grassmann) fermionic field and $b$ a boson field) where the Lagrangian obeys:
\begin{eqnarray}
L &=&\tilde{J}\sum_{\langle i;j\rangle} \left(|\chi_{ij}|^2 +|\Delta_{ij}|^2\right) + \sum_{is} f^{\dagger}_{is}(\partial_{\tau}-i\lambda_i)f_{is} \\ \nonumber
&-& \tilde{J} \left[ \sum_{\langle i;j\rangle} \chi_{ij}^* \left(\sum_s f^{\dagger}_{is} f_{js}\right) +h.c.\right]\\ \nonumber
&+& \tilde{J} \left[ \sum_{\langle i;j\rangle} \Delta_{ij}\left(f_{i\uparrow}^{\dagger} f_{j\downarrow}^{\dagger}-f_{i\downarrow}^{\dagger} f_{j\uparrow}^{\dagger}\right)+h.c.\right]\\ \nonumber
&+&\sum_i b_i^*(\partial_{\tau}-i\lambda_i+\mu)b_i -\sum_{i;j} t_{ij} b_i b_j^* f^{\dagger}_{is} f_{js},
\end{eqnarray}
where $\chi_{ij}$ represents {\it fermion (spinon) hopping} and $\Delta_{ij}$ represents {\it fermion} (spinon) pairing\footnote{They correspond to the two ways of representing the Heisenberg exchange interaction.}, and 
$\tilde{J}\sim J$\footnote{A discussion on the precise value of $\tilde{J}$ is given in Refs. \cite{Lee,Lee2}.}. Now, it is relevant to observe that the Lagrangian is invariant under a local $U(1)$ (gauge) transformation:
\begin{eqnarray}
f_{is} &\rightarrow& e^{i\theta_i} f_{is},\hskip 0.5cm b_i\rightarrow e^{i\theta_i} b_i \\ \nonumber
\chi_{ij} &\rightarrow& e^{-i\theta_i} \chi_{ij} e^{i\theta_j}, \hskip 0.5cm \Delta_{ij}\rightarrow e^{i\theta_i} \Delta_{ij} e^{i\theta_j}, \hskip 0.5cm \lambda_i \rightarrow \lambda_i +\partial_{\tau}\theta_i.
\end{eqnarray}
Note that fermions and bosons are coupled to the same gauge field.
The mean-field theory corresponds to the saddle-point solution to the functional integral\footnote{The saddle-point value of the Lagrange multiplier is $\lambda_i=\lambda$. For the half-filled case $\lambda=0=\mu$ and away from half-filling from half-filling one must satisfy $\sum_s \langle f^{\dagger}_{is} f_{is} \rangle = 1-\delta$.} ($\partial L/\partial \chi_{ij}^* =0$ and $\partial L/\partial \Delta_{ij}^*=0$):
\begin{equation}
\chi_{ij} =\sum_s \langle f^{\dagger}_{is} f_{js}\rangle \hskip 0.5cm \Delta_{ij}=\langle f_{i\uparrow} f_{j\downarrow} - f_{i\downarrow} f_{j\uparrow}\rangle.
\end{equation}
Similar to the renormalized mean-field theory, at half-filling, a variety of mean-field approaches give identical energy and dispersion \cite{Lee}, as a result of the underlying SU(2) particle-hole symmetry:
\begin{equation}
f^{\dagger}_{i\uparrow} \rightarrow \gamma_i f^{\dagger}_{i\uparrow} + \eta_i f_{i\downarrow} \hskip 0.5cm f_{i\downarrow} \rightarrow -\eta^*_i f_{i\uparrow}^{\dagger} + \gamma_i^* f_{i\downarrow}.
\end{equation}
Essentially, in the half-filled case, this approach suffers from the same problems as the renormalized
mean-field theory of Sec. 4.5. Note that at half-filling, there are no bosons and the theory is purely
that of fermions (similar to the renormalized mean-field theory).

On the other hand, by hole doping, one expects that the best state is the
projected d-wave superconductor, and that the theory becomes better defined. Essentially, the projected staggered flux state always lies above the projected d-wave state (and the energy difference goes to zero when $\delta\rightarrow 0$). The behavior of bosons is crucial for charge dynamics. Within the
mean-field theory bosons are free and condensed at $T_{BE}$. The bosons see the same band dispersion
as the fermions and condense at the bottom of the band minimum at low temperatures.
Assuming a weak three-dimensional hopping between `layers' one obtains a finite $T_c\sim T_{BE}$ that is proportional to the boson density $\delta$ \cite{Kotliar}. Below this temperature, we have electron pairing because the BCS order parameter $\langle d_{\vec{k}\uparrow} d_{-\vec{k}\downarrow} \rangle = b^2 \langle f_{\vec{k}\uparrow} f_{-\vec{k}\downarrow}\rangle \neq 0$ where $\langle b_i\rangle = b =\sqrt{\delta}$. 

This slave-boson approach supports the scenario by Anderson \cite{Anderson} that spin-singlet formation (RVB) turns into superconductivity via the Bose condensation of holons (vacant states).  Kotliar and Liu \cite{Kotliar} and Suzumura {\it et al.} \cite{Suzumura} found d-wave superconductivity in the mean-field theory. In the d-wave superconducting state, one gets: $\langle b_i\rangle \neq 0$, $\chi\neq 0$, and $\Delta\neq 0$. In the pseudogap phase, spin-singlet formation subsists but there is an incoherent charge motion: $\Delta$ and $\chi$ are nonzero while $b=0$. {\it Within this approach, $\Delta$ represents the pseudo-gap and the superconducting gap: this is a single gap scenario.}

It should be noted that the U(1) gauge theory (including Gaussian fluctuations around the saddle point) suffers from several limitations in the underdoped regime \cite{Lee}.

\subsection{SU(2) formulation}

Then, an SU(2) theory for underdoped high-$T_c$ cuprates taking into account Eq. (C.10) has been developed. At a general level, one can formulate an SU(2) gauge theory by introducing the fermion doublets:
\begin{equation}
\Phi_{i\uparrow}^{T} = (f_{i\uparrow}\ f_{i\downarrow}^{\dagger})\ \hskip 0.5cm 
\Phi_{i\downarrow}^{T} = (f_{i\downarrow}\  -f_{i\uparrow}^{\dagger}).
\end{equation}
Then, one can introduce an SU(2) version of the slave-boson theory \cite{WL}:
\begin{eqnarray}
d_{i\uparrow} &=& \frac{1}{\sqrt{2}}h_i^{\dagger}\psi_i = \frac{1}{\sqrt{2}}\left(b_{1i}^{\dagger} f_{i\uparrow} + b_{2i}^{\dagger}f_{i\downarrow}^{\dagger}\right) \\ \nonumber
d_{i\downarrow} &=& \frac{1}{\sqrt{2}} h_i^{\dagger}\bar{\psi} = \frac{1}{\sqrt{2}}\left(b_{1i}^{\dagger}f_{i\downarrow} - b^{\dagger}_{2i} f_{i\uparrow}^{\dagger}\right),
\end{eqnarray}
where we have introduced the (bosonic) vector $(h_i)^{\dagger}=(b_{1i}^{\dagger}\ b_{2i}^{\dagger})$ and similarly for the spin (spinon) part, and $\bar{\psi}=i\tau^2\psi^*$ where $\tau^2$ represents the second Pauli matrix. The $t-J$ model can now be written in terms of the fermion-boson fields. The local SU(2) singlets
must satisfy $(\psi^{\dagger}_i\vec{\tau}\psi_i + h^{\dagger}_i\vec{\tau}h_i)|phys\rangle=0$ to ensure
that the Hilbert space of the $t-J$ model is respected (operators $\psi^{\dagger}_i\vec{\tau}\psi_i$ that generate SU(2) local transformations vanish within the physical Hilbert space); $\tau_i$ are the Pauli matrices. On a given site, there are only three states that satisfy the above constraint: $f^{\dagger}_{\uparrow}|Vac\rangle$, $f^{\dagger}_{\downarrow}|Vac\rangle$, and $(1/\sqrt{2})(b_1^{\dagger}+b_2^{\dagger} f^{\dagger}_{\downarrow} f^{\dagger}_{\uparrow})|Vac\rangle$, corresponding to a spin-up a spin-down fermion, and a vacancy, respectively (again, $|Vac\rangle$ refers to the vacuum). The constraint leads to:
\begin{equation}
\sum_s \langle f^{\dagger}_{is}f_{is}\rangle + \langle b^{\dagger}_{1i} b_{i1} \rangle - \langle b^{\dagger}_{2i}b_{2i}\rangle =1.
\end{equation}
Within the SU(2) gauge theory, the density of fermions $\sum_s 
\langle f^{\dagger}_{is} f_{is}\rangle$ is not necessarily equal to $1-\delta$, where the hole doping is defined as $\langle b^{\dagger}_{1i} b_{1i} + b^{\dagger}_{2i} b_{2i}\rangle =\delta$. This is because
a vacancy in the $t-J$ model may be represented by an empty site with a $b_1$ boson or by two fermions (spinons) in a singlet state with a $b_2$ boson. It describes the physical idea that adding up a spin-up fermion or removing a spin-down fermion give the same state (after projection to the subspace of singly occupied fermions)\footnote{Annihilating a spin-up electron is produced by applying $f_{\uparrow}$ onto the vacancy state.}. 

In the superconducting phase, one must satisfy $\langle b_{2i}\rangle=0$ such that the SU(2) approach agrees with the U(1) slave-boson theory. 

The mean-field Hamiltonian takes the form:
\begin{eqnarray}
H = \sum_{\langle i;j\rangle} \frac{3}{8} J\left[\frac{1}{2}\hbox{Tr}(U_{ij}^{\dagger}U_{ij}) +(\psi^{\dagger}_iU_{ij}\psi_j+h.c.)\right] \\ \nonumber
-\frac{1}{2}\sum_{\langle i;j\rangle} t(h_i^{\dagger} U_{ij} h_j + h.c.) - \mu \sum_i h^{\dagger}_i h_i 
\\ \nonumber
+\sum_i a_0^l(\psi^{\dagger}_i\tau^l \psi_i +h_i^{\dagger}\tau^l h_i).
\end{eqnarray}
The averaged constraint $\langle \psi^{\dagger}_i\vec{\tau}\psi_i\rangle=0$ has been enforced through
the Lagangian multipliers $a_0^l$, and $\vec{a}_0$ is invariant under spin rotation, and
\[ U_{ij}=
 \left( \begin{array}{cc}
-\chi_{ij}^* & \Delta_{ij} \\
\Delta_{ij}^* & \chi_{ij} \end{array} \right).\] 
In this SU(2) formulation the d-wave superconducting phase is described by the following mean-field
Ansatz (by analogy to the U(1) slave-boson mean-field theory) \cite{WL}:
\begin{eqnarray}
U_{i,i+\hat{x}} &=& -\chi\tau^3 +\Delta \tau^1,\hskip 0.5cm U_{i,i+\hat{y}} = -\chi \tau^3-\Delta \tau^1, \\ \nonumber
a_0^3 &\neq& 0,\hskip 0.5cm  a_0^{1,2}=0,\hskip 0.5cm \langle b_1\rangle \neq 0,\hskip 0.5cm \langle b_2\rangle=0,
\end{eqnarray}
In the superconducting phase, the fermion and boson dispersion are given by $\pm E_F$ and $\pm E_b-\mu$ where
\begin{equation}
E_f =\sqrt{(\epsilon_f+a_0^3)^2+\eta_f^2},
\end{equation}
and
\begin{equation}
\epsilon_f = -\frac{3J}{4}(\cos k_x +\cos k_y)\chi,\hskip 0.5cm \eta_f=-\frac{3J}{4}(\cos k_x-\cos k_y)\Delta.
\end{equation}
Additionally, one obtains:
\begin{equation}
E_b=\sqrt{(\epsilon_b+a_0^3)^2+\eta_b^2},
\end{equation}
where
\begin{equation}
\epsilon_b = -2t(\cos k_x+\cos k_y)\chi,\hskip 0.5cm \eta_b = -2t(\cos k_x -\cos k_y)\Delta.
\end{equation}
The bosons condense (at $k=0$) at the bottom of the band minimum. 

Interestingly, the Fermi liquid phase is similar to the superconducting phase except that there is no fermion pairing $(\Delta=0)$. The Fermi liquid contains boson condensation. In this case, the electron
Green's function $\langle d^{\dagger} d\rangle \propto \langle \psi^{\dagger} h h^{\dagger} \psi\rangle$ is proportional to the fermion Green's function $\langle \psi^{\dagger} \psi\rangle$. Thus the electron spectral function contains a $\delta$-peak in the Fermi liquid phase.

\subsection{Spinon-holon binding in the Pseudogap phase}

Within the $SU(2)$ slave-boson approach, the ``pseudogap'' phase is
described by a staggered flux liquid (sfL) phase\footnote{The physics is similar to non-interacting electrons on a lattice pierced by magnetic flux lines whose directions alternate in a checkerboard pattern.}, where \cite{Affleck}:
\begin{eqnarray}
U_{i,i+\hat{x}} &=& -\chi\tau^3 -i(-1)^{i_x+i_y}\Delta,\hskip 0.5cm U_{i,i+\hat{y}} = -\chi\tau^3+i(-1)^{i_x+i_y}\Delta, \\ \nonumber
a_0^l &=& 0,\hskip 0.5cm \langle b_{1,2}\rangle = 0.
\end{eqnarray}
In the sfL phase, the fermion and boson dispersions are given by $\pm E_f$ and $\pm E_b-\mu$, but
now $a_0^3=0$. Since $a_0^3=0$ we must have $\sum_s \langle f^{\dagger}_{is} f_{is} \rangle =1$ and
as a result $\langle b^{\dagger}_1 b_1\rangle = \langle b^{\dagger}_2 b_2\rangle = \delta/2$.  In fact, the sfL phase and the d-wave pairing phase of the fermions are connected via a site-dependent SU(2) transformation that is explicitly given in Refs. \cite{Lee,WL}. Moreover, due to the U(1) gauge fluctuations, the antiferromagnetic spin fluctuations in the sfL phase are as strong as those of a nested
Fermi surface despite the presence of a pseudogap. In practice, the sfL phase resembles an exotic 
mixed Spin Density Wave (SDW) + superconducting (SC) state.

Now, it is relevant to focus on the electron spectral function in the pseudogap phase (that corresponds to
the sfL phase in the SU(2) slave-boson approach). At the mean-field level, the electron Green's function is given by the product of the fermion and boson Green's function, so the electron spectral function is a convolution of the boson and fermion spectral functions. 

In the pseudogap phase above $T_c$, a first approximation generally used in the literature consists to argue that the boson is almost condensed, so the boson spectral function contains a peak at ${k}=0$;
the weight of this peak is $\sim \delta$ and the width is $\sim T$. This
leads to the following mean-field BCS-like electron Green's function:
\begin{equation}
G_{el} = \frac{\delta}{2}\left(\frac{u_{\vec{k}}^2}{\omega-E_f} +\frac{v_{\vec{k}}^2}{\omega+E_f}\right) + 
G_{inc},
\end{equation} 
where $u_{\vec{k}}$ and $v_{\vec{k}}$ are the BCS coherence factors:
\begin{equation}
u_{\vec{k}} =\sqrt{\frac{E_f+\epsilon_f}{2E_f}}\hbox{sgn}(\eta_f),\hskip 0.5cm v_{\vec{k}} = \sqrt{\frac{E_f-\epsilon_f}{2E_f}}.
\end{equation}
The peak in the electron spectral function crosses zero energy at four points $(\pm \pi/2,\pm \pi/2)$. Thus, the mean-field sfL phase has four Fermi points (similar to the d-wave phase). 

The second approximation consists to argue that the gauge fluctuations produce
a strong attraction between bosons and fermions due to the fluctuations around the mean-field state.
This attempts to bind bosons and fermions and
form {\it Fermi arcs}. Then, one can provide the following ansatz for the ``renormalized''  electron Green's function \cite{Lee}:
\begin{equation}
G(\vec{k},\omega) = \frac{1}{G_{el}^{-1}(\vec{k},\omega) +V(\vec{k})},
\end{equation}
where
\begin{equation}
V(\vec{k}) = U +2t(\cos k_x+\cos k_y).
\end{equation}
The first term comes from the fluctuations of $a_0^l$ which is approximated by an on-site attraction:
$\delta H =- U d^{\dagger} d = -U/2(\psi^{\dagger} h)(h^{\dagger}\psi)$. The second term is due to
the fuctuations of $|\chi_{ij}|$ which induces $\delta H =-2t d^{\dagger}_j d_i$ (this reproduces the
original hopping). The value of $U$ then is fixed through the LSR $\int_0^{\infty} d\omega/(2\pi)\int d^2\vec{k}/(2\pi)^2\Im m G =\delta$. Within this scenario, the antinodal directions survives the binding energy $V(\vec{k})$. On the other hand $V(\vec{k})$ produces a negative chemical potential resulting in small hole pockets \cite{Lee}. 

A recent progress in this direction has been done in Refs. \cite{Ribeiro,Bieri}.

It should be noted that the nodal-antinodal dichotomy emerging from the SU(2) save-boson analysis
is distinct from that arising in the weak-$U$ limit from the RG approach; in the latter case, the antinodal regions are subject to prominent Umklapp scatterings that totally suppress the charge compressibility in the antinodal regions and produce a 2-gap scenario for underdoped cuprates (see Sec. 4). Additionally, in the latter scenario, the ``pairing glue'' around the nodal directions would stem from an Andreev-type scattering mechanism generated by the hot (antinodal) regions \cite{Furukawa,KarynAndreev,Geshkenbein}. Finally, one must bear in mind that within the  weak-coupling RG analysis, the appearance of a truncated Fermi surface strongly depends on the form of the tight-binding Hamiltonian (one requires a next-nearest neighbor hopping) while details of the electron band structure are rather unimportant for large interactions.

\subsection{Theory of Quasiparticles in the superconducting state}

Finally, we find it useful to discuss the superfluid density in the d-wave superconducting state, following
Lee and Wen \cite{superfluid}. We will  also address the theoretical predictions from the $SU(2)$ gauge
theory \cite{WenLee}.

Let us start with a superconductor where the elementary excitations in the superconducting state 
are well defined quasiparticles with dispersion:
\begin{equation}
E(\vec{k}) = [(\epsilon_{\vec k} -\mu)^2 +\Delta_{\vec{k}}^2]^{1/2},
\end{equation}
where $\epsilon_{\vec{k}}$ represents the band dispersion and 
the d-wave gap is defined as:
\begin{equation}
\Delta_{\vec{k}} = \frac{\Delta_0}{2}(\cos k_x - \cos k_y).
\end{equation} 
{\it Here, following, Lee and Wen \cite{superfluid}, we assume that there is a single gap scenario which is maximum of $\Delta_0$ at $(0,\pi)$}.
The important point is that in the d-wave superconducting state, there are four nodal points. In the vicinity
of the node near $(\pi/2,\pi/2)$, one has the anisotropic Dirac spectrum:
\begin{equation}
E(\vec{k}) = (v_F^2 k_1^2 +v_{\perp}^2 k_2^2)^{1/2},
\end{equation}
where $k_1=(k_x+k_y-\pi)/\sqrt{2}$, $k_2=(k_x-k_y)/\sqrt{2}$, $v_{\perp}\propto \Delta_0$, and $v_F$ is the Fermi velocity. The presence of a vector potential shifts the quasiparticle spectrum:
\begin{equation}
\label{C28}
E(\vec{k},\vec{A}) = E(\vec{k}) -\frac{1}{c}\vec{j}(\vec{k})\cdot\vec{A},
\end{equation}
where $\vec{j}(\vec{k})$ is the current carried by the normal state quasiparticle with momentum 
$\vec{k}$. This equation is correct to first order in the vector potential $\vec{A}$.
In Ref. \cite{superfluid}, $\vec{j}$ is assumed to be $-e \vec{v}_F = -e\partial_{\vec{k}} \epsilon$.
The authors of Ref. \cite{Millisetal} have introduced the Fermi liquid correction to the quasiparticle current \cite{alpha}:
\begin{equation}
\label{C29}
\vec{j}(\vec{k}) = -e\alpha \vec{v}_F.
\end{equation}
$\alpha\sim 1$ is related to the appropriate Landau parameter(s).
The superfluid tensor is defined by: $\vec{j}_s^{\mu} = -\frac{e^2}{c} \frac{\rho^{\mu\nu}_s}{m}\vec{A}_{\nu}$. It can be decomposed as:
\begin{equation}
\rho^{\mu\nu}_s = \rho_s(T=0)\delta_{\mu\nu} - \rho^{\mu\nu}_n.
\end{equation}
where $\rho_s(T=0)$ is directly measured by the weight of the Drude peak in the normal phase and by the London penetration depth in the superconducting phase. Experiments \cite{Uemura} report that
$\rho_s(T=0)\propto \delta$ in agreement with the renormalized mean-field theory as well as the slave-boson
theory. On the other hand, $\rho^{\mu\nu}_n$ is given by the quasiparticle response to the $\vec{A}$ field.

This can be evaluated by writing the free energy (at finite T), as
\begin{equation}
F(\vec{A},T) =-k_B T \sum_{\vec{k},s} \ln(1+e^{-\beta E(\vec{k},\vec{A})}),
\end{equation}
and differentiating twice with respect to $\vec{A}$. Exploiting the definition of the current carried by the quasiparticles, we get:
\begin{equation}
\frac{\rho_n^{\mu\nu}}{m} = -2\sum_{\vec{k}} \frac{dE}{dA_{\mu}} \frac{dE}{dA_{\nu}} \frac{\partial f}{\partial E}.
\end{equation}
$f$ is the Fermi-Dirac distribution. Using Eq. (\ref{C28}) and (\ref{C29}), one gets:
\begin{equation}
\frac{\rho_s(T)}{m} = \frac{\rho_s(T=0)}{m} -aT,
\end{equation}
where \cite{Millisetal}:
\begin{equation}
a=\frac{2\ln 2}{\pi}\alpha^2 \frac{v_F}{v_{\perp}}.
\end{equation}
This is the result of Eq. (72). The interaction between quasiparticles has been neglected in (C.31); this is justified for small temperature and $\vec{A}$ because the density of states of quasiparticles vanishes
linearly with energy, in contrast to the case of a Fermi liquid. {\it For small doping $\delta$, the quasiparticle excitation is an effective way of destroying the superconducting state by driving $\rho_s$ to zero}. 

An approximate expression of $T_c$ then can be given by extrapolating $\rho_s=0$. Assuming that
the gap or $v_{\perp}$ is independent of $\delta$ for small $\delta$, then one predicts that $T_c$ follows $\rho_s(T=0)\propto \delta$. This agres with the Uemura {\it et al.} plot \cite{Uemura}.
 It is relevant to observe that this scenario for $T_c\propto \delta$
is independent of the mechanism that produces superconductivity (Bose condensation, Kosterlitz-Thouless transition,...); it just relies on the existence of a gap with d-wave symmetry in the
superconducting state, thus quasiparticles proliferate at finite temperature. The proximity to the Mott
state appears through $\rho_s(T=0)\propto \delta$.  

In the slave-boson approach, the electron is decomposed into a fermion and a boson. The charge
is carried by $\delta$ bosons. The difficulty is that at the mean-field level, the superconducting state
is described by a condensation of bosons and the quasiparticle dispersion is given by the fermion
(spinon) dispersion. Since the vector potential $\vec{A}$ couples directly only to the bosons, the shift
in the quasiparticle spectrum is reduced and $\alpha\ll 1$. In the original formulation, $\alpha=\delta$.
Within the SU(2) slave-boson approach, by invoking the spin-charge recombination mechanism (through the ladder diagram), Wen and Lee have shown that it is maybe possible to find 
$\rho_s(T=0)\sim \delta$ whereas $\alpha\sim 1$ is large. (This experimental fact may also be consistent with the fact that the Fermi arcs only become superconducting below $T_c$ in the underdoped regime.)

\section{Optical Lattices}

\subsection{Fermionic Atom}

It should be noted that the ground state electronic structure of alkali atoms is relatively simple: all electrons but one occupy closed shells, and the remaining one is in an s orbital in a higher shell. The electrons have no angular momentum, and the main coupling between the nuclear spin and the electron in the absence of magnetic field is the hyperfine interaction. The quantum number for the total spin $F=I\pm 1/2$, where $I$ represents the nuclear spin, behaves according to the usual rules for addition of angular momentum. The splitting between the levels $F=I+1/2$ and $F=I-1/2$ is given by $\delta_f=(I+1/2)w$ where $w$ is defined from the hyperfine interaction $H_{hf}=w\vec{I}.\vec{S}$. For example, $^{40}$K has a nuclear spin $I=4$; it is instructive to observe that the experiments performed in Refs. \cite{Esslinger,Blochnew} have been done with two magnetic sub-levels of the $F=9/2$ hyperfine manifold. In the case of $^6$Li, the nuclear spin obeys $I=1$ and therefore the total spin can take the two allowed values $F=1/2$ and $F=3/2$; experiments can be, {\it e.g.}, realized in the two lowest hyperfine states. 

\subsection{Standing wave}

In this Appendix, we briefly remind how two counterpropagating laser beams produce a 1D standing wave. First, when an atom is subject to an electric field $\vec{E}(\vec{x},t)$ this gives rise to a dipole type Hamiltonian $H_{dip}=-\vec{\mu}.\vec{E}(\vec{x},t)$ +h.c.; $\mu$ is the dipole operator of the atom. The electric dipole moment is proportional to the electric field: $\langle \vec{\mu} \rangle \sim \alpha \vec{E}$. The energy of an atom in an electric field may be calculated using perturbation theory and a change in energy due to the electric field is $\sim -\alpha |\vec{E}|^2$. The motion of an atom is thus governed
by the Hamiltonian $H=\vec{p}^2/2m +V(\vec{x})$ and the pseudo-potential $V(\vec{x})\propto |E(\vec{x},t)|^2$. It is interesting to remind that the polarizabilities of alkali atoms are larger than that of hydrogen by factors with range from 30 to 90.

A 1D standing wave can be produced out of two running counter-propagating waves with the same intensity, as follows. Let $(\hat{x}_1,\hat{x}_2,\hat{x}_3)$ be three unit vectors pointing along $x,y,z$
direction, respectively. The electric field of the two waves is given by, say, $\vec{E}_1 \propto e^{ikx} \hat{x}_3$ and $\vec{E}_2 \propto e^{-ikx} \hat{x}_3$. The sum of the two electric fields is $\vec{E}_1+\vec{E}_2 \propto \cos(kx) \hat{x}_3$, and this produces a standing wave in $x$ direction; this creates an optical
potential $V(x)\propto \cos^2(kx)$ in one dimension with periodicity $\lambda/2$, where $\lambda=2\pi/k$ is the wave length. Using one (two) further pair(s) of laser beams propagating in $y$ ($y$ and $z$)  
direction(s), a 2D (3D) periodic trapping then can be realized. The depth of the lattice in each direction is controlled by the laser intensity.

\subsection{XXZ model for the plaquette system at 3/8 filling}

Let us consider two plaquettes at sites $R$ and $R+1$. The effective XY term is given by exchanging
$|4_R,2_{R+1}\rangle \leftrightarrow |2_R,4_{R+1}\rangle$ throough tunneling to the virtual state $|3_R,3_{R+1}\rangle$; we have used the notations $|2 \rangle = |\Psi_2 \rangle$ and $|4\rangle = |\Psi_4\rangle$. This results in the XY term: the prefactor for this process is proportional to $g^2 t'^{2}/\Delta_b$
and the parameter $g$ is essentially independent of interactions. In the dilute limit --- with either a few
$|\Psi_2\rangle$ or $|\Psi_4\rangle$ states --- the dominant term is this XY process resulting in d-wave superfluidity. Close to the 3/8 filling, one also gets an Ising exchange term. However, the prefactor is not
exactly equal to $g^2 t'^{2}/\Delta_b$. For example, one also has to take into account {\it ferromagnetic} interactions of the form $|4_{R} 4_{R+1}\rangle \leftrightarrow |4_R 4_{R+1}\rangle$ mediated by the $|5_R,3_{R+1}\rangle$ state, and $|2_R,2_{R+1}\rangle \leftrightarrow |2_R,2_{R+1}\rangle$ mediated by $|3_R,1_{R+1}\rangle$. Those processes result in a non-monotonous behavior of the coupling $g_z$ \cite{Rey}, as
a function $U$.  For quite weak interactions, one expects $g_z<g$. At a general level, d-wave superfluidity requires that $\Delta_b>0$ and that $g\gg g_z$; this implies $0\ll U<2.7t$ \cite{Rey}.


\begin{thebibliography}{99}

\bibitem{BM}
J. G. Bednorz and K. A. M\" uller, Z. Phys. B {\bf 64}, 189 (1986).

\bibitem{Anderson}
P. W. Anderson, Science {\bf 235}, 1196 (1987).

\bibitem{Mott}
N. F. Mott, Prof. Phys. Soc. (London) A {\bf 62}, 416 (1949).

\bibitem{Affleck}
I. Affleck and J. B. Marston, Phys. Rev. B {\bf 37}, 3774 (1988); Phys. Rev. B {\bf 39}, 11538 (1989).

\bibitem{phase1}
D. A. Wollmann, D. J. van Harlingen, W. C. Lee, D. M. Ginsberg, and A. J. Leggett, Phys. Rev. Lett. 
{\bf 71}, 2134 (1993).

\bibitem{phase2}
C. C. Tsuei, J. R. Kirtley, C. C. Chi, L. S. Yu-Jahnes, A. Gupta, T. Shaw, J. Z. Sun, and M. B. Ketchen, 
Phys. Rev. Lett. {\bf 73}, 593 (1994).

\bibitem{Davis}
See, {\it e.g.}, K. McElroy {\it et al.}, Phys. Rev. Lett. {\bf 94}, 197005 (2005).

\bibitem{BCS}
J. Bardeen, L. N. Cooper, and J. R. Schrieffer, Phys. Rev. {\bf 108}, 1175 (1957).

\bibitem{Alloul}
H. Alloul, T. Ohno, and P. Mendels, Phys. Rev. Lett. {\bf 63}, 1700 (1989).

\bibitem{Johnston}
D. C. Johnston, Phys. Rev. Lett. {\bf 62}, 957 (1989).

\bibitem{Andersonetal}
P. W. Anderson, P. A. Lee, M. Randeria, T. M. Rice, N. Trivedi, and F. C. Zhang, J. Phys. Condens. Matter
{\bf 16}, R755 (2004).

\bibitem{Zhang}
F. C. Zhang, C. Gros, T. M. Rice, and H. Shiba, Supercond. Sci. Technol. {\bf 1}, 36 (1988).

\bibitem{Ogata}
M. Ogata and H. Fukuyama, Rep. Prog. Phys. {\bf 71}, 036501 (2008).

\bibitem{Lee}
P. A. Lee, N. Nagaosa, and X. G. Wen, Rev. Mod. Phys. {\bf 78}, 17 (2006).

\bibitem{Norman}
M. R. Norman, H. Ding, M. Randeria, J. C. Campuzano, T. Yokoya, T. Takeuchi, T. Takahashi, T. Mochiku, K. Kadowaki, P. Guptasarma, and D. G. Hinks, Nature {\bf 392}, 157 (1998).

\bibitem{Kapitulnik}
D. S. Marshall, D. S. Dessau, A. G. Loeser, C. H. Park, A. V. Matsuura, J. N. Eckstein, I. Bozovic, P. Fournier, A. Kapitulnik, and Z.-X. Shen, Phys. Rev. Lett. {\bf 76}, 4841 (1996).

\bibitem{Kallin}
M. R. Norman, D. Pines, and C. Kallin, Adv. Phys. {\bf 54}, 715 (2005).

\bibitem{Kanigel}
A. Kanigel {\it et al.}, Nature Physics {\bf 2}, 447-451 (2006).

\bibitem{Ong}
Y. Wang, L. Li, and N. P. Ong, Phys. Rev. B {\bf 73}, 024510 (2006).

\bibitem{Timusk}
T. Timusk and B. Statt, Rep. Prog. Phys. {\bf 62}, 61 (1999). 

\bibitem{ladder1}
K. Kojima, A. Keren, G. M. Luke, B. Nachumi, W. D.  Wu, Y. J. Uemure, M. Azuma, and M. Takano,
Phys. Rev. Lett. {\bf 74}, 2812 (1995).

\bibitem{ladder2}
M. Uehara, T. Nagata, J. Akimitsu, H. Takahashi, N. Nori, and K. J. Kinoshita, J. Phys. Soc. Jpn. {\bf 65}, 2764 (1996).

\bibitem{Dagotto}
E. Dagotto and T. M. Rice, Science {\bf 271}, 618 (1996). 

\bibitem{largeU}
E. Dagotto, J. Riera, and D. J. Scalapino, Phys. Rev. B {\bf 45},  5744 (1992); T. M. Rice, S. Gopalan, and
M. Sigrist, Europhys. Lett. {\bf 23}, 445 (1993).

\bibitem{smallU}
M. Fabrizio, A. Parola, and E. Tosatti, Phys. Rev. B {\bf 46}, 3159 (1992); M. fabrizio, Phys. Rev. B {\bf 48}, 15838 (1993); D. V. Khveshchenko and T. M. Rice, Phys. Rev. B {\bf 50}, 252 (1994); H. J. Schulz,
Phys. Rev. B {\bf 53}, R2959 (1996).

\bibitem{Lin}
H. Lin, L. Balents, and M. Fisher, Phys. Rev. B {\bf 58}, 1794 (1998).

\bibitem{Leon}
 L. Balents and M. Fisher, Phys. Rev. B {\bf 53}, 12133 (1996).

\bibitem{UrsKaryn}
U. Ledermann and K. Le Hur, Phys. Rev. B {\bf 61}, 2497 (2000).

\bibitem{Frischmuth}
B. Frischmuth, S. Haas, G. Sierra, and T. M. Rice, Phys. Rev. B {\bf 55}, R3340 (1997).

\bibitem{UKM}
U. Ledermann, K. Le Hur, and T. M. Rice, Phys. Rev. B {\bf 62}, 16383 (2000).

\bibitem{Azuma}
M. Azuma, Z. Hiroi, M. Takano, K. Ishida, and Y. Kitaoka, Phys. Rev. Lett. {\bf 73}, 3463 (1994).

\bibitem{Maurice}
T. M. Rice, S. Haas, M. Sigrist, and F. C. Zhang, Phys. Rev. B {\bf 56}, 14655 (1997).

\bibitem{Scalapino}
S. R. White and D. J. Scalapino, Phys. Rev. B {\bf 57}, 3031 (1998).

\bibitem{JohnKaryn}
J. Hopkinson and K. Le Hur, Phys. Rev. B {\bf 69}, 245105 (2004).

\bibitem{Sudip}
S. Chakravarty, Phys. Rev. Lett. {\bf 77}, 4446 (1996).

\bibitem{Urs}
U. Ledermann, Phys. Rev. B {\bf 64}, 235102 (2001).

\bibitem{KohnLuttinger}
W. Kohn and J. M. Luttinger, Phys. Rev. Lett. {\bf 15}, 524 (1965).

\bibitem{Varma}
K. Miyake, S. Schmitt-Rink, and C. M. Varma, Phys. Rev. B {\bf 34}, 6554 (1986).

\bibitem{Scalapino1}
D. J. Scalapino, E. Loh, and J. E. Hirsch, Phys. Rev. B {\bf 34}, 8190 (1986).

\bibitem{Scalapino2}
N. E. Bickers, D. J. Scalapino, and R. T. Scalettar, Int. J. Mod. Phys. B {\bf 1}, 687 (1987).

\bibitem{Schulz}
H. J. Schulz, Europhys. Lett. {\bf 4}, 609 (1987).

\bibitem{Dzya}
I. Dzyaloshinskii, Sov. Phys. JETP {\bf 66}, 848 (1987).

\bibitem{Lederer}
P. Lederer, G. Montambaux, and D. Poilblanc, J. Phys. (Paris) {\bf 48}, 1613 (1987).

\bibitem{Honerkampetal}
C. Honerkamp, M. Salmhofer, N. Furukawa, and T. M. Rice, Phys. Rev. B {\bf 63}, 035109 (2001); C. Honerkamp, M. Salmhofer, and T. M. Rice, Euro. Phys. J. B. {\bf 27}, 127 (2002).

\bibitem{Furukawa}
N. Furukawa, T. M. Rice, and M. Salmhofer, Phys. Rev. Lett. {\bf 81}, 3195 (1998).

\bibitem{Lauchli}
A. L\" auchli, C. Honerkamp, and T. M. Rice, Phys. Rev. Lett. {\bf 92}, 037006 (2004).

\bibitem{Chubukov}
A. V. Chubukov and D. K. Morr, Phys. Rep. {\bf 288}, 355 (1997).

\bibitem{zhangnew}
K.-Y. Yang, T. M. Rice, F.-C.  Zhang, Phys. Rev. B {\bf 73}, 174501 (2006).

\bibitem{Yangnew}
K.-Y. Yang, H. B. Yang, T. M. Rice, M. Sigrist, and F.-C. Zhang, arXiv:0812.3045.

\bibitem{KTM}
R. M. Konik, T. M. Rice, and A.M. Tsvelik, Phys. Rev. Lett. {\bf 96}, 086407 (2006).

\bibitem{WL}
X. G. Wen and P. A. Lee,  Phys. Rev. Lett. {\bf 76}, 503 (1996).

\bibitem{Ferrero}
M. Ferrero, P. S. Cornaglia, L. De Leo, O. Parcollet, G. Kotliar, and A. Georges, arXiv:0806.4383.

\bibitem{Georgesnew}
A. N. Rubtsov, M. I. Katsnelson, A. I. Lichtenstein, and A. Georges, arXiv:0810.3819.

\bibitem{Andre}
A.-M.S. Tremblay, B. Kyung, and D. S\' en\' echal, Low Temperature Physics, {\bf 32}, 424-451 (2006).

\bibitem{Bloch}
I. Bloch, J. Dalibard, and W. Zwerger, Rev. Mod. Phys. {\bf 80}, 885 (2008).

\bibitem{fermion}
N. Strohmaier {\it et al.}, Phys. Rev. Lett. {\bf 99}, 220601 (2007).

\bibitem{Esslinger}
R. J\" ordens, N. Strohmaier, K. G\" unter, H. Moritz, and T. Esslinger, Nature {\bf 455}, 204-207 (2008).

\bibitem{Blochnew}
U. Schneider, L. Hackerm\" uller, S. Will, Th. Best, and I. Bloch, T. A. Costi, R. W. Helmes, D. Rasch, and A. Rosch, Science {\bf 322}, 1520-1525 (2008).

\bibitem{ST}
D. J. Scalapino and S. A. Trugman, Philos. Mag. B {\bf 74}, 607 (1996). 

\bibitem{Altman}
E. Altman and A. Auerbach, Phys. Rev. B {\bf 65}, 104508 (2002).

\bibitem{Kivelson}
H. Yao, W. F. Tsai, and S. A. Kivelson, Phys. Rev. B {\bf 76}, 161104(R) (2007).

\bibitem{Rey}
A. M. Rey, R. Sensarma, S. Foelling, M. Greiner, E. Demler, and M. D. Lukin, arXiv:0806.0166.

\bibitem{reviewS}
For a review, consult D. J. Scalapino, cond-mat/9908287.

\bibitem{Emery}
V. J. Emery, Synthetic Metals, {\bf 13}, 21 (1986).

\bibitem{Scal}
M. Cyrot, Solid State Comm., {\bf 60}, 253 (1986).

\bibitem{KarynMott}
K. Le Hur, Phys. Rev. B {\bf 63}, 165110 (2001).

\bibitem{Gutzwiller}
M. C. Gutzwiller, Phys. Rev. Lett. {\bf 10}, 159 (1963).

\bibitem{triplet}
M. Reigrotzki, H. Tsunetsugu, and T. M. Rice, J. Phys. Condens. Matter {\bf 6}, 9235 (1994).

\bibitem{Sigrist}
S. Wessel, M. Indergrand, A. L\" auchli, U. Ledermann, and M. Sigrist, Phys. Rev. B {\bf 67}, 184517 (2003).

\bibitem{Schulzpair}
H. J. Schulz,  arXiv:cond-mat/9807328.

\bibitem{Tsvelik}
A. O. Gogolin, A. A. Nersesyan, and A. M. Tsvelik, 
{\it Bosonization and Strongly Correlated Systems} (Cambridge University Press, 1999).

\bibitem{Giamarchi}
T. Giamarchi, Quantum Physics in One Dimension (Clarendon Press, Oxford, 2004). 

\bibitem{KSC}
K. Le Hur, S. Vishveshwara, and C. Bena, Phys. Rev. B {\bf 77}, 041406(R) (2008).

\bibitem{Julia}
J. S. Meyer, K. A. Matveev, and A. I. Larkin, Phys. Rev. Lett. {\bf 98}, 126404 (2007).

\bibitem{Shankar2}
R. Shankar, Phys. Rev. Lett. {\bf 92}, 333 (1980); Phys. Rev. Lett. {\bf 46}, 379 (1981).

\bibitem{ZhangS}
S.-C. Zhang, Science {\bf 275}, 1089 (1997).

\bibitem{Schulzso6}
H. J. Schulz, arXiv:cond-mat/9808167.

\bibitem{Demler}
E. Demler, W. Hanke, and S.-C. Zhang, Rev. Mod. Phys. {\bf 76}, 909-974 (2004).

\bibitem{SZH}
D. Scalapino, S. C. Zhang, and W. Hanke, Phys. Rev. B {\bf 51}(1), 443 (1998).

\bibitem{Konik}
R. Konik and A. Ludwig, Phys. Rev. B {\bf 64}, 155112 (2001).

\bibitem{Oshikawa}
M. Yamanaka, M. Oshikawa, and I. Affleck, Phys. Rev. Lett. {\bf 79}, 1110 (1997).

\bibitem{KarynAndreev}
K. Le Hur, Phys. Rev. B {\bf 64}, R060502 (2001).

\bibitem{Scalapino22}
S. R. White and D. J. Scalapino, Phys. Rev. B {\bf 57}, 3031 (1998).

\bibitem{Scalapino3}
S. R. White and D. J. Scalapino, Phys. Rev. B {\bf 55}, 6504 (1997).

\bibitem{Siller}
T. Siller, M. Troyer, T. M. Rice, and S. R. White, Phys. Rev. B {\bf 65}, 205109 (2002).

\bibitem{Chang}
M.-S. Chang and I. Affleck, arXiv:0706.3727.

\bibitem{Shankar}
R. Shankar, Rev. Mod. Phys. {\bf 66}, 129 (1994).

\bibitem{Lin2}
H. Lin, L. Balents, and M. Fisher, Phys. Rev. B {\bf 56}, 6569 (1997).

\bibitem{Haldane}
F. D. M. Haldane in J. R. Schrieffer, R. A. Broglia (Eds.), Proceedings of the International School
of Physics ``Enrico Fermi'', course 121, 1992, North Holland, New York, 1994. 

\bibitem{Pavarini}
E. Pavarini, I. Dasgupta, T. Saha-Dasgupta, O. Jepsen, and O. Andersen, Phys. Rev. Lett. {\bf 87}, 047003 (2001).

\bibitem{Metzner}
W. Metzner, J. Reiss, and D. Rohue, phys. stat. sol. (b) {\bf 243} 46 (2006); arXiv:cond-mat/0509412.

\bibitem{Read}
N. Read and S. Sachdev, Phys. Rev. Lett. {\bf 66}, 1773 (1991).

\bibitem{Puttika}
W. O. Putikka, M. U. Luchini, and R. R. P. Singh, cond-mat/9803141.

\bibitem{Geshkenbein}
V. Geshkenbein, L. Ioffe, and A. Larkin, Phys. Rev. B {\bf 55}(5), 3173 (1997).

\bibitem{Loram}
J. M. Loram, K. A. Mirza, J. R. Cooper {\it et al.}, Phys. Rev. Lett. {\bf 71}, 1740 (1993).

\bibitem{Dama}
A. Damascelli, Z. Hussain, and Z.-X. Shen, Rev. Mod. Phys. {\bf 75}, 473 (2003).

\bibitem{Tanaka}
K. Tanaka {\it et al.}, Science {\bf 314}, 1910 (2006).

\bibitem{Andreev}
G. Deutscher, Nature {\bf 397}, 410 (1999).

\bibitem{Raman}
M. Le Tacon {\it et al.}, Nature Physics {\bf 2}, 537 (2006).

\bibitem{EK}
V. Emery and S. Kivelson, Nature {\bf 374}, 434 (1995).

\bibitem{Uemura}
Y. J. Uemura {\it et al.}, Phys. Rev. Lett. {\bf 62}, 2317 (1989).

\bibitem{superfluid}
P. A. Lee and X. G. Wen, Phys. Rev. Lett. {\bf 78}, 4111 (1997).

\bibitem{Hardy}
W. N. Hardy, D. A. Bonn, D. C. Morgan, R. Liang, and K. Zhang, Phys. Rev. Lett. {\bf 70}, 3999 (1993).

\bibitem{Millisetal}
A. J. Millis, S. Girvin, L. Ioffe, and A. Larkin, J. Phys. Chem. Solids {\bf 59}, 1742 (1998).

\bibitem{Taillefer}
L. Taillefer, B. Lussier, R. Gagnon, K. Behnia, and H. Aubin, Phys. Rev. Lett. {\bf 79}, 483 (1997).

\bibitem{Randeria}
A. Paramekanti and M. Randeria, Phys. Rev. B {\bf 66}, 214517 (2002).

\bibitem{Lemberger}
J. Stajic {\it et al.}, Phys. Rev. B {\bf 68}, 024520 (2003).

\bibitem{BZA}
G. Baskaran, Z. Zou, and P. W. Anderson, Solid State Commun. {\bf 63}, 973 (1987).

\bibitem{Gros}
C. Gros, R. Joynt, and T. M. Rice, Phys. Rev. B {\bf 36}, 381 (1986).

\bibitem{Hirsch}
J. E. Hirsch, Phys. Rev. Lett. {\bf 54}, 1317 (1985).

\bibitem{Paramekanti}
A. Paramekanti, M. Randeria, and N. Trivedi, Phys. Rev. Lett. {\bf 87}, 217002 (2001).

\bibitem{Trivedi}
N. Trivedi and D. M. Ceperley, Phys. Rev. B {\bf 40}, 2737 (1989).

\bibitem{Vollhardt}
D. Vollhardt, Rev. Mod. Phys. {\bf 56}, 99 (1984).

\bibitem{Monthoux}
P.  Monthoux, A. Balatsky, and D. Pines, Phys. Rev. Lett. {\bf 67}, 3449 (1991).

\bibitem{Johnson}
P. D. Johnson {\it et al.}, Phys. Rev. Lett. {\bf 87}, 177007 (2001).

\bibitem{Sawatzky}
H. Eskes, M. B. J. Meinders, and G. Sawatzky, Phys. Rev. Lett. {\bf 87}, 227001 (2001).

\bibitem{Feng}
D. L. Feng {\it et al.}, Science {\bf 289}, 277 (2000).

\bibitem{Ding}
H. Ding {\it et al.}, Nature {\bf 382}, 51 (1996).

\bibitem{Tranquada}
Q. Li, M. H\" ucker, G. D. Gu, A. M. Tsvelik, and J. M. Tranquada, Phys. Rev. Lett.
{\bf 99}, 067001 (2007).

\bibitem{Berg}
E. Berg {\it et al.}, Phys. Rev. Lett. {\bf 99}, 127003 (2007) and references therein.

\bibitem{Yangstripe}
K.-Y. Yang, W.-Q. Chen, T. M. Rice, M. Sigrist, and F.-C. Zhang, arXiv:0807.3789.

\bibitem{Afflecketal}
I. Affleck, Z. Zou, T. Hsu, and P. W. Anderson, Phys. Rev. B {\bf 38}, 745 (1988).

\bibitem{Chakravarty}
S. Chakravarty, R. B. Laughlin, D. Morr, and C. Nayak, Phys. Rev. B {\bf 63}, 94503 (2001).

\bibitem{Kaul}
R. Kaul, Y.-B. Kim, S. Sachdev, and T. Senthil, Nature Physics {\bf 4}, 28-31 (2008).

\bibitem{LSR}
J. M. Luttinger and J. C. Ward, Phys. Rev. {\bf 118}, 1417 (1960); J. M. Luttinger, Phys. Rev. {\bf 119}, 1153 (1960).

\bibitem{KarynGreen}
K. Le Hur, Phys. Rev. B {\bf 74}, 165104 (2006).

\bibitem{AGD}
A. A. Abrikosov, L. P. Gorkov, and I. E. Dzyaloshinski, {\it Methods in Quantum Field Theory in Statistical Physics}, edited by R. A. Silverman, revised edition, Dover, New York; I. E. Dzyaloshinskii, Phys. Rev. B {\bf 68}, 85113 (2003).

\bibitem{tsv}
A. M. Tsvelik, {\it Quantum Field Theory in Condensed Matter Physics}, CUP, 2003.

\bibitem{RoschLut}
A. Rosch, Eur. Phys. J. B {\bf 59}, 495 (2007).

\bibitem{ProustN}
N. Doiron-Leyraud {\it et al.}, Nature {\bf 447}, 565 (2007).

\bibitem{Yang}
H. B. Yang {\it et al.}, Nature {\bf 456}, 77 (2008).

\bibitem{Bascones}
B. Valenzuela and E. Bascones, Phys. Rev. Lett. {\bf 98}, 17704 (2007).

\bibitem{Kohsaka}
Y. Kohsaka {\it et al.}, Nature {\bf 454}, 1072 (2008).

\bibitem{Ng}
T.-K. Ng, Phys. Rev. B {\bf 71}, 172509 (2005).

\bibitem{norman}
M. R. Norman, M. Randeria, H. Ding, and J. C. Campuzano, Phys. Rev. B {\bf 57}, R11093 (1998).

\bibitem{FranzMillis}
M. Franz and A. J. Millis, Phys. Rev. B {\bf 58}, 14572 (1998).

\bibitem{Normanarcs}
M. R. Norman {\it et al.}, Phys. Rev. B {\bf 76}, 174501 (2007).

\bibitem{Abdel}
M. Abdel-Jawad {\it et al.}, Nature Physics {\bf 2}, 821 (2006); M. Abdel-Jawad {\it et al.}, Phys. Rev. Lett. {\bf 99}, 107002 (2007).

\bibitem{TailleferN}
L. Taillefer, Nature Physics, {\bf 2}, 810 (2006).

\bibitem{McKenzie}
A. P. MacKenzie {\it et al.}, Phys. Rev. B {\bf 53}, 5848 (1996).

\bibitem{Proust}
C. Proust {\it et al.}, Phys. Rev. Lett. {\bf 89}, 147003 (2002).

\bibitem{Vignolle}
B. Vignolle {\it et al.}, Nature {\bf 455}, 952 (2008).

\bibitem{Zanchi}
D. Zanchi and H. J. Schulz, Europhys. Lett. {\bf 44}, 235 (1997).

\bibitem{Halboth}
C. J. Halboth and W. Metzner, Phys. Rev. B {\bf 61}, 7364 (2000); Phys. Rev. Lett. {\bf 85}, 5162 (2000).

\bibitem{Zanchi2}
D. Zanchi, Europhys. Lett. {\bf 55}, 376 (2001).

\bibitem{Honerkamp}
C. Honerkamp, Eur. Phys. J. B {\bf 21}, 81 (2001).

\bibitem{Katanin}
A. A. Katanin and A. P. Kampf, Phys. Rev. Lett. {\bf 93}, 106406 (2004).

\bibitem{Rohe}
D. Rohe and W. Metzner, Phys. Rev. B {\bf 71}, 115116 (2005).

\bibitem{Ossadnik}
M. Ossadnik, C. Honerkamp, T.M. Rice, and M. Sigrist, arXiv:0805.3489.

\bibitem{Nick}
P. A. Lee and N. Read, Phys. Rev. Lett. {\bf 58}, 2691 (1987).

\bibitem{Fermions1}
B. deMarco and D. S. Jin, Science {\bf 285}, 1703 (1999).

\bibitem{Fermions2}
A. G. Truscott, K. E. Strecker, W. I. McAlexander, G. B. Partridge, and R. G. Hulet, Science {\bf 291}, 2570 (2001).

\bibitem{Fermions3}
F. Schreck, L. Khaykovich, K. L. Corwin, G. Ferrari, T. Bourdel, J. Cubizolles, and C. Salomon,
Phys. Rev. Lett. {\bf 87}, 080403 (2001).

\bibitem{Fermions4}
Z. Hadzibabic, C. A. Stan, K. Dieckmann, S. Gupta, M. W. Zwierlein, A. G\" orlitz, and W. Ketterle, Phys. Rev. Lett. {\bf 88}, 160401 (2002).

\bibitem{Fermion1}
M. Greiner, C. A. Regal, and D. S. Jin, Nature {\bf 426}, 537 (2003).

\bibitem{Ketterle}
J. K. Chin, D. E. Miller, Y. Liu, C. Stan, W. Setiawan, C. Sanner, K. Xu, and W. Ketterle, Nature {\bf 443}, 961 (2006).

\bibitem{Walter}
W. Hofstetter, J. I. Cirac, P. Zoller, E. Demler, and M. D. Lukin, Phys. Rev. Lett. {\bf 89}, 220407 (2002).

\bibitem{Blochtrap}
I. Bloch, Nature Physics {\bf 1}, 24 (2005).

\bibitem{Jaksch}
D. Jaksch and P. Zoller, cond-mat/0410614.

\bibitem{Georges}
F. Werner, O. Parcollet, A. Georges, and S. R. Hassan, Phys. Rev. Lett. {\bf 95}, 056401 (2005).

\bibitem{kohl}
M. K\" ohl, Phys. Rev. A {\bf 73}, 031601(R) (2006); T. St\" oferle, H. Moritz, K. G\" unter, M. K\" ohl, and T. Esslinger, Phys. Rev. Lett. {\bf 96}, 030401 (2006).

\bibitem{Huse}
C. J. M. Mathy and D. Huse, arXiv:0805.1507 and arXiv:0903.0108.

\bibitem{MC}
R. Staudt {\it et al.}, Eur. Phys. J. B {\bf 17}, 411 (2000).

\bibitem{Achim}
A. Rosch, private discussion.

\bibitem{Georgeslight}
For a review, consult A. Georges, arXiv:cond-mat/0702122.

\bibitem{BlochS}
S. Trotzky {\it et al.}, Science {\bf 319}, 295 (2008).

\bibitem{GKL}
G. A. Fiete, K. Le Hur, and L. Balents, Phys. Rev. B {\bf 72}, 125416 (2005).

\bibitem{G}
G. A. Fiete, Rev. Mod. Phys. {\bf 79}, 801 (2007).

\bibitem{CK}
S. Chakravarty and S. A. Kivelson, Phys. Rev. B {\bf 64}, 064511 (2001).

\bibitem{Schumann}
R. Schumann, Annalen der Physik {\bf 11}, 49 (2001).

\bibitem{Paredes}
B. Paredes and I. Bloch, arXiv:0711.3796.

\bibitem{LiebWu}
E. H. Lieb and F. Y. Wu, Phys. Rev. Lett. {\bf 20}, 1145 (1968).

\bibitem{Ki}
H. Yao, W.-F. Tsai, and S. A. Kivelson, Phys. Rev. B {\bf 76}, 161104 (2007).

\bibitem{Stewart}
J. T. Stewart, J. P. Gaebler, and D. S. Jin, Nature {\bf 454}, 744 (2008).

\bibitem{Kurn}
D. M.  Stamper-Kurn {\it et al.}, Phys. Rev. Lett. {\bf 83}, 2876 (1999).

\bibitem{Altman2}
E. Altman, E. Demler, and M. Lukin, Phys. Rev. A {\bf 70}, 013603 (2004).

\bibitem{Blochp}
S. F\" olling {\it et al.}, Nature {\bf 434}, 481 (2005).

\bibitem{Greiner}
M. Greiner {\it et al.}, Nature {\bf 537}, 426 (2003).

\bibitem{Tinkham}
M. Tinkham, {\it Introduction to Superconductivity}, 2nd Ed., McGraw-Hill, NY, 1996.

\bibitem{Sarma}
M. R. Peterson, C. Zhang, S. Tewari, and S. Das Sarma,  Phys. Rev. Lett. {\bf 101}, 150406 (2008).

\bibitem{Ashwin}
L. Mathey, E. Altman, and A. Vishwanath, Phys. Rev. Lett. {\bf 100}, 240401 (2008).

\bibitem{Leggett}
A. J. Leggett, {\it Diatomic molecules and Cooper pairs, in Modern Trends in the Theory of Condensed Matter}, edited by A. Pekalski and J. Przystawa (Springer-Verlag, Berlin), 1980, p. 13-27.

\bibitem{and}
D. M. Eagles, Phys. Rev. {\bf 186}, 456 (1969).

\bibitem{NSR}
P. Nozi\` eres and S. Schmitt-Rink, J. Low Temp. Phys. {\bf 59}, 195 (1985).

\bibitem{Levin}
K. Levin and Q. Chen, arXix:cond-mat/0610006.

\bibitem{Levina}
Y. He {\it et al.}, Phys. Rev. B. {\bf 76}, 224516 (2007).

\bibitem{Levin3}
For a review, consult: Q. Chen, J. Stajic, S. Tan, and K. Levin, arXiv:cond-mat/0404274.

\bibitem{Thomas}
J. Kinast {\it et al.}, Science {\bf 307}, 1296 (2005).

\bibitem{Monthoux2}
P. Monthoux, D. Pines, and G. G. Lonzarich, Nature {\bf 450}, 1177 (2007).

\bibitem{Lohneysen}
H. v. L\" ohneysen, A. Rosch, M. Vojta, and P. W\" olfle, Rev. Mod. Phys. {\bf 79}, 1015 (2007).

\bibitem{Carbotte}
F. Marsiglio and J. Carbotte, {\it The Physics of Conventional and Unconventional Superconductors} edited by K.H. Bennemann and J.B. Ketterson (Springer-Verlag), 123 pages.

\bibitem{Jarrell}
Th. Maier, M. Jarrell, Th. Pruschke, and J. Keller, Phys. Rev. Lett. {\bf 85}, 1524 (2000).

\bibitem{Amit}
Y. Lubashevsky and A. Keren, Phys. Rev. B {\bf 78}, 020505(R) (2008).

\bibitem{Altmansu}
For a recent reference, consult L. Goren and E. Altman, arXiv:0810.2814, and references therein.

\bibitem{Yulli}
O. Yulli {\it et al.}, Phys. Rev. Lett. {\bf 101}, 057005 (2008).

\bibitem{singlet}
F. C. Zhang and T. M. Rice, Phys. Rev. B {\bf 37}, 3759 (1988).

\bibitem{Kondo}
T. Kondo {\it et al.}, Nature {\bf 457}, 296 (2009).

\bibitem{Rasch}
A. Rosch {\it et al.}, arXiv:0809.0505.

\bibitem{Lee2}
P. A. Lee, Rep. Prog. Phys. {\bf 71}, 012501 (2008).

\bibitem{Barnes}
S. E. Barnes, J. Phys. F: Met. Phys. {\bf 6}, 1475 (1976).

\bibitem{Coleman}
P. Coleman, Phys. Rev. B {\bf 29}, 3035 (1984).

\bibitem{Kotliar}
G. Kotliar and J. Liu, Phys. Rev. B {\bf 38}, 5142 (1988).

\bibitem{Suzumura}
Y. Suzumura {\it et al.}, J. Phys. Soc. Jpn. {\bf 57}, 2768 (1988). 

\bibitem{Ribeiro}
T. C. Ribeiro and X.-G. Wen, Phys. Rev. Lett. {\bf 95}, 057001 (2005).

\bibitem{Bieri}
S. Bieri and D. A. Ivanov, arXiv:0809.5230.

\bibitem{WenLee}
X.-G. Wen and P. A. Lee, Phys. Rev. Lett. {\bf 80}, 2193 (1998).

\bibitem{alpha}
A. I. Larkin, Sov. Phys. JETP {\bf 46}, 1478 (1965).


\end{thebibliography}
\end{document}